\documentclass[11pt,twoside,a4paper,cmspaper,final,collab]{cms-tdr}
\def\svnVersion{44e20a6-D}\def\svnDate{2026/09/09}\def\cmsCernNoTag{CERN-EP-2026-200}\def\cmsCernDate{\today}\def\cmsMessage{Submitted to the Journal of High Energy Physics}
\begin{document}\cmsNoteHeader{SUS-23-005}

\cmsNoteHeader{SUS-23-005}

\newcommand{\pa}{\ensuremath{\HepParticle{a}{}{}\xspace}}
\newcommand{\taum}{\ensuremath{\PGt_{\PGm}}\xspace}
\newcommand{\taue}{\ensuremath{\PGt_{\Pe}}\xspace}
\newcommand{\taul}{\ensuremath{\PGt_{\Pell}}\xspace}
\newcommand{\DEEPDITAU}{\textsc{DeepDiTau}\xspace}
\providecommand{\NA}{\ensuremath{\text{---}}}

\providecommand{\cmsTable}[1]{\resizebox{\textwidth}{!}{#1}}

\title{Search for a light pseudoscalar Higgs boson in final states with boosted muon pairs and tau lepton pairs in proton-proton collisions at  \texorpdfstring{$\sqrt{s} = 13\TeV$}{sqrt(s) = 13 TeV}}

\abstract{
A search is presented for a light pseudoscalar Higgs boson, \pa, using LHC proton-proton data at $\sqrt{s} = 13\TeV$ collected by the CMS experiment from 2016 to 2018, corresponding to an integrated luminosity of 138\fbinv. The analysis targets the process $\PH \to \pa\pa \to \PGm\PGm\PGt\PGt$, where the scalar \PH can be either the 125\GeV Higgs boson or a heavier state. Because of the large mass difference between \PH and \pa, the two tau leptons are highly Lorentz boosted and collimated, often failing standard CMS tau lepton reconstruction. To address this, dedicated ditau reconstruction techniques, including a deep neural network, are developed to improve the efficiency for events in which at least one tau lepton decays to hadrons. No significant excess over the standard model (SM) background is observed. Model-independent upper limits at the 95\% confidence level are set on the product of the Higgs boson production cross section and branching fraction, divided by the SM cross section. For a Higgs boson mass ($m_{\PH}$) equal to 125\GeV, these limits range from $5 \times 10^{-5}$ to $2.7 \times 10^{-4}$ for pseudoscalar masses between 3.6 and 21\GeV. For the first time using LHC data, upper limits are also obtained in final states with tau leptons for a hypothetical scalar boson with a mass up to 1\TeV. For $m_{\PH} = 1\TeV$, the limits range from $3.5 \times 10^{-4}$ to $1.5 \times 10^{-3}$ for pseudoscalar masses between 3.6 and 50\GeV. In addition, limits are set for SM extensions involving two-Higgs-doublet plus singlet models for $m_{\PH} = 125\GeV$, extending previous LHC results.
}

\hypersetup{
pdfauthor={CMS Collaboration},
pdftitle={Search for a light pseudoscalar Higgs boson in final states with boosted muon pairs and tau lepton pairs in proton-proton collisions at sqrt(s) = 13 TeV},
pdfsubject={CMS},
pdfkeywords={CMS, Higgs, BSM}}

\maketitle 

\section{Introduction}
Studies of the properties of the standard model (SM) Higgs boson can be used to constrain models that extend the Higgs sector beyond the SM~\cite{CMS:2024eka, CMS:2024bua, CMS:2023vzh, ATLAS:2023yqk, CMS:2023gjz, CMS:2022wpo, ATLAS:2024yzu, ATLAS:2024wfv, ATLAS:2024gth, ATLAS:2024vxc, ATLAS:2023qpu, ATLAS:2023owm, ATLAS:2023oaq, ATLAS:2023pwa,Curtin:2013fra}.
Examples include the two-Higgs-doublet model (2HDM)~\cite{Branco:2011iw} with a scalar singlet (2HDM+S)~\cite{Curtin:2013fra} and the next-to-minimal supersymmetric SM (NMSSM)~\cite{Ellwanger:2009dp}.
Specialized techniques are required to investigate models in which Higgs boson decays occur via undetected modes, defined as modes that are not probed in standard analyses focusing either on decays to SM particle pairs or on invisible decays that leave no trace in the detector.
The most recent combination using all CMS data at 13\TeV~\cite{CMS:2022dwd,CMS:2026nce} considers models in which the Higgs sector contains only doublets and singlets, with couplings otherwise allowed to vary relative to their SM values.
Those analyses set a 95\% confidence level (\CL) upper limit on the Higgs boson branching fraction~($\mathcal{B}$) to undetected modes of $<$ 0.16--0.25.
Among the most prominent possibilities for Higgs boson decays via modes not predicted by the SM~\cite{Dermisek:2005ar, Dermisek:2006wr} are decays of the type $\PH\to \pa\pa$~\cite{Chang:2008cw}, where \PH is the SM Higgs boson and \pa is a lighter neutral pseudoscalar boson.
Similar decays of heavier scalar states are also predicted in extended Higgs sector models.
This process arises naturally in the NMSSM~\cite{Ellwanger:2009dp, Dermisek:2005ar, Dermisek:2006wr}, in 2HDM models~\cite{King:2012tr, Celis:2013rcs, Grinstein:2013npa, Coleppa:2013dya, Chen:2013rba, Craig:2013hca, Wang:2013sha, Curtin:2013fra, Cao:2013gba, Christensen:2013dra, Cerdeno:2013cz, Chalons:2012qe, Ahriche:2013vqa, Baglio:2014nea, Dumont:2014wha}, and in some models with axion-like particles that couple to the SM Higgs boson~\cite{Davoudiasl:2021haa,Bauer:2017ris}.

Searches using the full $\sqrt{s} = 13\TeV$ data set~\cite{ATLAS:2021hbr,CMS:2024zfv,CMS:2022fyt,CMS:2022xxa,CMS:2024uru,CMS:2024jyb} place stringent upper limits on the product of the cross section and branching fraction for the process $\PH\to\pa\pa$ to four-fermion final states.
Some of these searches are interpreted in the context of 2HDM+S~\cite{CMS:2024uru,CMS:2024zfv}, allowing constraints to be placed on $\mathcal{B}(\PH\to\pa\pa)$.
There are four types of 2HDM, and thus of 2HDM+S, distinguished by the couplings of SM fermions to the two doublet fields ($\Phi_1$ and $\Phi_2$) and by the dependence of their properties on the ratio of the vacuum expectation values of the two Higgs doublets (\tanb)~\cite{Curtin:2013fra}.
In particular, the NMSSM corresponds to the Type-II 2HDM+S, while in the Type-III 2HDM+S only the charged leptons couple to $\Phi_1$, leading to enhanced rates for pseudoscalar decays to leptonic final states, especially at large values of \tanb.
For the Type-II (NMSSM-like) 2HDM+S model at \tanb= 2, the branching fraction $\mathcal{B}(\PH\to\pa\pa)$ is constrained to be less than 40--100\% for $m_{\pa}$ = 20--60\GeV using the $\PQb\PQb\PQb\PQb$ final state~\cite{CMS:2024zfv} in associated production with a \PW boson, and less than 5--20\% for $m_{\pa}$ = 12--60\GeV using the $\PQb\PQb \PGt\PGt$ and $\PQb\PQb \PGm\PGm$ final states~\cite{CMS:2024uru}.
In the Type-III model at the same value of \tanb = 2, where pseudoscalar decays to leptons are favored, similarly strong upper limits are obtained by searches in the $\PQb\PQb\PGt\PGt$ and $\PQb\PQb\PGm\PGm$ final states, while the $\PQb\PQb\PQb\PQb$ search in the associated production with a \PW boson is not competitive~\cite{Summary2HDMSRun2}.

\begin{figure}[htbp]
  \centering
  \includegraphics[width=0.5\textwidth]{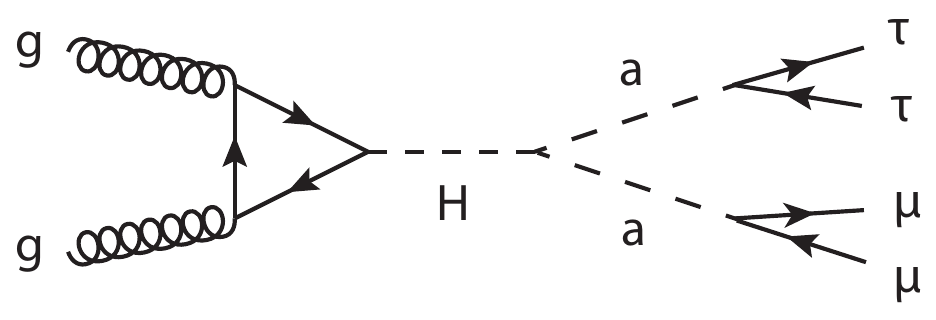}
  \caption{Feynman diagram of a Higgs boson produced via gluon fusion and decaying into two pseudoscalar bosons, one of which subsequently decays into two muons and the other decays into two tau leptons.}
  \label{fig:diagram_haa}
\end{figure}

The analysis presented here extends the CMS 8\TeV search~\cite{Khachatryan:2017mnf} and the 13\TeV search~\cite{CMS:2020ffa}, based on an integrated luminosity of 35.9\fbinv, by considering the dimuon plus ditau ($\PGm\PGm\PGt\PGt$) final state arising from $\PH\to\pa\pa\to\PGm\PGm\PGt\PGt$, where the Higgs boson is produced via the dominant gluon-gluon fusion (ggF, as shown in Fig.~\ref{fig:diagram_haa}) and vector boson fusion (VBF) modes~\cite{LHCHiggsCrossSectionWorkingGroup:2016ypw}.
The presence of additional neutral scalar bosons, also denoted \PH, with masses $m_{\PH}$ exceeding 125\GeV, is motivated in generic 2HDM+S scenarios~\cite{Curtin:2013fra,Bernon:2015qea}.
Such an \PH boson generically has a large branching fraction to any kinematically accessible pair of lighter bosons~\cite{Bernon:2014nxa,Bernon:2015qea}.
To test the existence of a heavy neutral scalar in addition to the SM Higgs boson, this paper also considers $\PH\to\pa\pa\to\PGm\PGm\PGt\PGt$ decays with $m_{\PH}$ ranging from 250--1000\GeV.  
This analysis targets pseudoscalar bosons with masses that are small when compared to the scalar mass ($m_{\PH}$), as shown in Table~\ref{tab:a_mass_h_mass}.
At approximately twice the tau lepton mass, the minimum pseudoscalar mass ($m_{\pa}$) considered is the lowest value kinematically accessible via the $\pa\to\PGt\PGt$ decay.
The upper end of the tested $m_{\pa}$ range depends on the $m_{\PH}$ value to ensure that the pseudoscalar decay products are boosted and that the analysis acceptance is thereby maximized. Additionally, requiring $m_{\pa} \leq 50\GeV$ avoids large Drell--Yan backgrounds.

The large Lorentz boost of the \pa boson causes its decay products to overlap.
Different techniques are employed to reconstruct the collimated tau lepton pair, depending on the final state.
When each \PGt decays to a muon or electron and neutrinos, the standard CMS muon or electron identification provides sufficient efficiency and sensitivity for the boosted topology.
The same boosted tau lepton pair reconstruction technique as in Ref.~\cite{CMS:2020ffa} is used to target the decay $\pa\to\taum\tauh$, where \taum denotes a tau lepton decaying to a muon and neutrinos, and \tauh denotes a tau lepton decaying to one or more hadrons and a neutrino.
In this analysis, the technique is extended to tau lepton pairs in which one tau lepton decays to an electron (\taue) and neutrinos and the other decays hadronically, $\pa\to\taue\tauh$.
Additionally, a deep neural network is developed to identify a collimated tau lepton pair from the decay $\pa\to\tauh\tauh$, in which each tau lepton decays into one or more hadrons and a neutrino. This approach addresses inefficiencies in the standard hadronic \PGt reconstruction in the boosted regime.

The effectiveness of these new techniques enables, for the first time, a search for the decay of an SM Higgs boson or a heavier neutral scalar to $\pa\pa$ in the $\PGm\PGm\PGt\PGt$ final state at low $m_{\pa}$ that combines information from the $\taum\taue$, $\taum\tauh$, $\taue\tauh$, and $\tauh\tauh$ channels, covering 94\% of the theoretically accessible $\pa\to\PGt\PGt$ decays.
The search is performed using an unbinned parameterized maximum likelihood fit to the signal and background contributions of the two-dimensional distribution of the $\PGm\PGm$ invariant mass ($m_{\PGm\PGm}$) and the four-body visible mass ($m_{\PGm\PGm\PGt\PGt}$).

\begin{table}[htbp]
  \centering
  \topcaption{Values of the scalar mass $m_{\PH}$ and pseudoscalar mass $m_{\pa}$ considered in the search for $\PH\to\pa\pa\to\PGm\PGm\PGt\PGt$.
  The third column lists the generated pseudoscalar mass points or ranges; ranges are given together with the step size in braces.
  For $m_{\PH}$ = 125\GeV, the lightest neutral scalar corresponds to the SM Higgs boson.}
  \begin{tabular}{ccc}
    $m_{\PH}$ (\GeVns) & $m_{\pa}$ search range (\GeVns) & Generated $m_{\pa}$ \{step size\} (\GeVns) \\
    \hline
    125   & 3.6--21  & 3.6, 4--21 \{1\} \\
    250   & 3.6--25  &   5--20 \{5\} \\
    500   & 3.6--30  &   5--20 \{5\} \\
    750   & 3.6--35  &  10--30 \{5\} \\
    1000  & 3.6--50  &  10--50 \{10\} \\
  \end{tabular}
  \label{tab:a_mass_h_mass}
\end{table}

This paper is organized as follows.
A brief description of the CMS detector is given in Section~\ref{sec:detector}.
Section~\ref{sec:samples} summarizes the data set and simulated signal samples used in this analysis.
Section~\ref{sec:reco} describes the object identification algorithms, including those for the $\taum\tauh$ and $\taue\tauh$ pairs.
The performance of the $\tauh\tauh$ identification is detailed in Section~\ref{sec:deepditau}.
Section~\ref{sec:selection} describes the event selection.
The background and signal models used in the two-dimensional unbinned fit are described in Section~\ref{sec:bkg}, and the treatment of systematic uncertainties is discussed in Section~\ref{sec:unc}.
The model-independent results, as well as their interpretation in the context of the four 2HDM+S types, are presented in Section~\ref{sec:results}.
The paper is summarized in Section~\ref{sec:summary}.  Tabulated results are provided in the HEPData record~\cite{hepdata} for this analysis.

\section{The CMS detector}
\label{sec:detector}

The central feature of the CMS apparatus is a superconducting solenoid with an internal diameter of 6\unit{m}, providing a magnetic field of 3.8\unit{T}.
Within the solenoid volume are a silicon pixel and strip tracker, a lead tungstate crystal electromagnetic calorimeter (ECAL), and a brass-and-scintillator hadron calorimeter, each composed of a barrel and two endcap sections.
Forward calorimeters extend the pseudorapidity ($\eta$) coverage provided by the barrel and endcap detectors.
Muons are measured in gas-ionization detectors embedded in the steel flux-return yoke outside the solenoid.
Events of interest are selected using a two-tiered trigger system~\cite{Khachatryan:2016bia}.
The first level, composed of custom hardware processors, uses information from the calorimeters and muon detectors to select events at a rate of around 100\unit{kHz} within a time interval of less than 4\mus~\cite{CMS:2020cmk}.
The second level, known as the high-level trigger, consists of a farm of processors running a version of the full event reconstruction software optimized for fast processing and reduces the event rate to around 1\unit{kHz} before data storage~\cite{Khachatryan:2016bia,CMS:2024aqx}.
A more detailed description of the CMS detector, together with a definition of the coordinate system used and the relevant kinematic variables, can be found in Refs.~\cite{Chatrchyan:2008zzk,CMS:2023gfb}.

\section{Data and signal simulation}
\label{sec:samples}

The data set used in this analysis consists of LHC proton-proton $(\Pp\Pp)$ collisions collected by the CMS detector in 2016, 2017, and 2018 at $\sqrt{s} = 13\TeV$, corresponding to integrated luminosities of 36.3, 41.5, and 59.8\fbinv, respectively.

The acceptance and reconstruction efficiency of the $\PH\to\pa\pa\to\PGm\PGm\PGt\PGt$ process are evaluated using simulated events.
The signal process is generated at leading order with \MGvATNLO version 2.6.7~\cite{mg_amcnlo}.
The \PYTHIA~8.230~\cite{Sjostrand:2014zea} generator is used for parton showering and hadronization, and the underlying event is simulated with the CP5 tune~\cite{CMS:2019csb}.
The NNPDF3.1 set~\cite{Ball:2017nwa} at next-to-next-to-leading order in perturbative quantum chromodynamics (QCD) is used to model the parton distribution functions (PDFs) for all processes.
Samples are generated for $m_{\pa}$ in the ranges listed in Table~\ref{tab:a_mass_h_mass}, with step sizes that depend on $m_{\PH}$.
The ggF Higgs boson production process is simulated for each sample, and the resulting signal yields are scaled to the sum of the expected yields from the ggF and VBF processes.
The VBF process contributes 8\% to the expected signal yield of the $m_{\PH}$ = 125\GeV hypothesis.
A correction to the signal acceptance calculated from the ggF simulations is derived from a coarse grid of VBF signal simulations.
The correction, which ranges from 0.5\% for $m_{\PH} = 125\GeV$ to 3.0\% for $m_{\PH} = 1000\GeV$, arises from small differences in the analysis acceptance between ggF and VBF events and is applied as a function of $m_{\PH}$ and $m_{\pa}$, with an uncertainty of 0.5\%.
This correction primarily stems from differences in the transverse momentum (\pt) spectra of the generated \PH and \pa bosons.
These differences have a negligible effect on the shapes of the reconstructed pseudoscalar mass distributions used to discriminate signal from background.
The vector boson and top quark associated Higgs boson production modes ($\PW\PH$, $\PZ\PH$, and $\PQt\PQt\PH$) account for 5.0\% or less of the total Higgs boson production, with the fraction decreasing as $m_{\PH}$ increases~\cite{LHCHiggsCrossSectionWorkingGroup:2016ypw}.
They do not significantly improve the sensitivity of this search because of their lower cross sections and reduced acceptance and are therefore not included.

For all processes, the detector response is simulated using a detailed description of the CMS detector based on the \GEANTfour package~\cite{GEANT}, and event reconstruction is performed with the same algorithms as those used for data.
The simulated samples include additional interactions per bunch crossing (pileup), and the events are weighted such that the distribution of the number of interactions per bunch crossing matches that measured in data, which has an average of about 27--38 interactions per bunch crossing in the 2016--2018 data-taking period.

\section{Object reconstruction}
\label{sec:reco}

Using information from all CMS subdetectors, the particle-flow (PF) algorithm aims to reconstruct and identify each individual particle in an event~\cite{Sirunyan:2017ulk}.
The particles are classified into mutually exclusive categories: charged hadrons, neutral hadrons, photons, muons, and electrons.
Jets and \tauh candidates are reconstructed algorithmically using the PF candidates as inputs.
The missing transverse momentum vector (\ptvecmiss) is computed as the negative vector sum of the transverse momenta of all the PF candidates in an event, and its magnitude is denoted as \ptmiss. The \ptvecmiss is modified to account for corrections to the energy scale of the reconstructed jets in the event.
The primary $\Pp\Pp$ interaction vertex is defined as the reconstructed vertex with the largest value of summed physics-object $\pt^{2}$.
The physics objects considered in this determination are the jets, clustered using a jet-finding algorithm~\cite{Cacciari:2008gp,Cacciari:2011ma} from the charged-particle tracks associated with the vertex, together with the corresponding associated \ptvecmiss, taken as the negative vector sum of the \pt of those jets~\cite{CMS-TDR-15-02}.
Additional identification criteria are applied to reconstructed muons, electrons, photons, \tauh candidates, jets, and \ptvecmiss to reduce the rate of misidentified objects.
This section summarizes the standard CMS reconstruction and identification of muons, electrons, jets, and \tauh candidates used in this analysis. Dedicated analysis-specific techniques for the reconstruction and identification of highly boosted tau-lepton pairs are also discussed.

\subsection{Muons}
\label{sec:reco-muons}

Muons are reconstructed within $\abs{\eta} < 2.4$~\cite{Sirunyan:2018fpa}.
Their reconstruction combines information from both the tracker and the muon system.
Muons are selected from among the reconstructed muon track candidates by applying minimal requirements on the track components in the muon system and requiring consistency with small energy deposits in the calorimeters.

The isolation of individual muons is defined relative to their transverse momentum ($\pt^{\PGm}$) as the sum of the \pt of charged hadrons and neutral particles within a cone around the muon direction at the interaction vertex with radius $\Delta R = \sqrt{(\Delta\eta)^2 + (\Delta\phi)^2} < 0.4$, where $\phi$ is the azimuthal angle in radians:
\begin{equation}
    I^{\PGm} = \left( \sum \pt^{\text{charged}} + \max \left[0, \sum \pt^{\text{neutral}} + \sum \pt^{\PGg} - \pt^\mathrm{PU} \right] \right) / \pt^{\PGm}.
\end{equation}
Here, $\sum \pt^{\textrm{charged}}$ is the scalar \pt sum of charged hadrons originating from the primary vertex.
The terms $\sum \pt^{\textrm{neutral}}$ and $\sum \pt^{\PGg}$ represent the scalar \pt sums of neutral hadrons and photons, respectively.
The neutral contribution to the isolation from pileup interactions, $\pt^{\textrm{PU}}$, is estimated as $0.5\sum_i \pt^{\textrm{PU},i}$, where $i$ runs over the charged hadrons originating from pileup vertices, and the factor of 0.5 corrects for the ratio of charged to neutral particle contributions in the isolation cone.
Muons are considered isolated if $I^{\PGm} < 0.25$.

\subsection{Electrons}
\label{sec:reco-electrons}
Electrons are reconstructed within the geometrical acceptance $\abs{\eta} < 2.5$.
Their reconstruction combines information from clusters of energy deposits in the ECAL and the corresponding trajectories in the tracker~\cite{Khachatryan:2015hwa}.
Electron identification relies on the electromagnetic shower shape and other observables based on tracker and calorimeter information.
The selection criteria depend on \pt, \abs{\eta}, and a categorization based on observables sensitive to the amount of bremsstrahlung radiation emitted along the particle trajectory in the tracker.
The requirement on the relative isolation of reconstructed electrons depends on the final state and is discussed in Section~\ref{sec:selection}.

\subsection{Jets}
\label{sec:reco-jets}

Jets are reconstructed from PF candidates.
The anti-\kt jet clustering algorithm~\cite{Cacciari:2008gp, Cacciari:2011ma} with a distance parameter of 0.4 (AK4) is used.
The standard jet energy corrections~\cite{Chatrchyan:2011ds} are applied.
To reject jets originating from pileup collisions, a multivariate (MVA) jet identification algorithm~\cite{CMS-DP-2020-020} is applied.
This algorithm exploits differences in the shapes of energy deposits within the jet cone between pileup jets and jets originating from quarks or gluons.
In this analysis, reconstructed jets that originate from a tau lepton are identified using a customized deep neural network (DNN), as described in Section~\ref{sec:deepditau}.

\subsection{Hadronic tau lepton decays}
\label{sec:reco-tau}

Standard, isolated hadronically decaying tau leptons are reconstructed and identified within $\abs{\eta} < 2.3$ using the hadron-plus-strips (HPS) algorithm~\cite{Sirunyan:2018pgf}.
The algorithm targets the main decay modes by selecting PF candidates containing one charged hadron and up to two neutral pions, or three charged hadrons.
The HPS algorithm is seeded by the jets described in Section~\ref{sec:reco-jets}.
The \tauh candidates are reconstructed on the basis of the number of tracks and the number of ECAL strips with energy deposits in the $\eta$-$\phi$ plane.

This analysis uses a specialized $\taum\tauh$ and $\taue\tauh$ reconstruction algorithm that follows the same HPS method described above, but with a modified jet seed.
This method is designed to reconstruct boosted $\taum\tauh$ ($\taue\tauh$) objects, in which the tau lepton decaying to a muon (electron) overlaps with the hadronic decay products of the other tau lepton.
This approach has high reconstruction efficiency and a low misidentification probability.

As first developed in Refs.~\cite{Khachatryan:2017mnf} and~\cite{CMS:2020ffa} for the $\taum\tauh$ case, a joint reconstruction of the \tauh candidate and a nearby muon or electron is performed.
Jets that seed the \tauh reconstruction are first modified by removing muons (electrons) with $\pt>3\GeV\,(\pt>10\GeV$) that pass minimal identification requirements from their constituents.
The \tauh candidates reconstructed using these modified jets are required to have $\pt>10\GeV$, where the reconstructed $\pt^{\tauh}$ corresponds to the visible portion of the tau lepton decay.
Any \tauh candidate with a distance of closest approach greater than 0.5\unit{cm} in the $z$ direction from the primary vertex is rejected to suppress candidates arising from constituents that do not originate from the primary vertex.

The full $\taum\tauh$ and $\taue\tauh$ identification procedure includes the modified HPS algorithm described above, together with an isolation requirement on the \tauh candidate.
The isolation of a \tauh candidate is computed using an MVA discriminant~\cite{Sirunyan:2018pgf}, trained to distinguish genuine \tauh candidates from jets.
The discriminant is calculated using PF candidates in the region around the \tauh candidate defined by $\Delta R < 0.8$, excluding the overlapping muon or electron.
The \tauh candidates are required to pass a selection on the MVA discriminant output as a function of $\pt^{\tauh}$ that maintains an efficiency of approximately 80\%.
A loose discriminant to reject muons or electrons~\cite{Sirunyan:2018pgf} is also applied.

\begin{figure}
  \centering
  \includegraphics[width=0.49\textwidth]{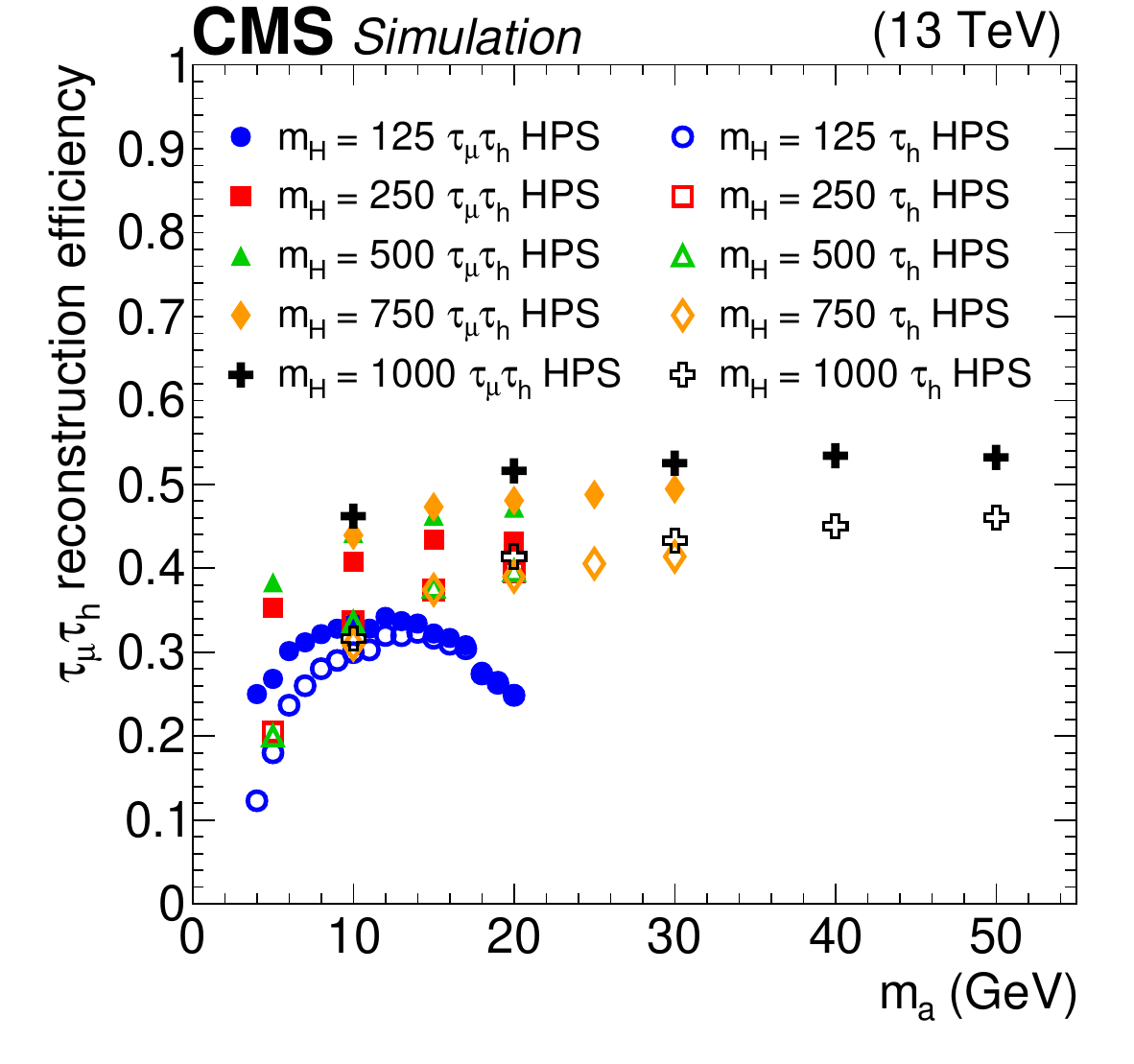}
  \includegraphics[width=0.49\textwidth]{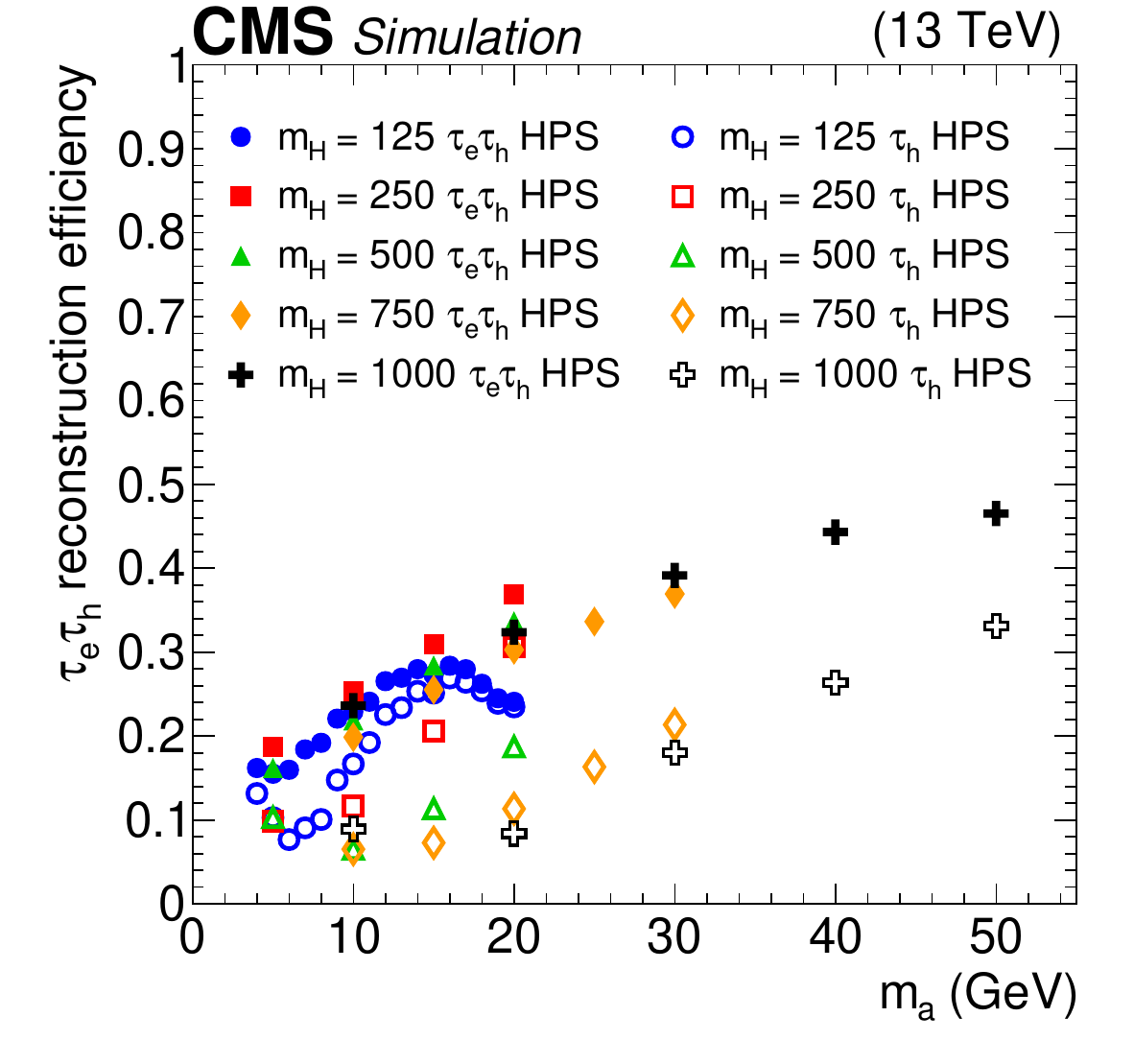}
  \caption{The $\PGt\PGt$ reconstruction efficiency of the standard \PGt HPS algorithm (open markers) and the modified $\taum\tauh$ (left) or $\taue\tauh$ (right) HPS algorithm (solid markers) as a function of $m_{\pa}$ for $m_{\PH}$ = 125, 250, 500, 750, and 1000\GeV.
    Because of the requirement $\Delta R(\taul, \tauh) < 0.8$, the efficiency for $m_{\PH}$ = 125\GeV decreases above $m_{\pa}$ = 12\GeV in the $\taum\tauh$ channel and above $m_{\pa}$ = 14\GeV in the $\taue\tauh$ channel, corresponding to the region in which the $\PGt\PGt$ topology becomes increasingly resolved.
    In the right plot, the dip in the standard HPS efficiency for the low $m_{\pa}$ region is due to two competing effects: the requirement of $\pt^{\Pe} >$ 10\GeV, which loses efficiency with increasing $m_{\pa}$, and the requirement of passing an anti-electron discriminant, which increases in efficiency as $m_{\pa}$ increases. This analysis deploys the modified HPS reconstruction which shows a significantly higher efficiency than the standard \PGt HPS scenario.}
  \label{fig:CleanedEff}
\end{figure}

The modified $\taum\tauh$ and $\taue\tauh$ reconstruction and identification algorithm significantly improves the signal efficiency across the full range of Higgs boson and pseudoscalar hypotheses considered, as shown in Fig.~\ref{fig:CleanedEff}.  For example, for the $\taum\tauh$ channel, efficiencies are larger by factors of up to 2, 1.9, and 1.5 for $m_{\PH}$ = 125, 500, and 1000\GeV, respectively, while for the $\taue\tauh$ channel, efficiencies are larger by factors of up to 2, 3, and 4 for $m_{\PH}$ = 125, 500, and 1000\GeV, respectively.
The efficiency of the $\taum\tauh$ ($\taue\tauh$) reconstruction and identification is measured in events with a muon (electron) passing the identification criteria described in Section~\ref{sec:reco-muons} (\ref{sec:reco-electrons}).  It is defined as the fraction of those events that also have a \tauh candidate passing either the standard \tauh HPS reconstruction or the $\taum\tauh$ ($\taue\tauh$) HPS reconstruction, as well as the MVA isolation discriminant.
The muon (electron) and \tauh candidate must have \pt greater than the thresholds described above.
The increase in efficiency arises from two main factors: the modification of the jets that seed the $\taum\tauh$ ($\taue\tauh$) reconstruction and the exclusion of the muon (electron) energy from the MVA isolation discriminant.
At low $m_{\pa}$, where the pseudoscalar decay products are highly collimated ($\Delta R_{\PGt\PGt} < 0.4$), the jet seed modification is the dominant contributor to the efficiency increase because of the larger Lorentz boost.
For more widely separated decay products ($0.4 < \Delta R_{\PGt\PGt} < 0.8$), the change in the MVA discriminant becomes the primary source of the efficiency gain.
The reduced efficiency at low $m_{\pa}$, observed for both the standard and modified HPS reconstruction techniques, is caused by the large Lorentz boost, which makes the muon or electron nearly collinear with a charged hadron from the \tauh candidate.
At lower Lorentz boost, the muon or electron and the \tauh candidate are more widely separated.
In this case, the efficiency is reduced by the requirement of a boosted topology, especially for $m_{\PH}$ = 125\GeV.
For higher \PH boson masses, the efficiency is less affected by an increase in $m_{\pa}$ because the corresponding reduction in Lorentz boost is generally not sufficient to separate the tau leptons from a pseudoscalar decay beyond the selection requirement of $\Delta R_{\PGt\PGt} < 0.8$.
That requirement is motivated by the $\Delta R$ = 0.8 cone size for calculating the \tauh isolation discriminant.

The combined efficiencies of the reconstruction, identification, and isolation requirements for \tauh candidates are measured in two bins of \tauh \pt using a ``tag-and-probe'' technique~\cite{Sirunyan:2018pgf} applied to an inclusive sample of $\PZ\to\taum\tauh$ events.
These efficiencies are measured in both data and simulation.
The data-to-simulation efficiency ratios are used as scale factors to correct the simulated event yields.
Their values are listed in Table~\ref{tab:tauidsfsummary}.

\begin{table}[!htp]
  \centering
  \topcaption{Measured scale factors and their uncertainties for \tauh candidates.
  The 2016 data comprise two eras with different strip tracker running conditions~\cite{CMS:2025bbk}.}
  \begin{tabular}{lcc}
    Eras      & $10 < \pt^{\tauh} < 20\GeV$     & $\pt^{\tauh} > 20\GeV$  \\ \noalign{\vskip 1pt}\hline
    2016 (a)   & 0.95 $\pm$ 0.25           & 0.95 $\pm$ 0.07  \\
    2016 (b)   & 1.03 $\pm$ 0.24           & 1.03 $\pm$ 0.07  \\
    2017          & 0.99 $\pm$ 0.29           & 0.97 $\pm$ 0.07  \\
    2018          & 1.07 $\pm$ 0.27           & 1.02 $\pm$ 0.06  \\
  \end{tabular}
  \label{tab:tauidsfsummary}
\end{table}

These scale factors are found to be independent of $\abs{\eta^{\tauh}}$.
For $10 < \pt^{\tauh} < 20\GeV$, the $\PZ\to\taum\tauh$ data sample contains significant QCD and $\PW$ + jets backgrounds, leading to a larger uncertainty on the scale factor than for $\pt^{\tauh}>20\GeV$. The simulated $\PW$ + jets distribution is rescaled using a control region in data as in Ref.~\cite{Sirunyan:2018pgf}, and its shape uncertainty is propagated.

For the $\tauh\tauh$ final state, the highly boosted decay products result in overlapping $\tauh\tauh$ candidates, for which neither the standard \tauh reconstruction approach nor the modified method described above is efficient.
To address this, a dedicated deep neural network tagger, \DEEPDITAU (Section~\ref{sec:deepditau}), is developed to determine the probability that a reconstructed AK4 jet originates from a $\tauh\tauh$ pair.

\section{\DEEPDITAU for tagging hadronic decays of the tau lepton pair}
\label{sec:deepditau}

Conventional $\tauh$ reconstruction methods in CMS are optimized for single, isolated hadronic tau lepton decays.
This analysis targets a boosted topology in which the decay products of tau lepton pairs frequently overlap and are reconstructed as single AK4 jets.
The \DEEPDITAU algorithm was developed for this analysis to target the large ditau branching fraction to hadrons ($\tauh\tauh$), which was not included in previous analyses.
The \DEEPDITAU algorithm is based on a DNN trained to classify AK4 jets into three categories of overlapping objects.
Its output is a vector of three normalized scores corresponding to light jets (udsc and gluon), jets originating from a bottom quark (\cPqb jets), and $\tauh\tauh$ jets, which sum to unity for each AK4 jet.
The score in the $\tauh\tauh$ category is used as the selection criterion in this analysis. The \DEEPDITAU algorithm extends the concepts of deep-learning-based jet flavor and $\tauh$ identification techniques~\cite{Bols:2020bkb,CMS:2022prd} to the reconstruction of highly boosted ditau topologies.

The \DEEPDITAU algorithm is trained on a sample of 3.2 million AK4 jets, including light jets, \cPqb jets, and $\tauh\tauh$ jets from simulated QCD, \ttbar, and signal events, respectively.
The signal event simulation is as described in Section~\ref{sec:samples}.
As in the signal simulation, the QCD and \ttbar samples use \PYTHIA~8.230 for parton showering, hadronization, and modeling of the underlying event, with the CP5 tune~\cite{CMS:2019csb} used for the latter.
The QCD hard-scattering process is simulated at leading order in \alpS using \PYTHIA~8.230, whereas the \ttbar process is simulated at next-to-leading order using \MGvATNLO version 2.6.7.
To ensure a balanced training sample, excess light jets and \cPqb jets are removed at random, resulting in equal numbers of light jets, \cPqb jets, and $\tauh\tauh$ jets.

The \DEEPDITAU tagger uses up to 284 input features, including 30 variables describing global jet properties and information from constituent particles: up to ten charged hadrons, each described by 18 variables, including position and track quality measurements; up to ten neutral hadrons, each described by five variables, including kinematic variables; and up to four photons, each described by six variables, including kinematic variables and variables related to pileup mitigation and quality requirements used to select high-quality prompt photons. The distributions of the input features are normalized to have a mean of zero and a standard deviation of one, except for the jet $\eta$ and jet $\phi$, which are scaled linearly from zero to one, and the jet \pt, which is transformed logarithmically.
The jet particle-flow constituents are transformed such that their features are expressed relative to the jet axis.
During training, weights are applied to the jets such that the \pt and $\eta$ spectra of light and \cPqb jets match those of the $\tauh\tauh$ jets.

The constituent PF objects of each jet are concatenated with the global jet input features and passed to a fully connected neural network, in which each node is connected to all nodes in the adjacent layers.
The network has a depth of eight layers, each with a width of 256 nodes.
Batch normalization is applied so that the inputs to each layer have a mean of zero and a standard deviation of unity.
Dropout layers with a dropout rate of 0.1 are inserted between dense layers during the training stage to mitigate overtraining.
The neural network is optimized by minimizing a cross-entropy loss function, with 10\% of the sample reserved for validation.

\begin{figure}[htb]
    \centering
    \includegraphics[width=0.6\textwidth]{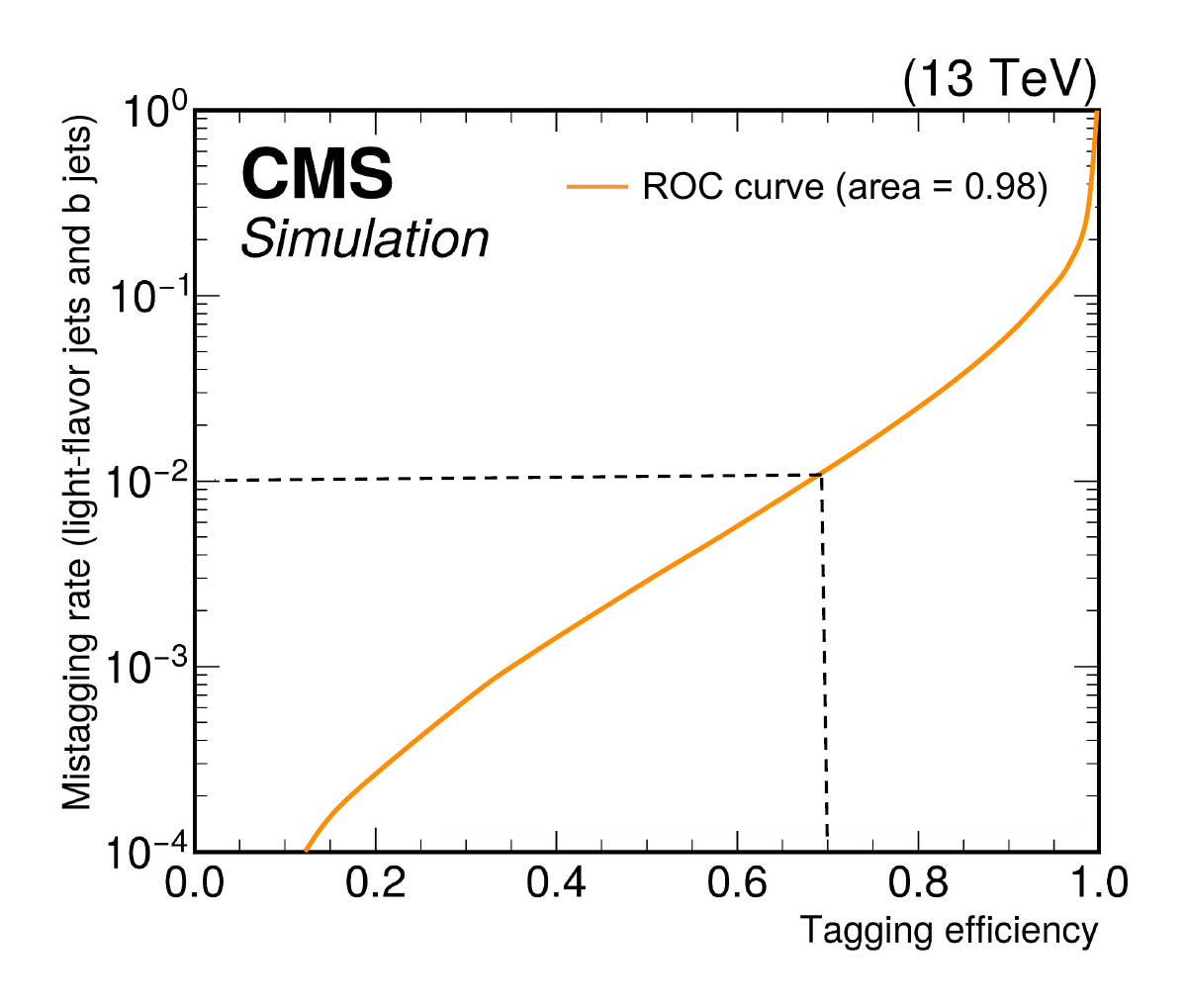}
    \caption{Receiver operating characteristic curve for the \DEEPDITAU classifier.
      The tagging efficiency corresponds to the fraction of $\tauh\tauh$ jets correctly identified as $\tauh\tauh$ jets, while the mistagging rate corresponds to the fraction of light jets and \cPqb jets misidentified as $\tauh\tauh$ jets. The dashed lines show that the medium working point used in this analysis corresponds to a tagging efficiency of 70\% and a mistagging rate of approximately 1\%.}
    \label{fig:roc_TauHTauH}
\end{figure}

The receiver operating characteristic curve for the \DEEPDITAU classifier is presented in Fig.~\ref{fig:roc_TauHTauH}. Each reconstructed jet is required to pass the medium working point of the \DEEPDITAU discriminant, corresponding to a tagging efficiency of 70\% and a mistagging rate of approximately 1\%, which suppresses jets originating from both light- and heavy-flavor quarks.
To suppress the effect of initial-state radiation, the selected jet must satisfy $\pt > 30\GeV$ and $\abs{\eta} < 2.4$.  A version of the \DEEPDITAU classifier trained only on highly boosted $m_{\PH}, m_{\pa}$ mass pairs shows unchanged performance with respect to the baseline classifier.

\section{Event selection}
\label{sec:selection}

Collision events are selected using a trigger that requires the presence of an isolated muon with $\pt > 24\GeV$ in 2016 and 2018 or $\pt > 27\GeV$ in 2017~\cite{Khachatryan:2016bia}.
The trigger efficiencies are measured in both data and simulation using the tag-and-probe technique.
A global trigger scale factor of 0.97--0.99, depending on the year, is applied to the simulation~\cite{CMS:2024aqx}.
The event is required to contain two isolated oppositely charged muons with $\Delta R < 1.0$ and $3 < m_{\PGm\PGm} < 55\GeV$.
The leading muon, matched to the muon that triggered the event, must have $\pt > 26\GeV$ in 2016 and 2018 or $\pt > 29\GeV$ in 2017.
The second muon must have $\pt > 3\GeV$.
These muons form a $\PGm\PGm$ pair that constitutes one of the pseudoscalar candidates.
If more than one pair of muons in the event satisfies these criteria, the pair with the smallest $\Delta R$ is chosen as the $\pa\to\PGm\PGm$ candidate.
The second pseudoscalar candidate is selected through its decay to an oppositely charged (OS) \PGt pair.
Taking the \PGt decays into account, this analysis considers the $\taum\taue$, $\taum\tauh$, $\taue\tauh$, and $\tauh\tauh$ final states, and an explicit OS requirement is imposed for all channels except for the $\tauh\tauh$ channel.

The tau lepton selections in the four final-state channels are summarized in Table~\ref{tab:di-tau-selection}.
The $\taum\taue$ selection requires one identified muon with $\pt > 3\GeV$ and one electron with $\pt > 10\GeV$, with the additional requirement that $\Delta R_{\taum\taue} < 1.0$.
The electron is required to pass a standard identification criterion that includes a requirement on isolation from nearby hadrons~\cite{CMS:2020uim}.
The $\taum\tauh$ ($\taue\tauh$) selection requires one identified muon (electron) with $\pt > 3$ (10) \GeV, with no isolation requirement included in the identification criteria, and one \tauh candidate with $\pt > 10\GeV$, reconstructed as described in Section~\ref{sec:reco-tau}.
The reconstructed muon (electron) corresponds to the visible portion of the \taum (\taue) decay.
The two \PGt candidates in the $\taum\tauh$ and $\taue\tauh$ channels must satisfy $\Delta R_{\PGt\PGt} < 0.8$.
In the $\taum\taue$, $\taum\tauh$, and $\taue\tauh$ channels, if more than one pair of \PGt candidates in the event satisfies these criteria, the pair with the smallest $\Delta R$ is chosen as the $\pa\to\PGt\PGt$ candidate.
In the $\taum\tauh$ and $\taue\tauh$ channels, the $\Delta R_{\PGt\PGt}$ requirement is driven by the isolation discriminant of the modified HPS algorithm.
Because the $\taum\taue$ channel has a lower background from jets misreconstructed as tau leptons, the $\Delta R_{\PGt\PGt}$ requirement is loosened to increase the signal selection efficiency.
Overall, this requirement ensures a boosted topology.
In addition to the selection of the ditau pair, a minimum separation in $\Delta R$ is required between the tau candidates and the muons identified as the direct pseudoscalar decay products, as detailed in Table~\ref{tab:di-tau-selection}.
Finally, in the $\taum\tauh$ channel only, the transverse momentum of the nontriggering muon ($\PGm_{2}$) in the $\pa\to\PGm\PGm$ pair is required to exceed that of the $\taum$ candidate to eliminate remaining $\PZ \to \PGm\PGm$ events.
These selections, together with the corresponding selection of the $\PGm\PGm$ pair, reduce the combinatorial background in which the wrong combination of leptons is assigned to the pseudoscalar candidates.
The reconstruction of the $\tauh\tauh$ final state is discussed in Section~\ref{sec:deepditau}.

\begin{table}[htbp]
\topcaption{Tau lepton selection criteria in the four different channels. In the columns labeled $\Delta R_{\PGm\taul}$ and $\Delta R_{\PGm\tauh}$, \PGm refers to either of the two muons reconstructed as the $\pa\to\PGm\PGm$ candidate. In the column labeled $\Delta R_{\PGm\taul}$, \taul refers to either the \taum or \taue candidate. In the final column, $\PGm_{2}$ refers to the nontriggering muon in the $\pa\to\PGm\PGm$ pair. A dash (\NA) means that the selection in the column header does not apply to the channel in that row.}
\label{tab:di-tau-selection}
\centering
\cmsTable{
\begin{tabular}{lccccccccc}
Channel &
\begin{tabular}{@{}c@{}}\taum \PT\\(\GeVns)\end{tabular} &
\begin{tabular}{@{}c@{}}\taue \PT\\(\GeVns)\end{tabular} &
\begin{tabular}{@{}c@{}}\taue\\isolation\end{tabular} &
\begin{tabular}{@{}c@{}}\tauh \PT\\(\GeVns)\end{tabular} &
\begin{tabular}{@{}c@{}}\tauh\\reconstruction\end{tabular} &
$\Delta R_{\PGt\PGt}$ & $\Delta R_{\PGm\taul}$ & $\Delta R_{\PGm\tauh}$ &
\begin{tabular}{@{}c@{}}Additional\\requirements\end{tabular} \\
\hline
$\taum\taue$ & $>$3 & $>$10 & Yes & \NA & \NA & $<$1.0 & $>$0.4 & \NA & \NA \\
$\taum\tauh$ & $>$3 & \NA & \NA & $>$10 & Section~\ref{sec:reco-tau} & $<$0.8 & $>$0.4 & $>$0.8 &
  \begin{tabular}{@{}c@{}}\PT($\PGm_{2}$) $>$\\\PT(\taum)\end{tabular} \\
$\taue\tauh$ & \NA & $>$10 & No & $>$10 & Section~\ref{sec:reco-tau} & $<$0.8 & $>$0.4 & $>$0.8 & \NA \\
$\tauh\tauh$ & \NA & \NA & \NA &
  \begin{tabular}{@{}c@{}}$>$30\\(jet, cf.\\Section~\ref{sec:reco-jets})\end{tabular} &
  Section~\ref{sec:deepditau} & \NA & \NA &
  \begin{tabular}{@{}c@{}}$>$0.4\\(\PGm, jet)\end{tabular} & \NA \\
\end{tabular}
}
\end{table}

\section{Background and signal modeling}
\label{sec:bkg}

The main source of background in this search is Drell-Yan $\PGm\PGm$ production in association with at least one jet that is misidentified as the $\PGt\PGt$ candidate.
This background is determined directly from data.
The predicted two-dimensional distribution of $m_{\PGm\PGm\PGt\PGt}$ versus $m_{\PGm\PGm}$ in the signal region (SR) resulting from SM background processes is estimated from a sideband region by reweighting the sideband data events using a tight-to-loose ratio.
The background shape is parametrized by a two-dimensional unbinned maximum likelihood fit~\cite{CMS:2024onh} to the reweighted sideband data.
For the $\tauh\tauh$ channel, the four-momentum of the jet that seeds the \DEEPDITAU object is used to reconstruct the four-body mass.

\begin{figure}[htb]
    \centering
    \includegraphics[width=0.8\textwidth]{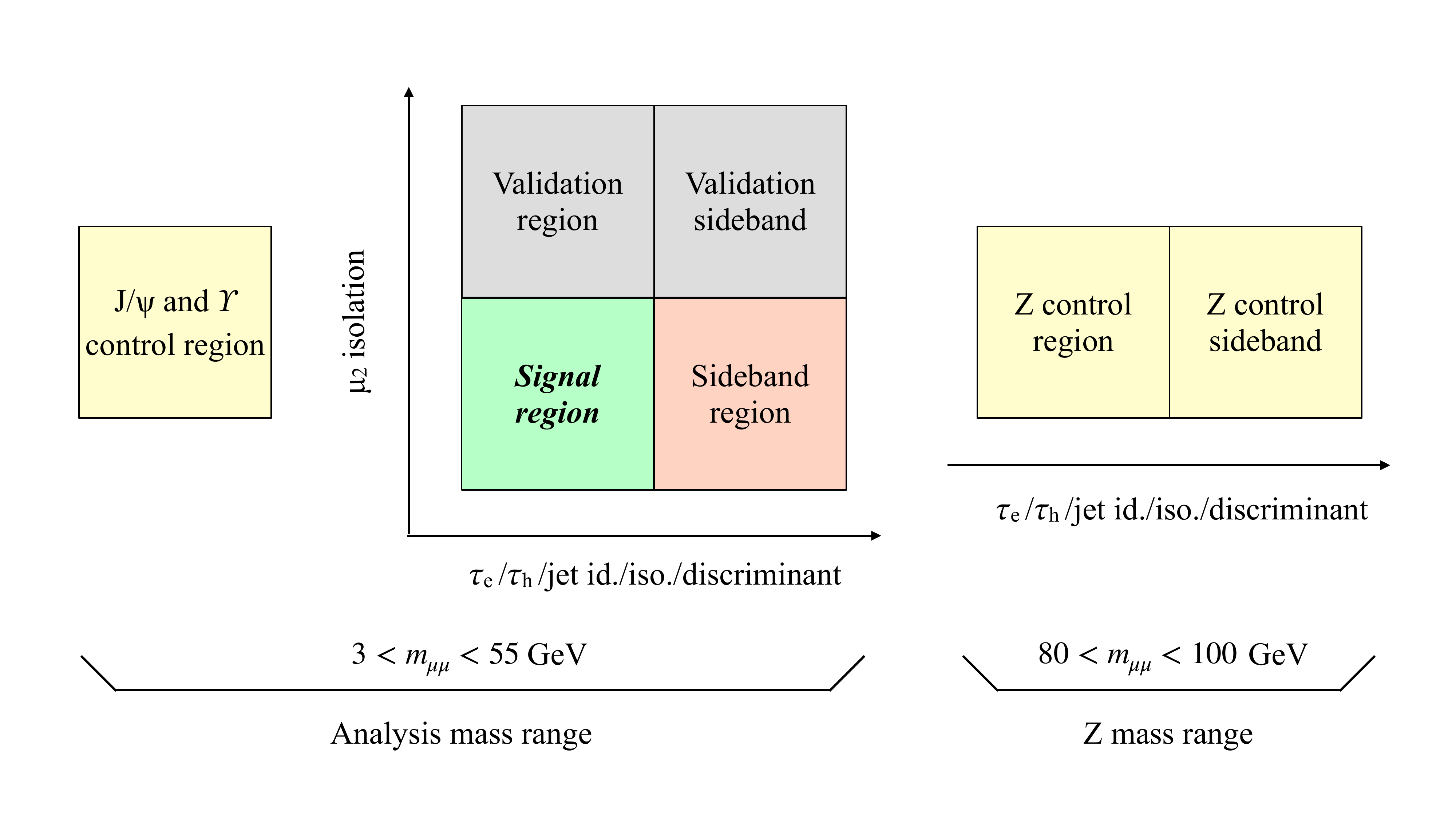}
    \caption{
      Schematic of the analysis strategy.
      In the \PZ boson mass range, the tight-to-loose ratio is derived from the $\PZ\to\PGm\PGm$+jet control region and sideband.
      In the analysis mass range, events with two isolated muons and no loosely selected $\PGt\PGt$ candidates enter the control region.
      Events that additionally contain a loosely selected $\PGt\PGt$ candidate are further categorized into the signal region or sideband, as described in the text.
      The sideband data, weighted by the tight-to-loose ratio, are used to estimate the background in the signal region.
      In addition, by inverting the isolation requirement on the muon that did not trigger the event, two analogous regions are defined for validating the tight-to-loose method.
    }
    \label{fig:regions}
\end{figure}

Events considered for categorization into the SR and sideband are required to contain a candidate consistent with $\pa\to\PGm\PGm$ and a loosely selected $\PGt\PGt$ candidate passing the \pt and $\Delta R$~requirements listed in Table~\ref{tab:di-tau-selection}.
These events must also satisfy the \taue identification requirement without isolation in the $\taue\tauh$ channel, a loosened electron identification requirement in the $\taum\taue$ channel, or a loosened \tauh isolation requirement in the $\taum\tauh$ and $\taue\tauh$ channels.
The SR and sideband are then defined according to whether the electron passes or fails the tighter identification (Section~\ref{sec:reco-electrons}) and isolation requirements (Section~\ref{sec:selection}) in the $\taum\taue$ channel, whether the \tauh candidate passes or fails the tighter isolation requirement in the $\taum\tauh$ and $\taue\tauh$ channels (Section~\ref{sec:reco-tau}), or whether the jet passes or fails the \DEEPDITAU discriminant in the $\tauh\tauh$ channel.
A schematic of the SR and sideband is shown in Fig.~\ref{fig:regions}.

The tight-to-loose ratio is derived from a $\PZ\to\PGm\PGm$+jet control region (CR) and sideband, shown in Fig.~\ref{fig:regions}.
The $\PZ\to\PGm\PGm$+jet control sample is collected using a single-muon trigger, with the additional requirement of two isolated oppositely charged muons having an invariant mass of $80 < m_{\PGm\PGm} < 100\GeV$.
The CR and sideband are defined in the same way as the SR and sideband, with the exception of the $m_{\PGm\PGm}$ requirement.
The tight-to-loose ratio is defined as the number of jet candidates passing the tight selection divided by the number passing the loose selection, measured in bins of lepton or jet \pt.
For \tauh candidates, the measurement is also performed separately for each \tauh decay mode.
In general, the decay mode with three tracks has a lower tight-to-loose ratio than the modes with a single track.
Residual contributions from diboson processes, estimated from simulation, are subtracted from the data.
The tight-to-loose ratio is measured as a function of \pt to be 0.2--0.6 for the $\tauh\tauh$ channel, and 0.05--0.35 as a function of \pt and \PGt decay mode for the $\taum\tauh$ and $\taue\tauh$ channels.

The tight-to-loose method is validated in a CR and sideband that are orthogonal to the SR and sideband by inverting the isolation requirement on the muon from the $\PGm\PGm$ pair that did not trigger the event, $\PGm_{2}$.
These regions are shown in Fig.~\ref{fig:regions}.

The background shape contains prominent $\PGm\PGm$ resonances in the mass range examined: \JPsi, \Pgy, \PgUa, \PgUb, and \PgUc.
In the $m_{\PGm\PGm}$~distribution, the known resonance peaks appear on top of the Drell--Yan continuum.
Probability density functions (pdfs) are constructed in four regions of the $m_{\PGm\PGm}$ spectrum: $3 < m_{\PGm\PGm} < 8\GeV$, $8 < m_{\PGm\PGm} < 12\GeV$, $12 < m_{\PGm\PGm} < 30\GeV$ ($12 < m_{\PGm\PGm} < 35\GeV$ for $m_{\PH}$ = 750\GeV), and $30 < m_{\PGm\PGm} < 55\GeV$.
The fourth range is used only for $m_{\PH}$ = 1000\GeV, consistent with the signal processes listed in Table~\ref{tab:a_mass_h_mass}.
All channels follow these $m_{\PGm\PGm}$ regions, except for the $\taue\tauh$ channel, where the $m_{\PGm\PGm}$ range is adjusted depending on the Higgs boson mass hypothesis.
The pdfs consist of the sum of an exponential function and either zero, two, or three Voigt functions, defined as Breit--Wigner functions convolved with Gaussian resolution functions, to model the SM resonances in the respective regions.
An additional exponential function is included to model the rising continuum background near the \JPsi resonance and the extended tail in the lowest and highest $m_{\PGm\PGm}$ regions, respectively.
Although the \JPsi resonance lies outside the kinematically allowed search window for a signal resonance, it is included in the fit to provide a better description of the background near the \Pgy resonance.
In the $m_{\PGm\PGm\PGt\PGt}$ distribution, the $\PGm\PGm$+jet background appears as an exponentially falling distribution with a turn-on around 40--60\GeV because of the \pt requirements applied to the two reconstructed muons and the $\PGt\PGt$ candidate.

The $m_{\PGm\PGm\PGt\PGt}$ background distributions in each of the four $m_{\PGm\PGm}$ ranges are modeled with either an exponential function or the product of an error function and an exponential function when the kinematic threshold is significant.
To determine whether the error function is needed to describe the data, a background pdf containing both functional forms is fit to the sideband data.
If the fit does not converge or the parameters of the error function are poorly constrained, with relative uncertainties greater than 100\%, the error function component is omitted.
The fit ranges of $m_{\PGm\PGm\PGt\PGt}$ in all four channels and an additional constraint on the $m_{\PGm\PGm}$ ranges for the $\taue\tauh$ channel  are summarized in Table~\ref{tab:fitranges}.  Regions are chosen to provide sufficient statistics and improve the background fit quality, while retaining at least 90\% of the signal.  
In addition, for the $\taue\tauh$ channel, and for the $\taum\taue$ channel for $m_{\PH} = 750 \GeV$ in 2016 only, the low-mass dimuon region is excluded from the fit because of its negligible signal acceptance and correspondingly poor sensitivity.
Only the region in which the signal integral exceeds 0.02 events after normalization to the integrated luminosity is considered.
Accordingly, the range of $m_{\PGm\PGm}$ is adjusted depending on the Higgs boson mass.

\begin{table}[!htp]
  \centering
  \topcaption{Four-body mass ranges used in the two-dimensional fit.
  For the $\taue\tauh$~channel, an additional requirement is imposed on $m_{\PGm\PGm}$, as shown in the last row.
  All entries are given in GeV.}
  \begin{tabular}{lccccc}
    Channel & $m_{\PH}$ = 125 & $m_{\PH}$ = 250 & $m_{\PH}$ = 500 & $m_{\PH}$ = 750 & $m_{\PH}$ = 1000  \\ \hline
    $\taum\taue$ ($m_{\PGm\PGm\PGt\PGt}$)  & 60--150 &  80--1200 & 80--1200 & 80--1200 & 80--1200 \\
    $\taum\tauh$ ($m_{\PGm\PGm\PGt\PGt}$) & 60--150 &  130--280 & 150--1200 & 150--1200 & 150--1200 \\
    $\taue\tauh$ ($m_{\PGm\PGm\PGt\PGt}$) & 80--150 &  80--1200 & 80--1200 & 80--1200 & 80--1200 \\
    $\tauh\tauh$ ($m_{\PGm\PGm\PGt\PGt}$) & 80--150 &  130--280 & 250--600 & 350--800 & 450--1200 \\
    $\taue\tauh$ ($m_{\PGm\PGm}$) & 10--30 &  10--30 & 14--30 & 18--30 & 18--55 \\
  \end{tabular}
  \label{tab:fitranges}
\end{table}

The $m_{\PGm\PGm}$ and $m_{\PGm\PGm\PGt\PGt}$ pdfs are multiplied to construct a two-dimensional pdf.
Because $m_{\PGm\PGm\PGt\PGt}$ is weakly correlated with $m_{\PGm\PGm}$ in the background distribution, the parameters of the $m_{\PGm\PGm\PGt\PGt}$ background model are allowed to vary independently within each $m_{\PGm\PGm}$ range.

A CR is defined and included in the fit to constrain the parameters of the $\PGm\PGm$ resonances, the mean ($M$), width ($\Gamma$), and resolution ($\Sigma$), as well as the relative normalizations, $N_{i}/N_{j}$, where $i$ and $j$ denote a pair of background resonances.
The relative normalizations are constrained between the \JPsi and \Pgy resonances and between the \PgUa resonance and each of the \PgUb and \PgUc resonances.
The low-mass CR, shown in Fig.~\ref{fig:regions}, is defined by requiring the presence of a $\PGm\PGm$ pair and the absence of a tau lepton pair.
The resonance parameters are consistent between the signal and control regions and are therefore treated as identical.
All remaining parameters, including those of the $m_{\PGm\PGm}$ continuum ($\lambda^{i}_{\PGm\PGm}$, where $i$ indexes the used exponential components), the $m_{\PGm\PGm\PGt\PGt}$ continuum ($\lambda^{i}_{\PGm\PGm\PGt\PGt}$), the $m_{\PGm\PGm\PGt\PGt}$ error function shift ($a_{\mathrm{erf}}$) and scale ($b_{\mathrm{erf}}$), and the relative normalizations of the $m_{\PGm\PGm}$ resonances to the $m_{\PGm\PGm}$ continuum ($N_{\PgUa+\PgUb+\PgUc}/N_{\textrm{continuum}}$ and $N_{\JPsi+\Pgy}/N_{\textrm{continuum}}$), are allowed to vary independently within the uncertainties determined from a fit to the sideband data and are not constrained between regions.

Table~\ref{tab:parameters} summarizes the background model parameters and their relationships between the two fit regions used in the analysis.

\begin{table}[!htp]
  \centering
  \topcaption{The background model parameters and their relationships between the signal region and control region: shared, in which the parameters are identical in the indicated regions, and independent, in which the parameters vary freely and are not constrained by any other region.}
  \begin{tabular}{lcccc}
    Category & Parameters & Signal region & Control region \\ \hline
    $\PGm\PGm$ resonances & $M$, $\Sigma$, $\Gamma$  & \multicolumn{2}{c}{Shared} \\
    $\PGm\PGm$ continuum  & $\lambda^{i}_{\PGm\PGm}$                  & Independent & Independent \\
    $\PGm\PGm\PGt\PGt$  & $a_{\mathrm{erf}}$, $b_{\mathrm{erf}}$, $\lambda^{i}_{\PGm\PGm\PGt\PGt}$       & Independent & -- \\
    Normalizations        & $N_{\Pgy}/N_{\JPsi}$            & \multicolumn{2}{c}{Shared} \\
    & $N_{\PgUb}/N_{\PgUa}$         & \multicolumn{2}{c}{Shared} \\
    & $N_{\PgUc}/N_{\PgUa}$       & \multicolumn{2}{c}{Shared} \\
    & $N_{\PgUa+\PgUb+\PgUc}/N_{\textrm{continuum}}$            & Independent & Independent \\
    & $N_{\JPsi+\Pgy}/N_{\textrm{continuum}}$     & Independent & Independent \\ 
  \end{tabular}
  \label{tab:parameters}
\end{table}

The signal is characterized by a narrow resonance in $m_{\PGm\PGm}$ from the pseudoscalar decay and a broader distribution in $m_{\PGm\PGm\PGt\PGt}$ because of the invisible decay products of one of the pseudoscalar bosons.
For the signal processes, the correlation between the $m_{\PGm\PGm}$ and $m_{\PGm\PGm\PGt\PGt}$ distributions is small.
The model parameters are determined from fits to simulated signal samples.
Each generated distribution, corresponding to a given $m_{\PH}$ and $m_{\pa}$ hypothesis, is fit with a two-dimensional pdf constructed as the product of a Voigt function for $m_{\PGm\PGm}$ and a split-normal distribution, defined as a Gaussian-like function with different widths on either side of the peak, for $m_{\PGm\PGm\PGt\PGt}$.
For each parameter, a polynomial function is used to interpolate between the generated $m_{\pa}$ at each $m_{\PH}$ hypothesis (cf. Section~\ref{sec:samples}) and to extrapolate down to 3.6\GeV for $m_{\PH}$ of 250, 500, 750, and 1000\GeV, and up to 25, 30, and 35\GeV for $m_{\PH}$ of 250, 500, and 750\GeV, respectively, in order to cover the ranges listed in Table~\ref{tab:a_mass_h_mass}.
Within the extrapolation ranges, the polynomial functions vary negligibly within the fit uncertainties, except in the $\taum\taue$ and $\taue\tauh$ channels, which are therefore excluded from the fits in subsets of the mass ranges.

\begin{figure}[htbp]
  \centering
  \includegraphics[width=0.4\textwidth]{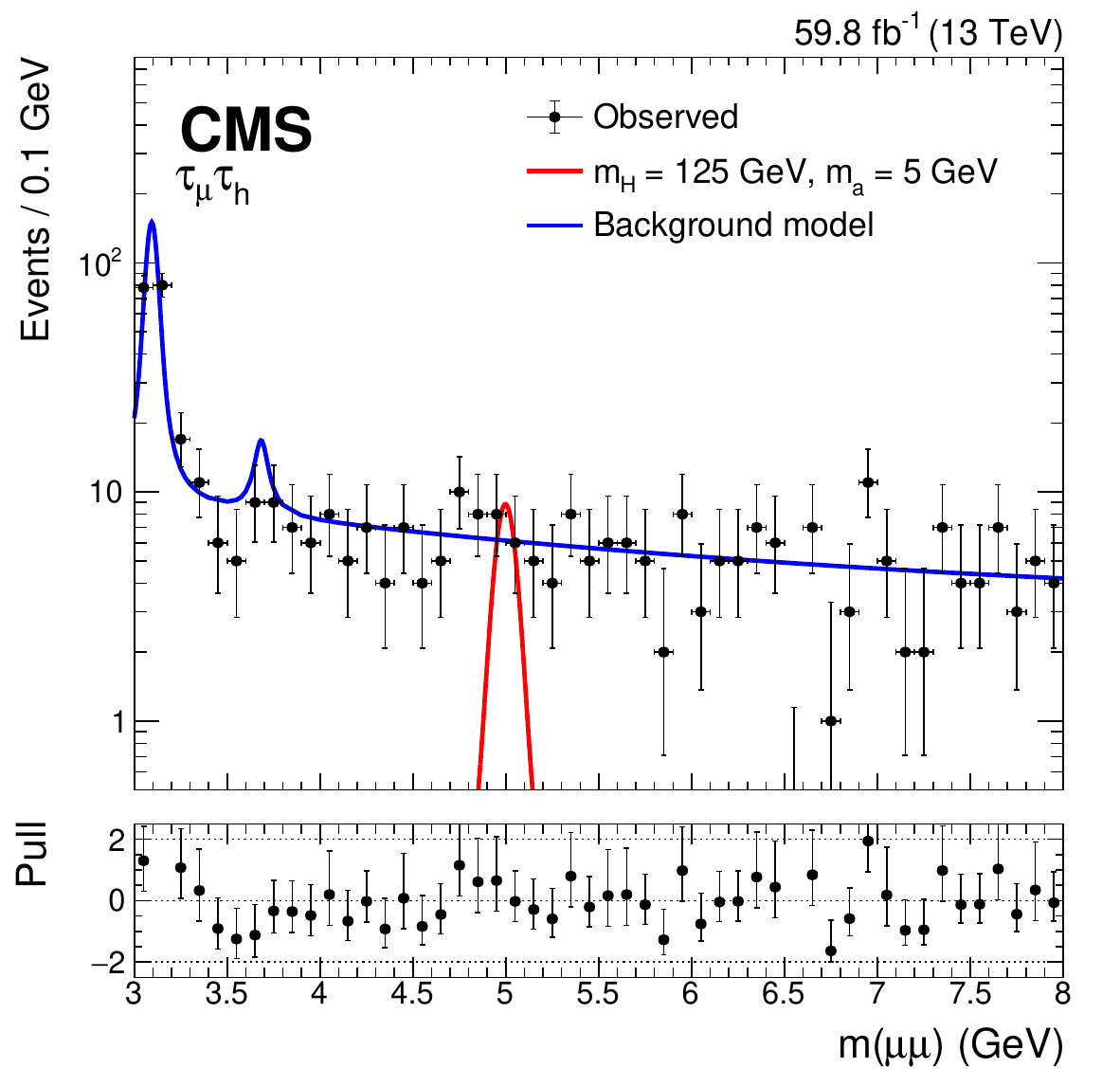}
  \includegraphics[width=0.4\textwidth]{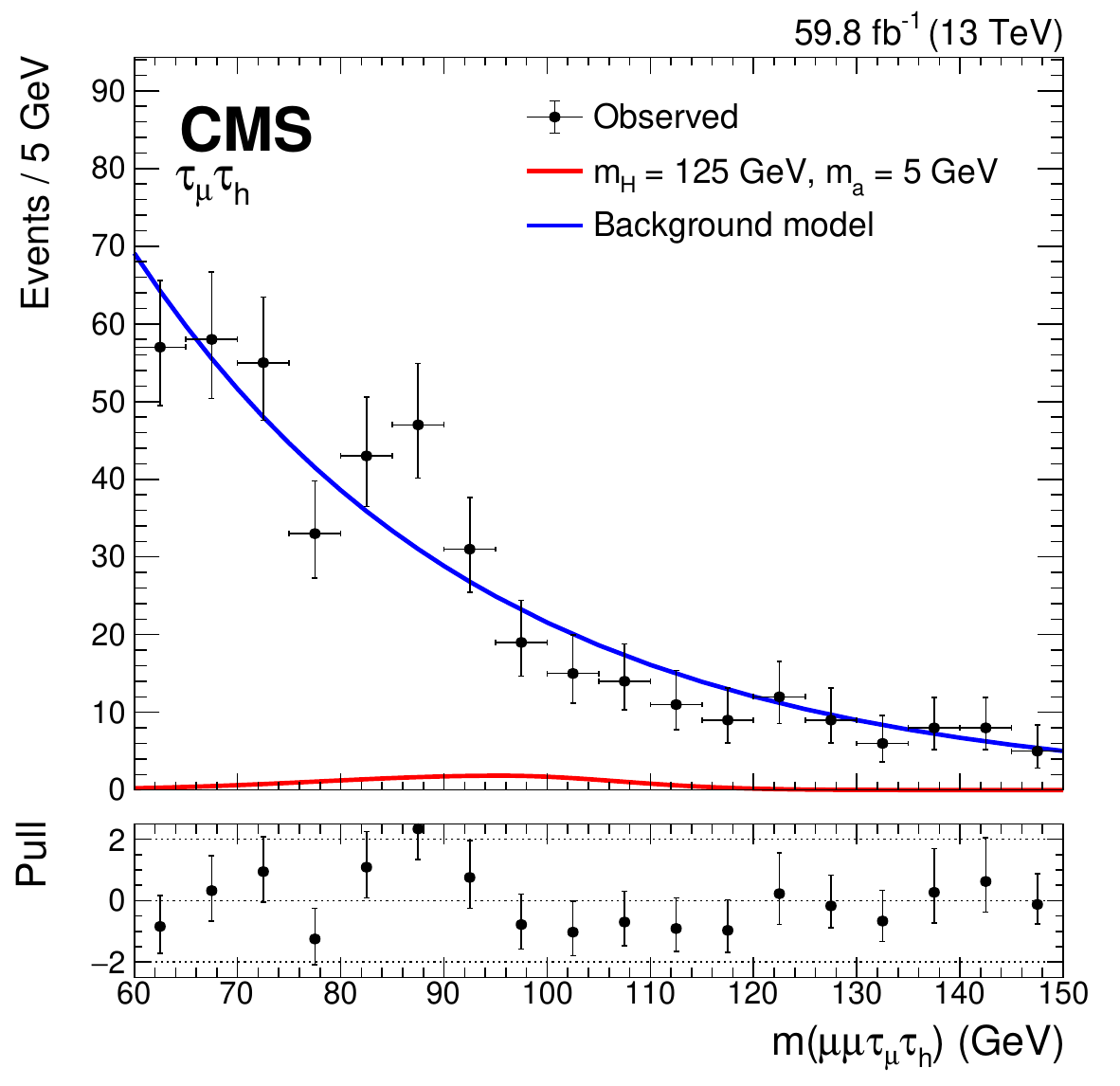} \\
  \includegraphics[width=0.4\textwidth]{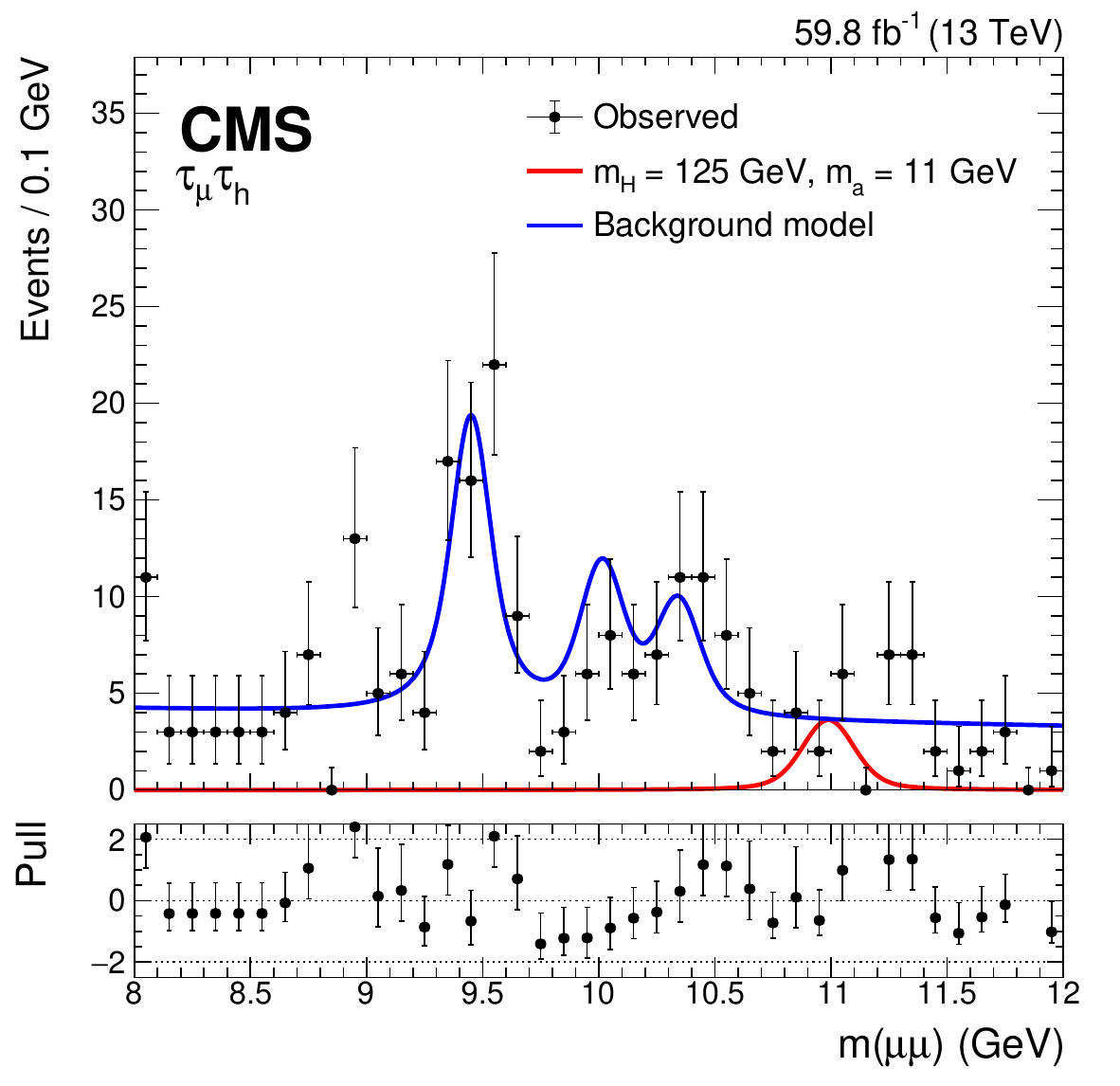} 
  \includegraphics[width=0.4\textwidth]{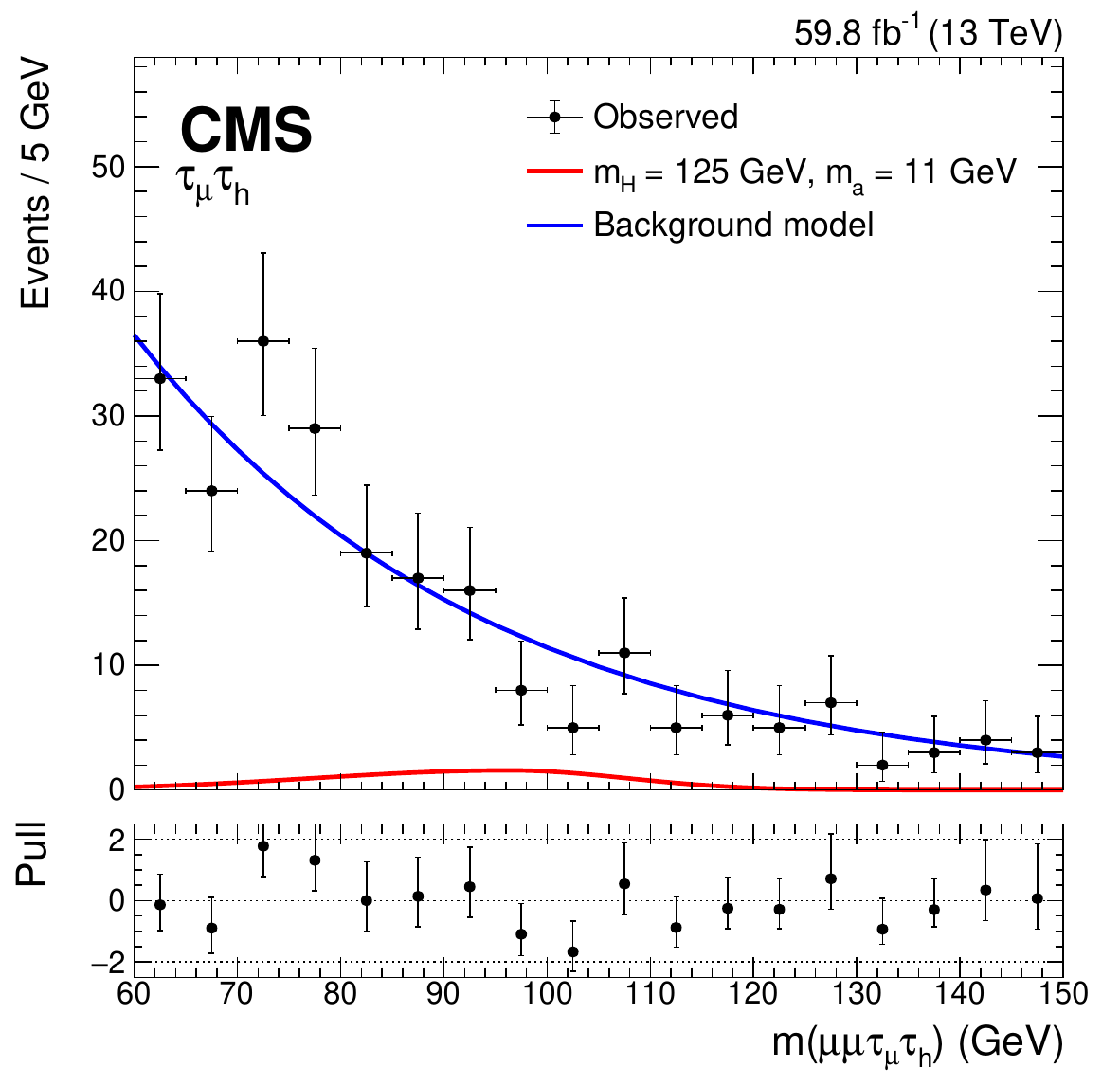} \\
  \includegraphics[width=0.4\textwidth]{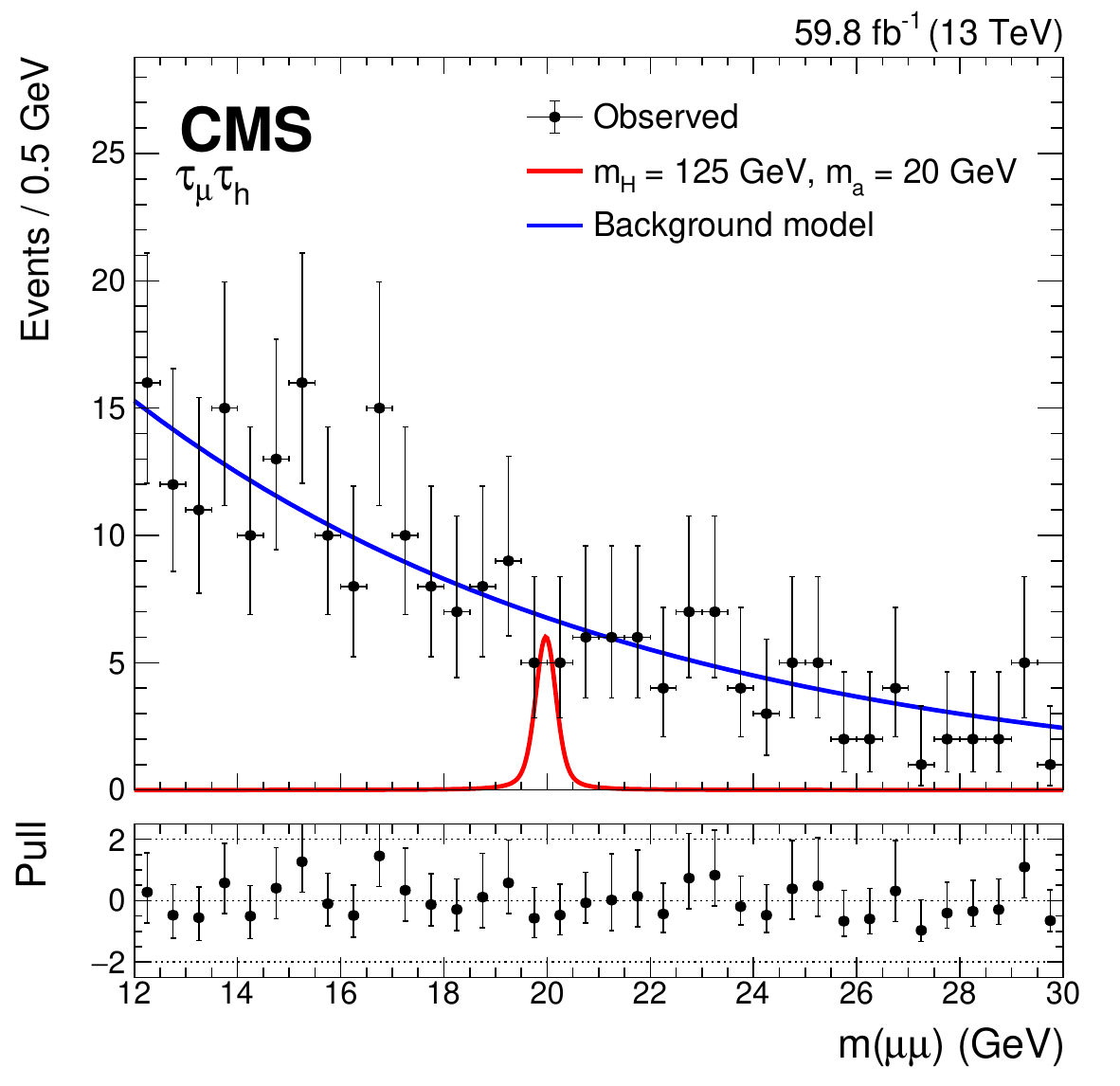} 
  \includegraphics[width=0.4\textwidth]{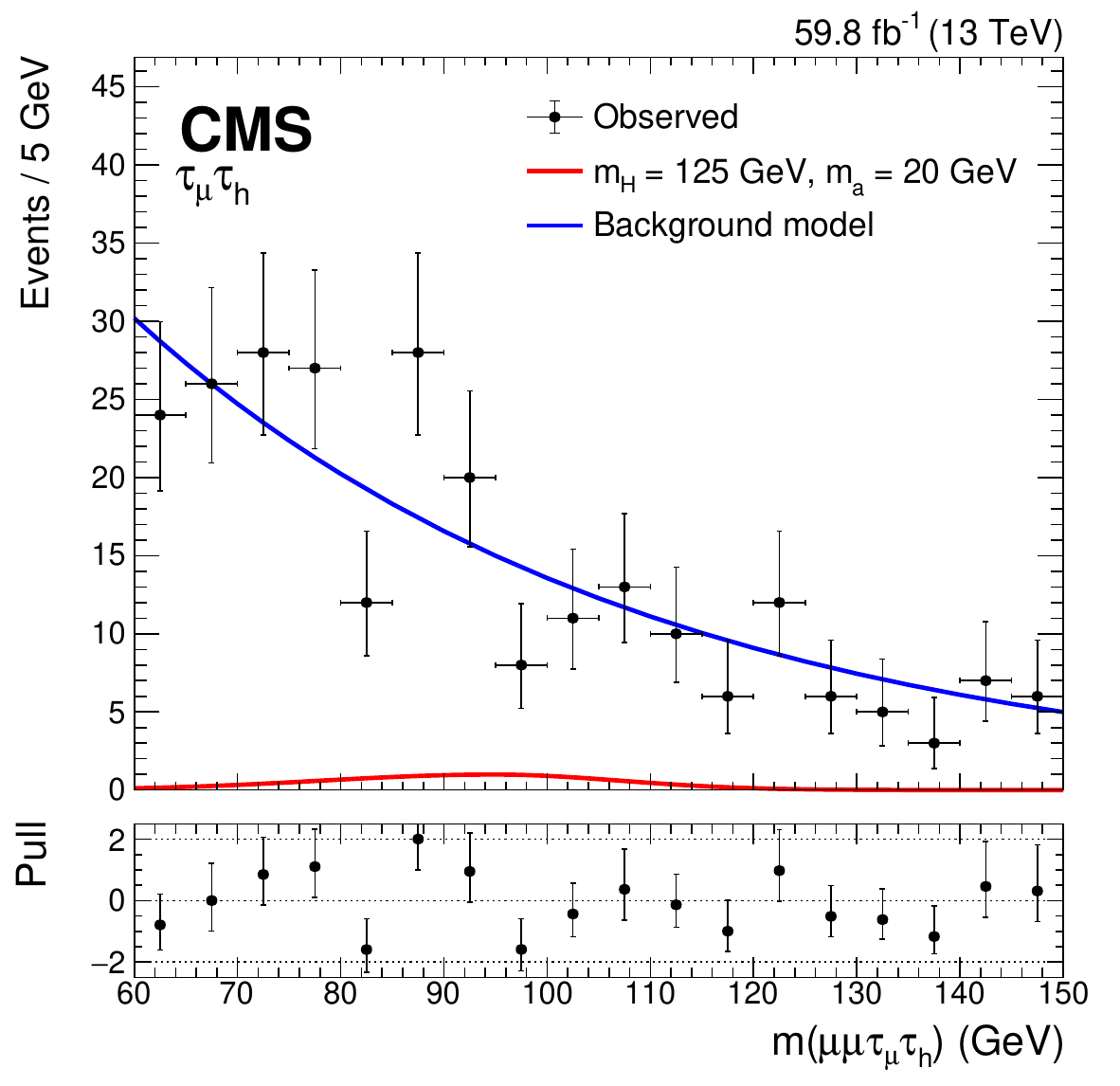} \\
  \caption{
  Projections of the post-fit two-dimensional background pdfs (blue) and the observed data onto the $m_{\PGm\PGm}$ (left) and four-body visible mass (right) axes for the $\taum\tauh$ channel for $m_{\PH}$ = 125\GeV using the 2018 data set.
  The rows correspond to pseudoscalar masses of $m_{\pa}$ = 5, 11, and 20\GeV, from top to bottom.
  Example signal distributions (red) are overlaid assuming $\mathcal{B}(\PH\to\pa\pa\to\PGm\PGm\PGt\PGt) = 5 \times 10^{-4}$.
  }
  
  \label{fig:postfitmh_H125}
\end{figure}

\begin{figure}[htbp]
  \centering
  \includegraphics[width=0.4\textwidth]{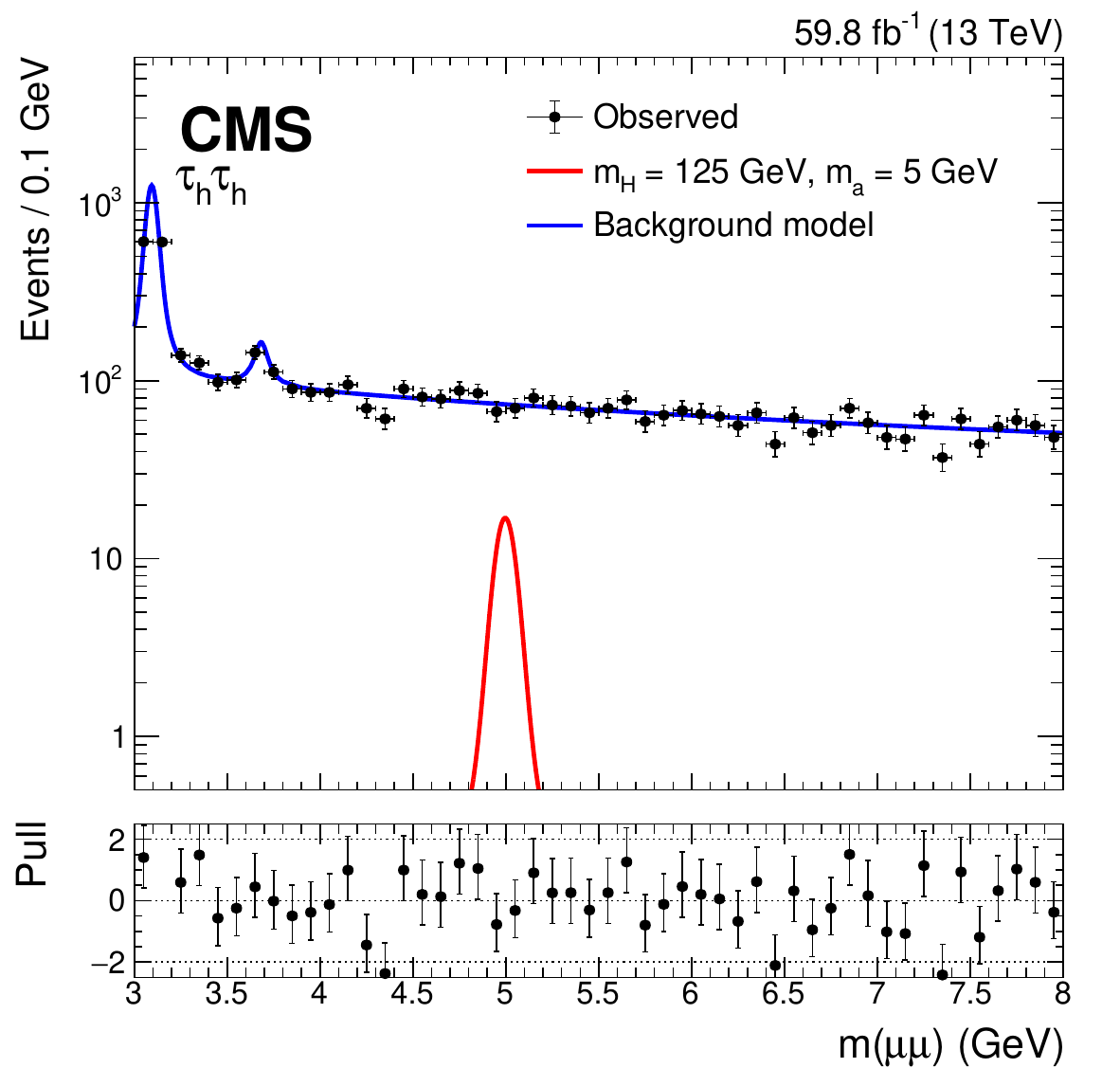}
  \includegraphics[width=0.4\textwidth]{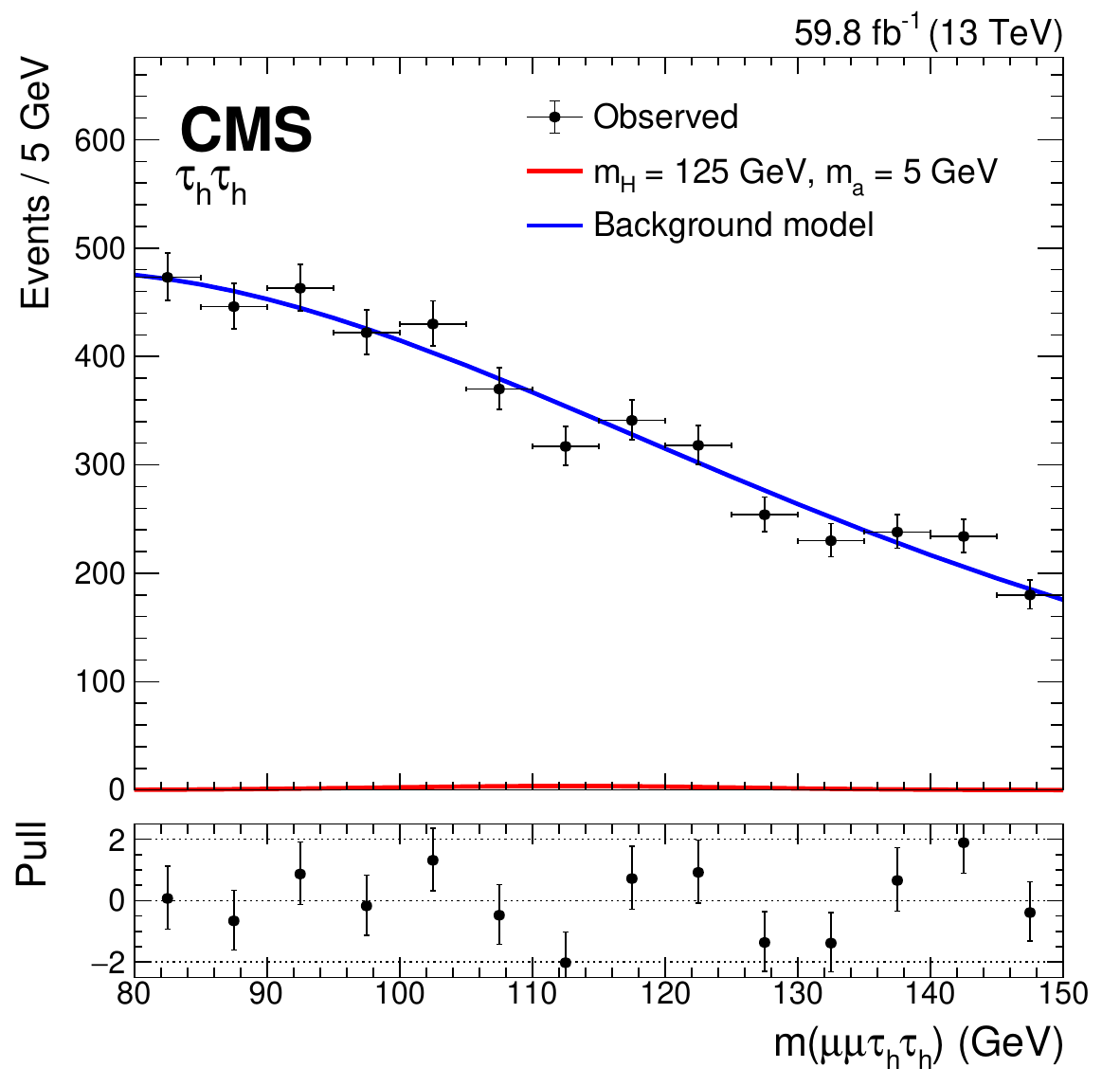} \\
  \includegraphics[width=0.4\textwidth]{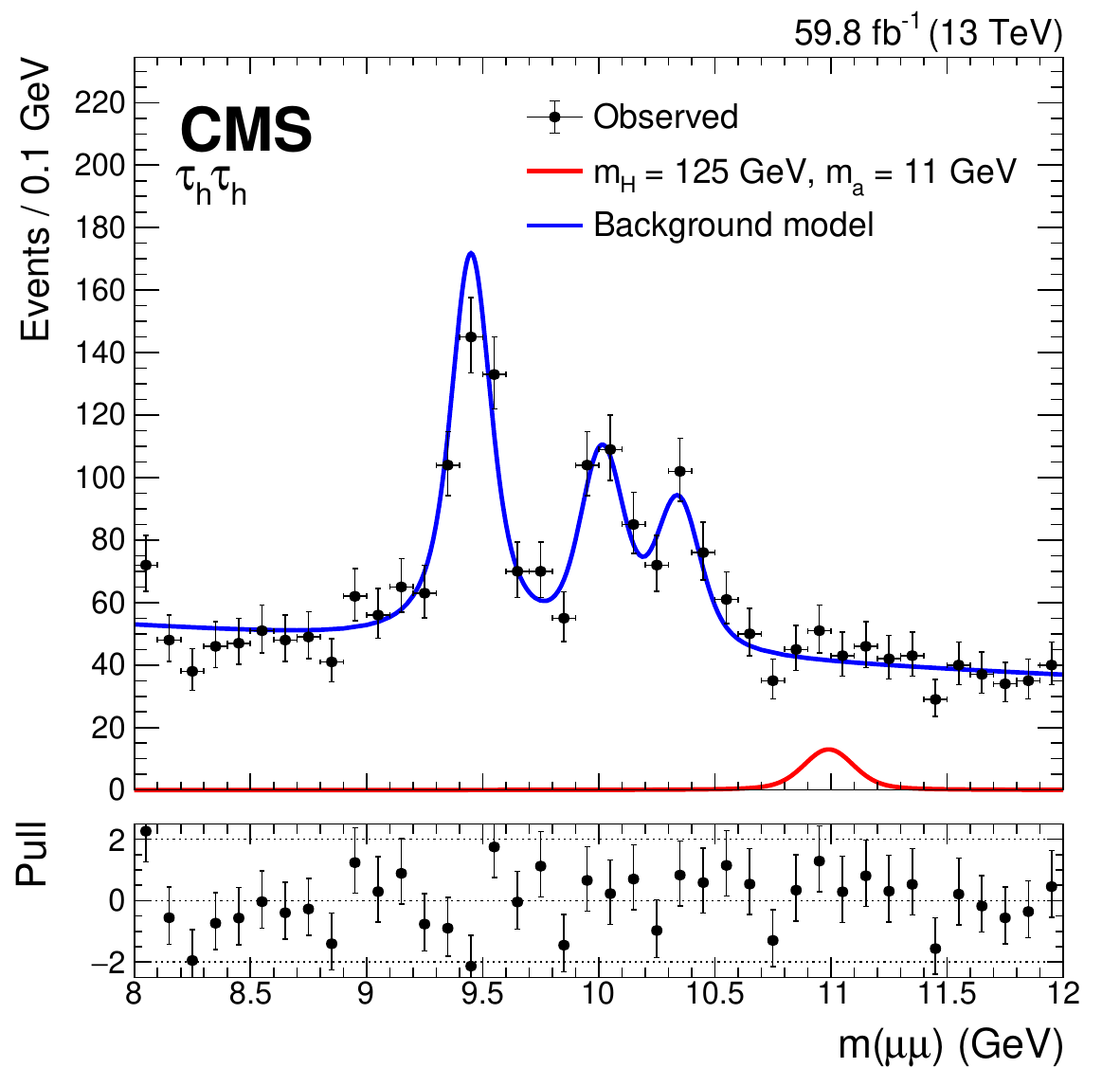} 
  \includegraphics[width=0.4\textwidth]{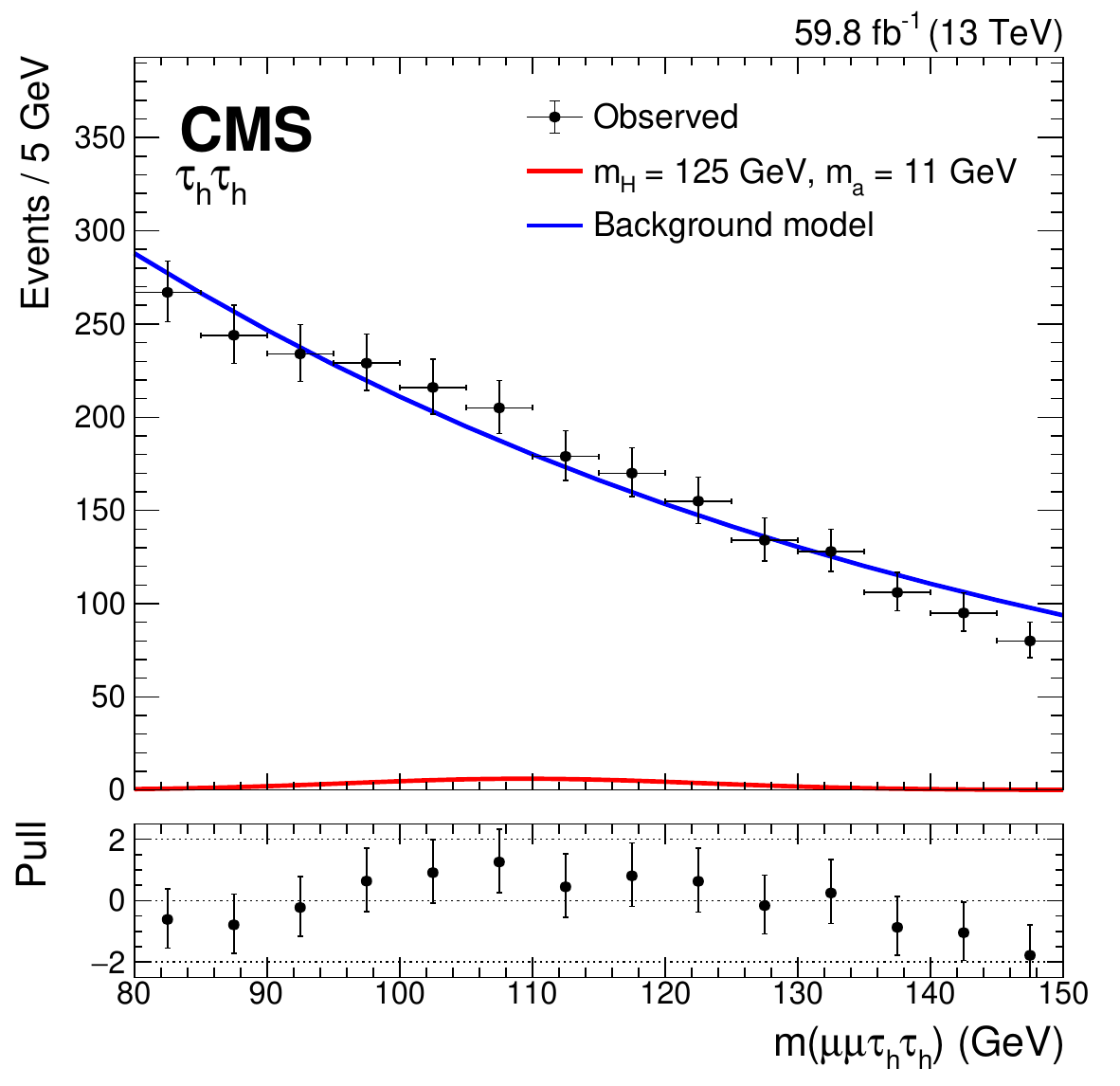} \\
  \includegraphics[width=0.4\textwidth]{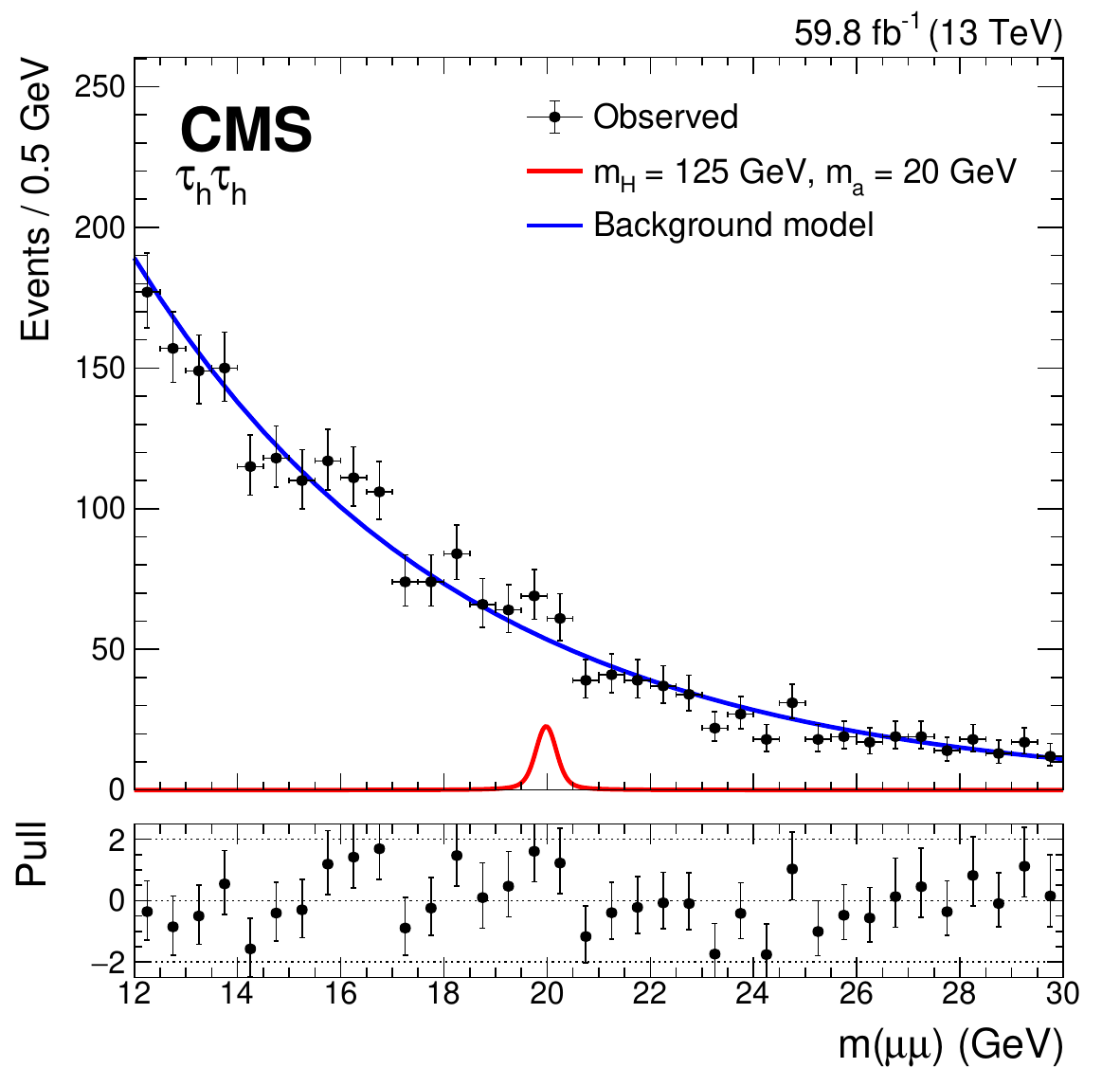} 
  \includegraphics[width=0.4\textwidth]{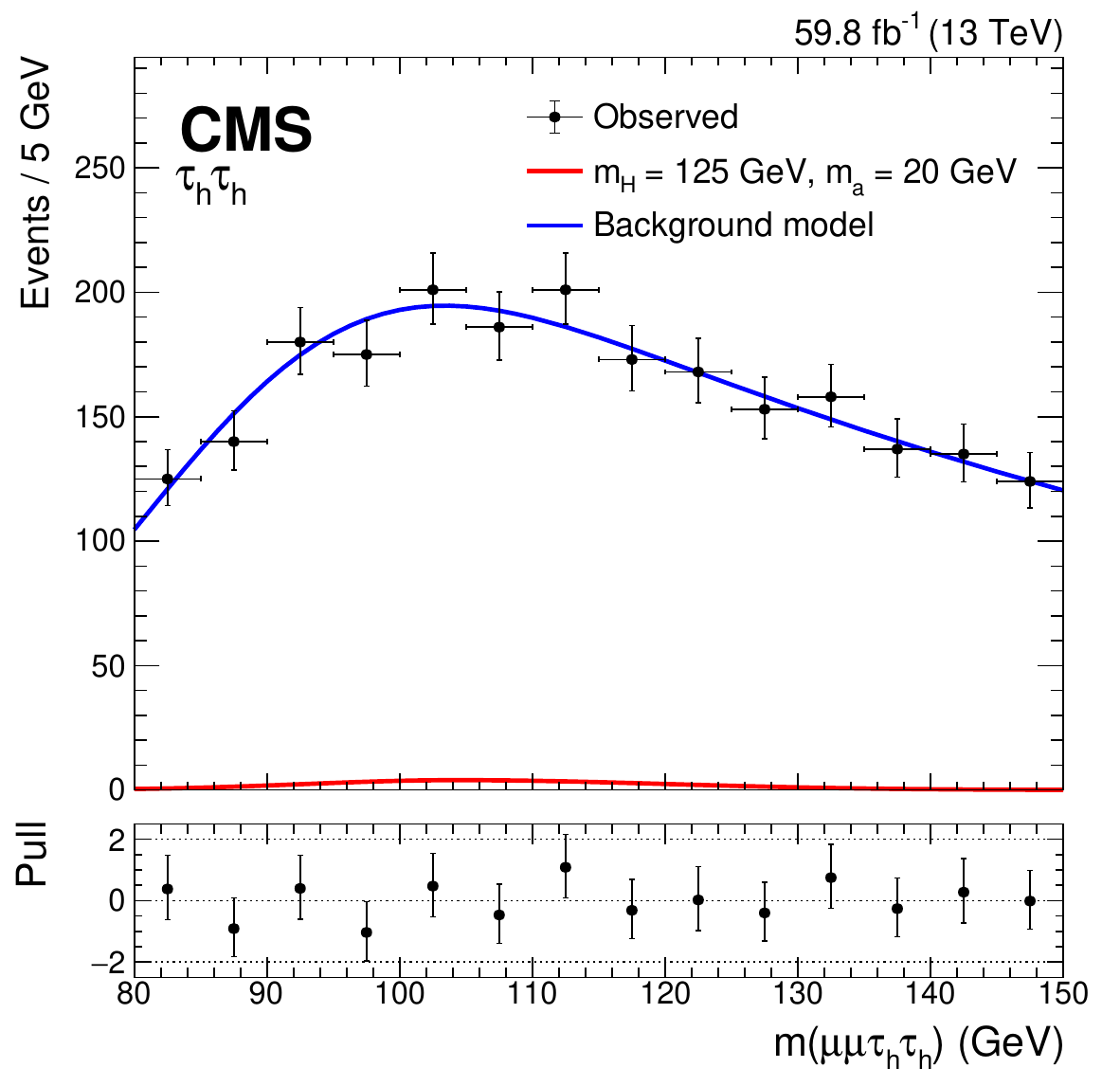} \\
  \caption{
  Projections of the post-fit two-dimensional background pdfs (blue) and the observed data onto the $m_{\PGm\PGm}$ (left) and four-body visible mass (right) axes for the $\tauh\tauh$ channel for $m_{\PH}$ = 125\GeV using the 2018 data set.
  The rows correspond to pseudoscalar masses of $m_{\pa}$ = 5, 11, and 20\GeV, from top to bottom.
  Example signal distributions (red) are overlaid assuming $\mathcal{B}(\PH\to\pa\pa\to\PGm\PGm\PGt\PGt) = 5 \times 10^{-4}$.
  }
  \label{fig:postfithh_H125}
\end{figure}

\begin{figure}[htbp]
  \centering
  \includegraphics[width=0.4\textwidth]{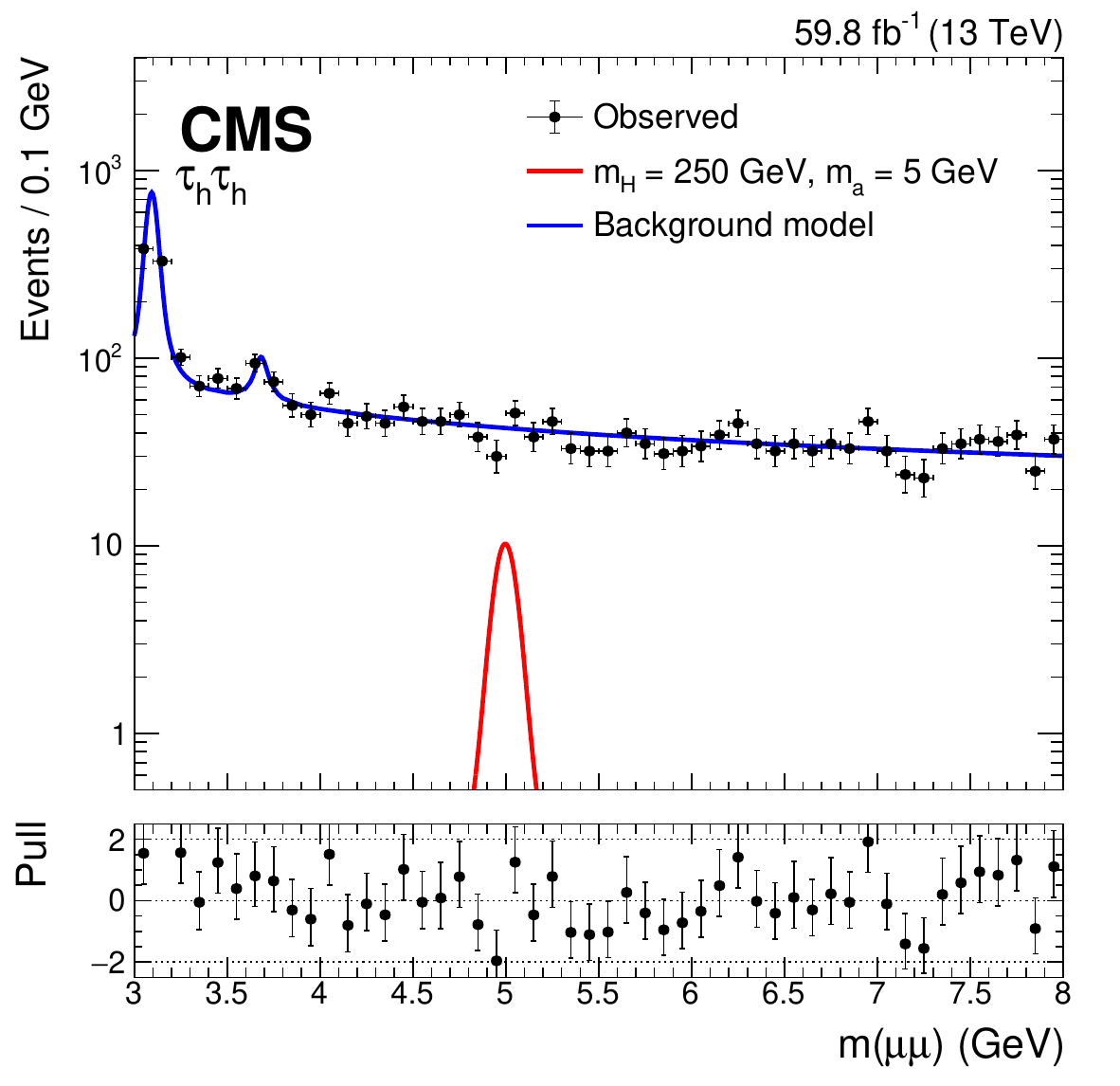}
  \includegraphics[width=0.4\textwidth]{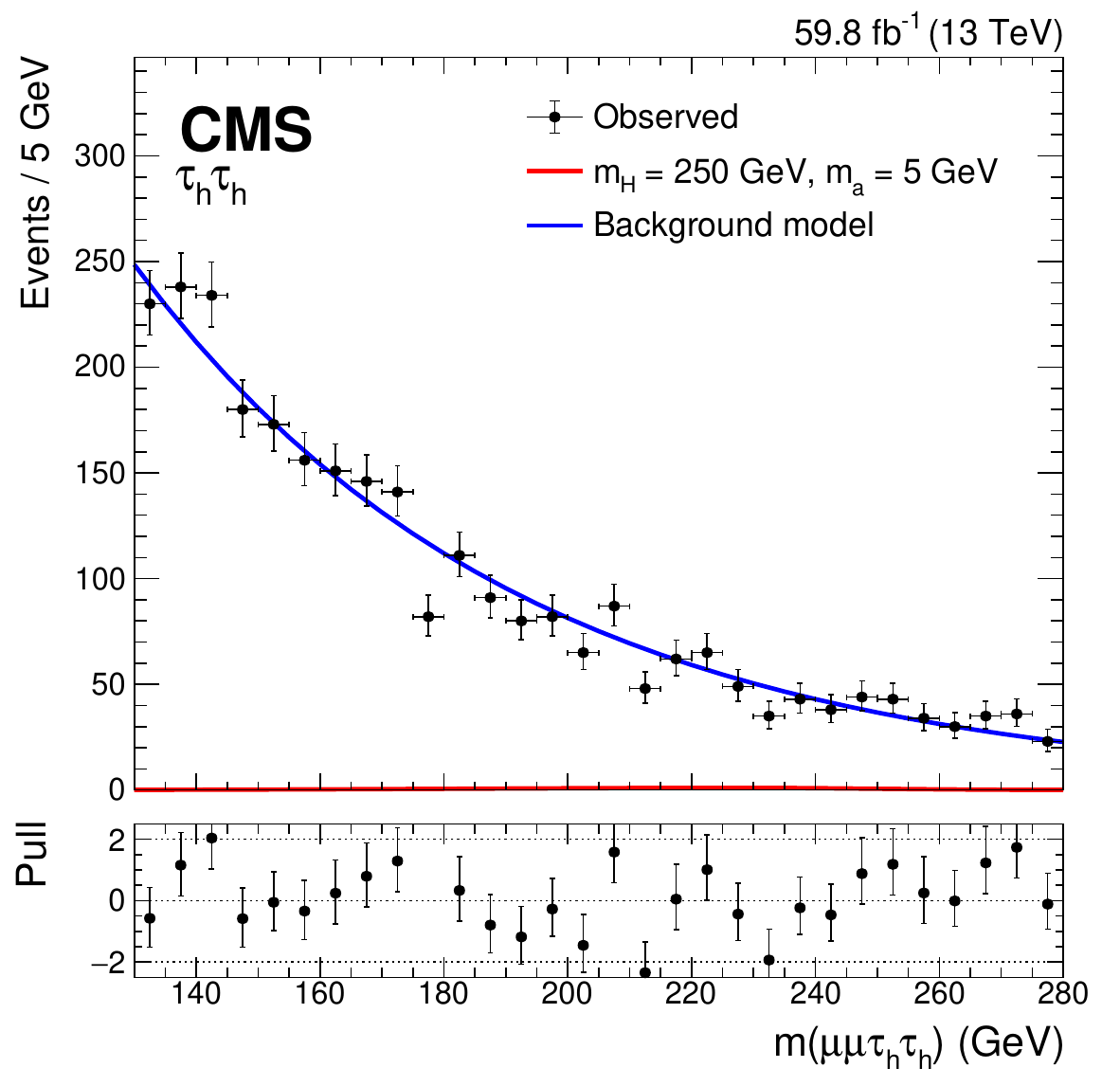} \\
  \includegraphics[width=0.4\textwidth]{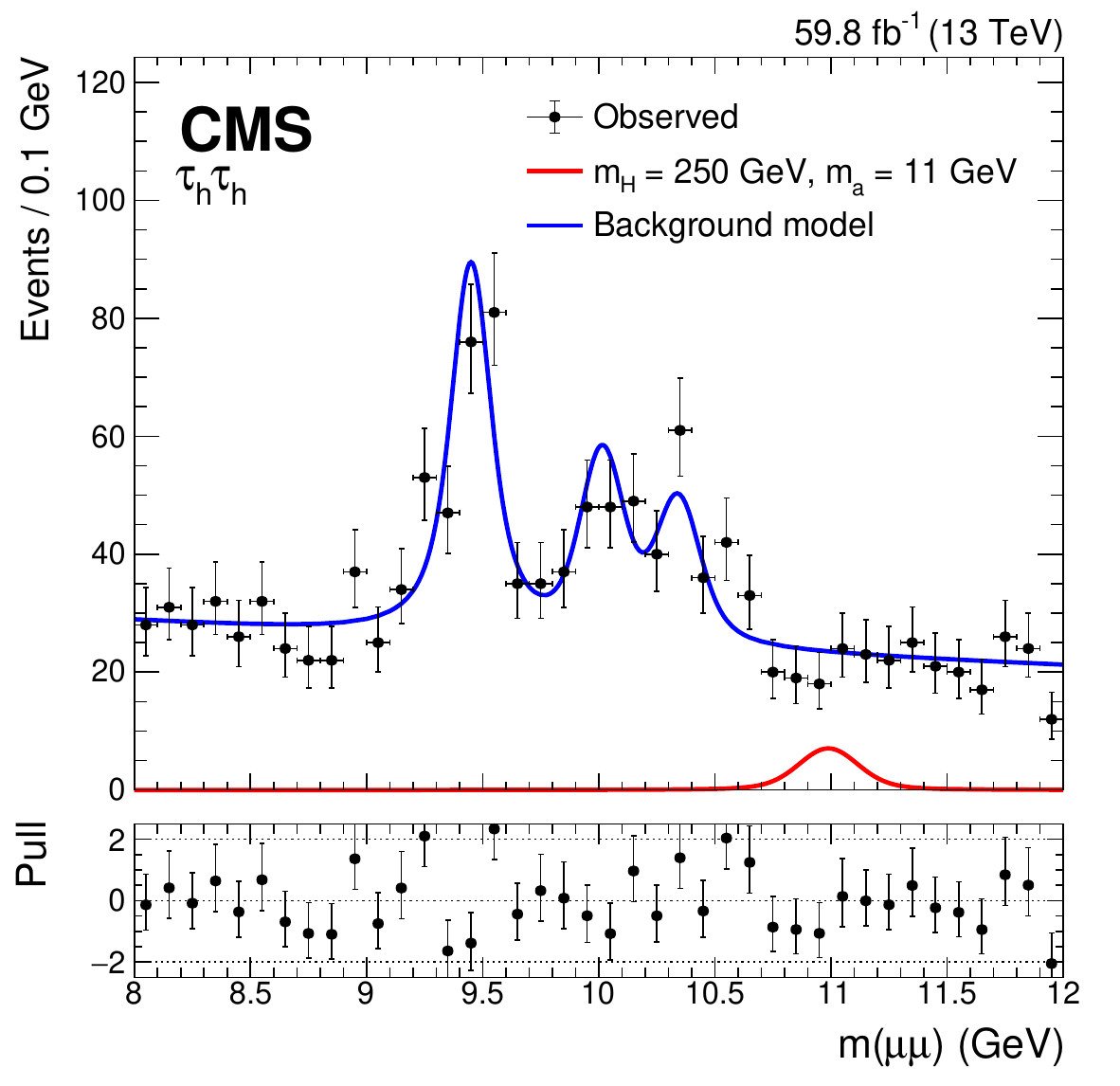} 
  \includegraphics[width=0.4\textwidth]{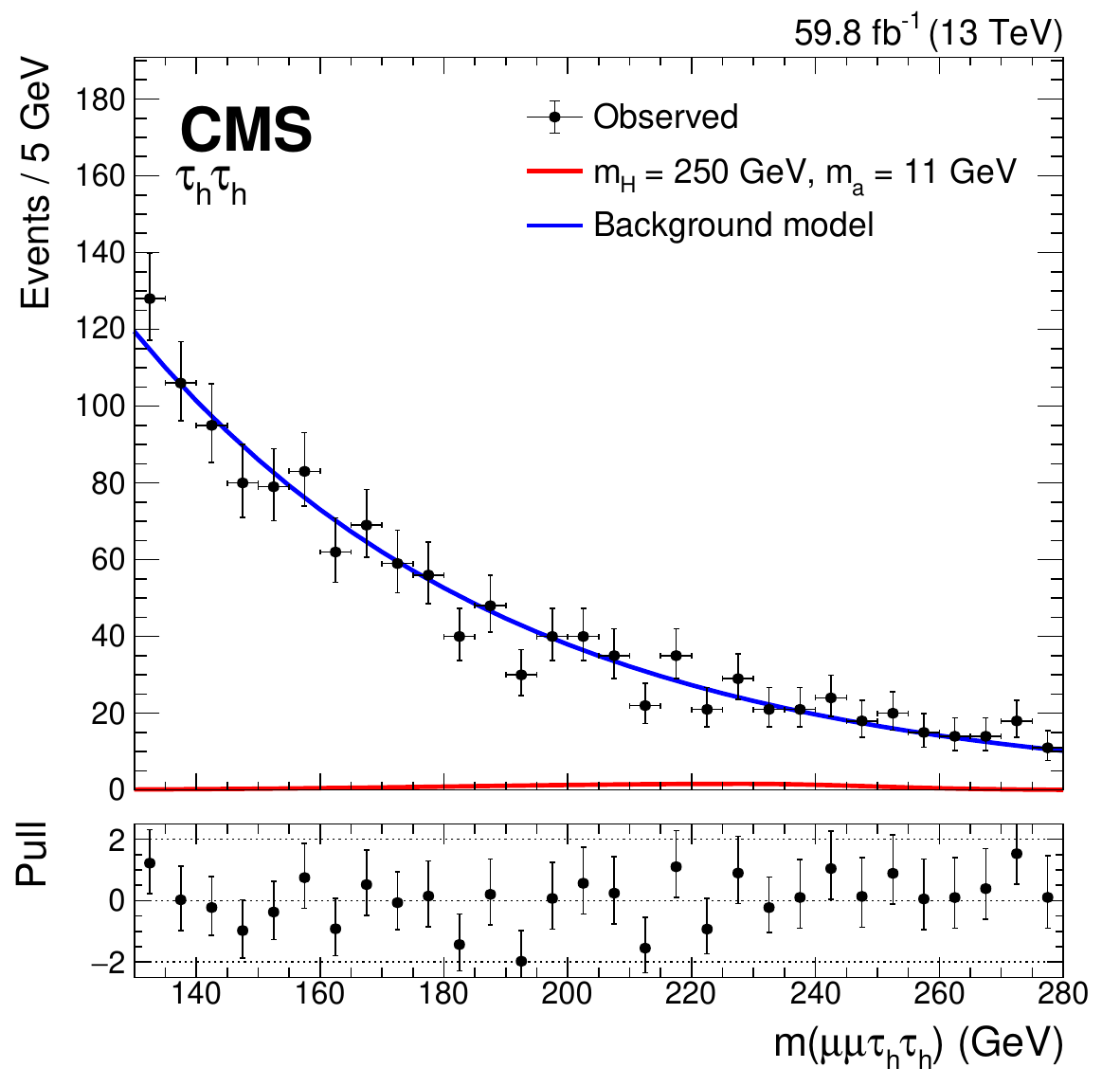} \\
  \includegraphics[width=0.4\textwidth]{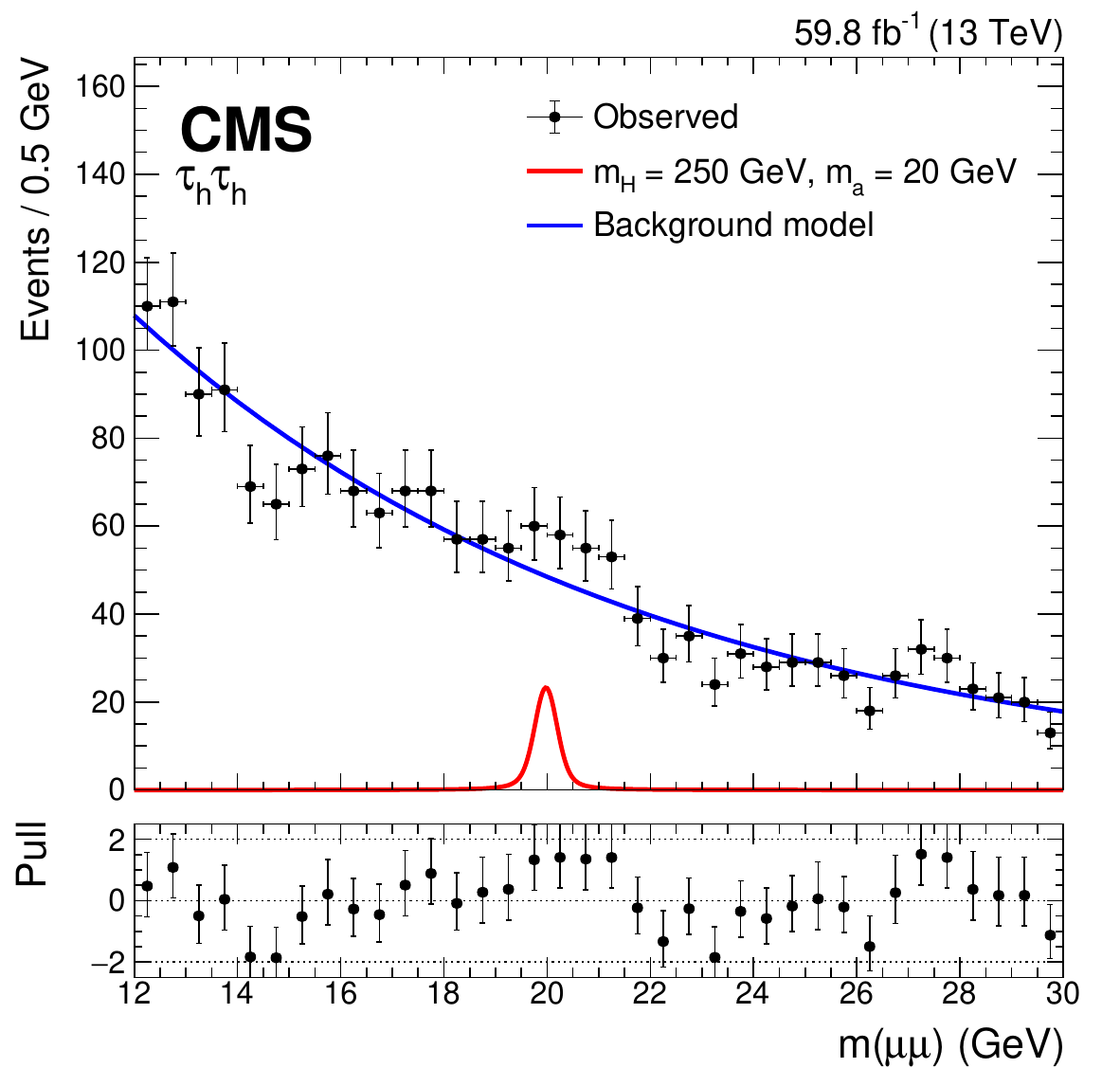} 
  \includegraphics[width=0.4\textwidth]{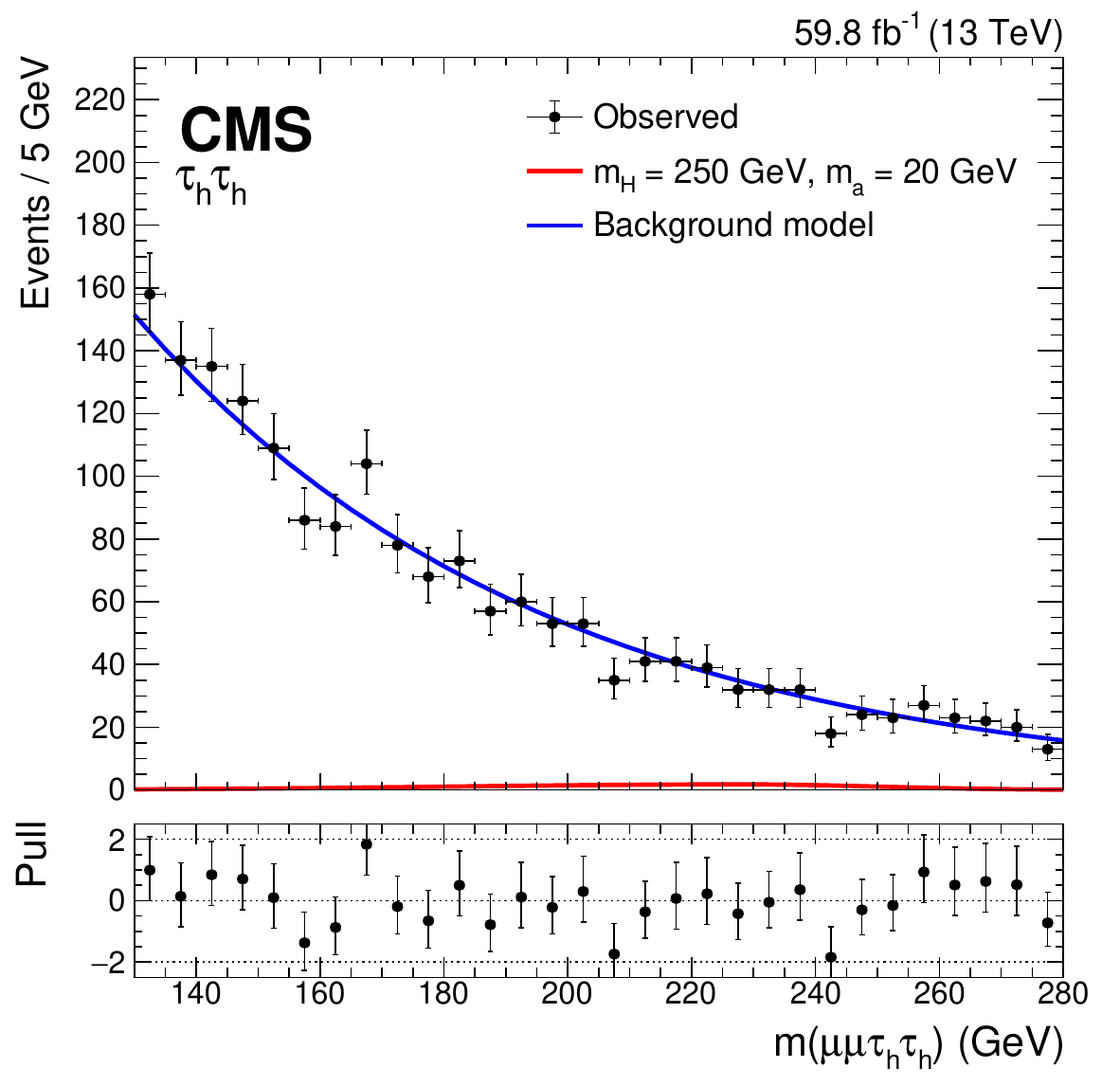} \\
  \caption{
  Projections of the post-fit two-dimensional background pdfs (blue) and the observed data onto the $m_{\PGm\PGm}$ (left) and four-body visible mass (right) axes for the $\tauh\tauh$ channel for $m_{\PH}$ = 250\GeV using the 2018 data set.
  The rows correspond to pseudoscalar masses of $m_{\pa}$ = 5, 11, and 20\GeV, from top to bottom.
  Example signal distributions (red) are overlaid assuming $\mathcal{B}(\PH\to\pa\pa\to\PGm\PGm\PGt\PGt) = 5 \times 10^{-4}$.
  }
  \label{fig:postfithh_H250}
\end{figure}

\begin{figure}[htbp]
  \centering
  \includegraphics[width=0.4\textwidth]{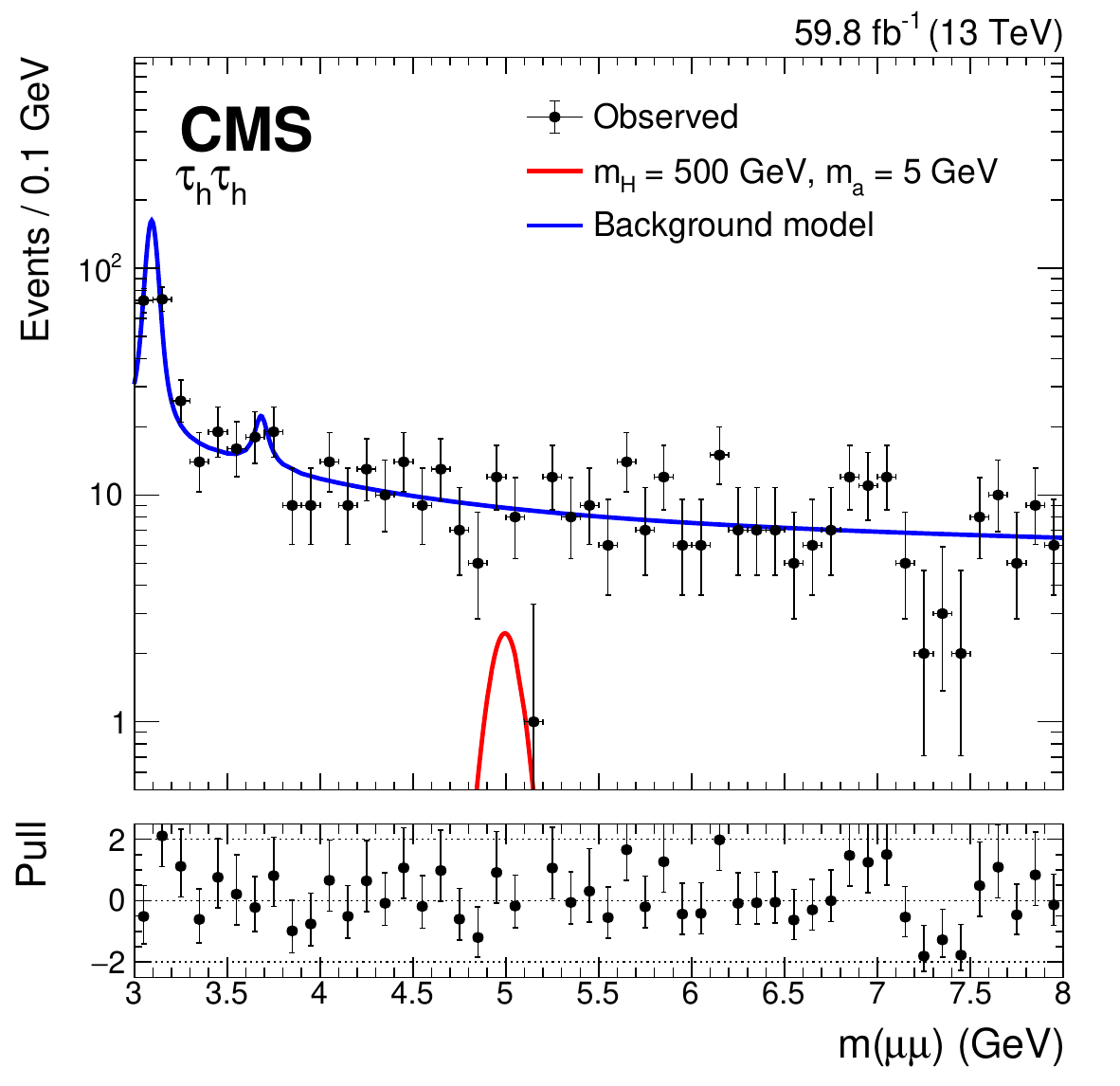}
  \includegraphics[width=0.4\textwidth]{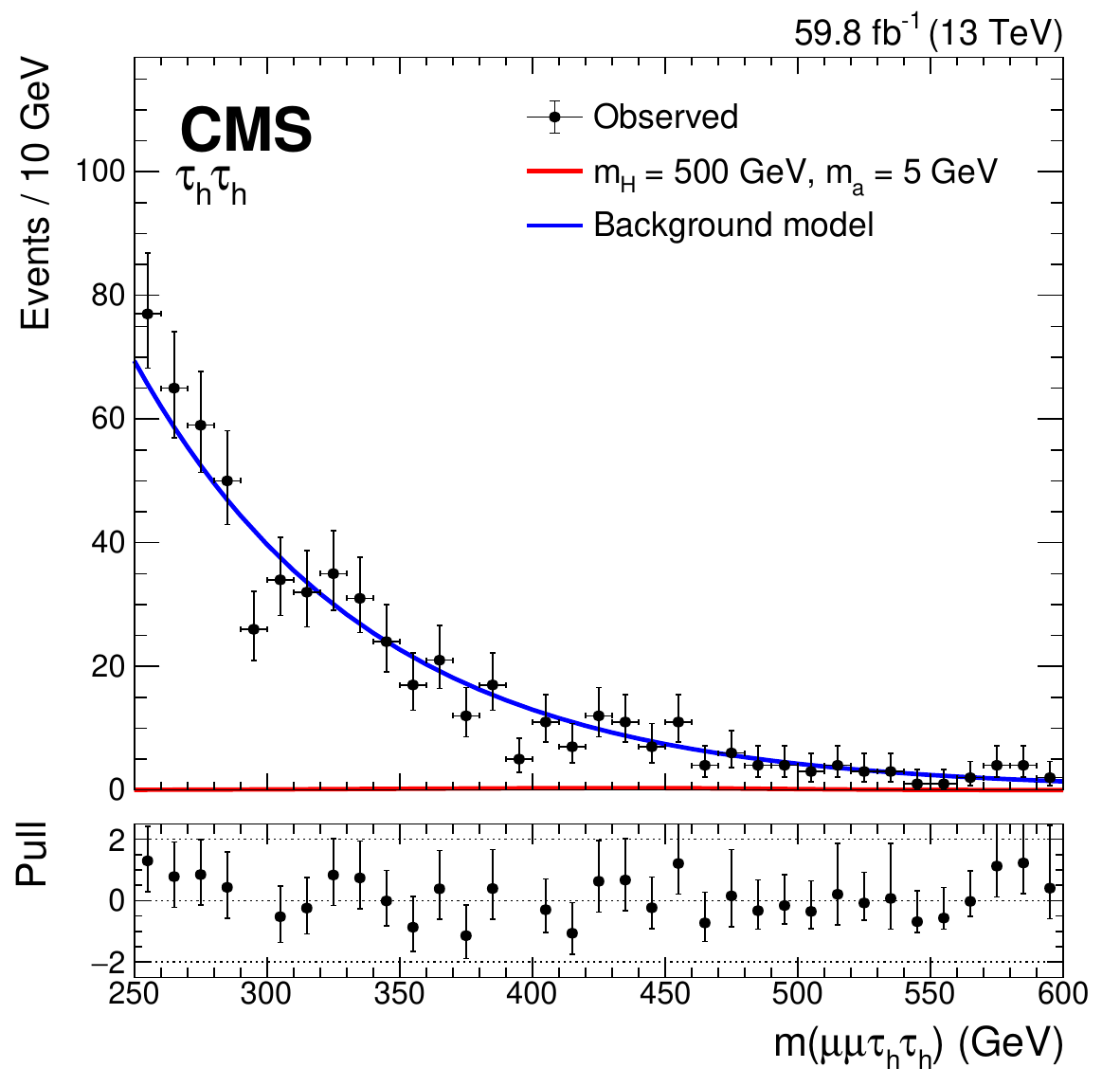} \\
  \includegraphics[width=0.4\textwidth]{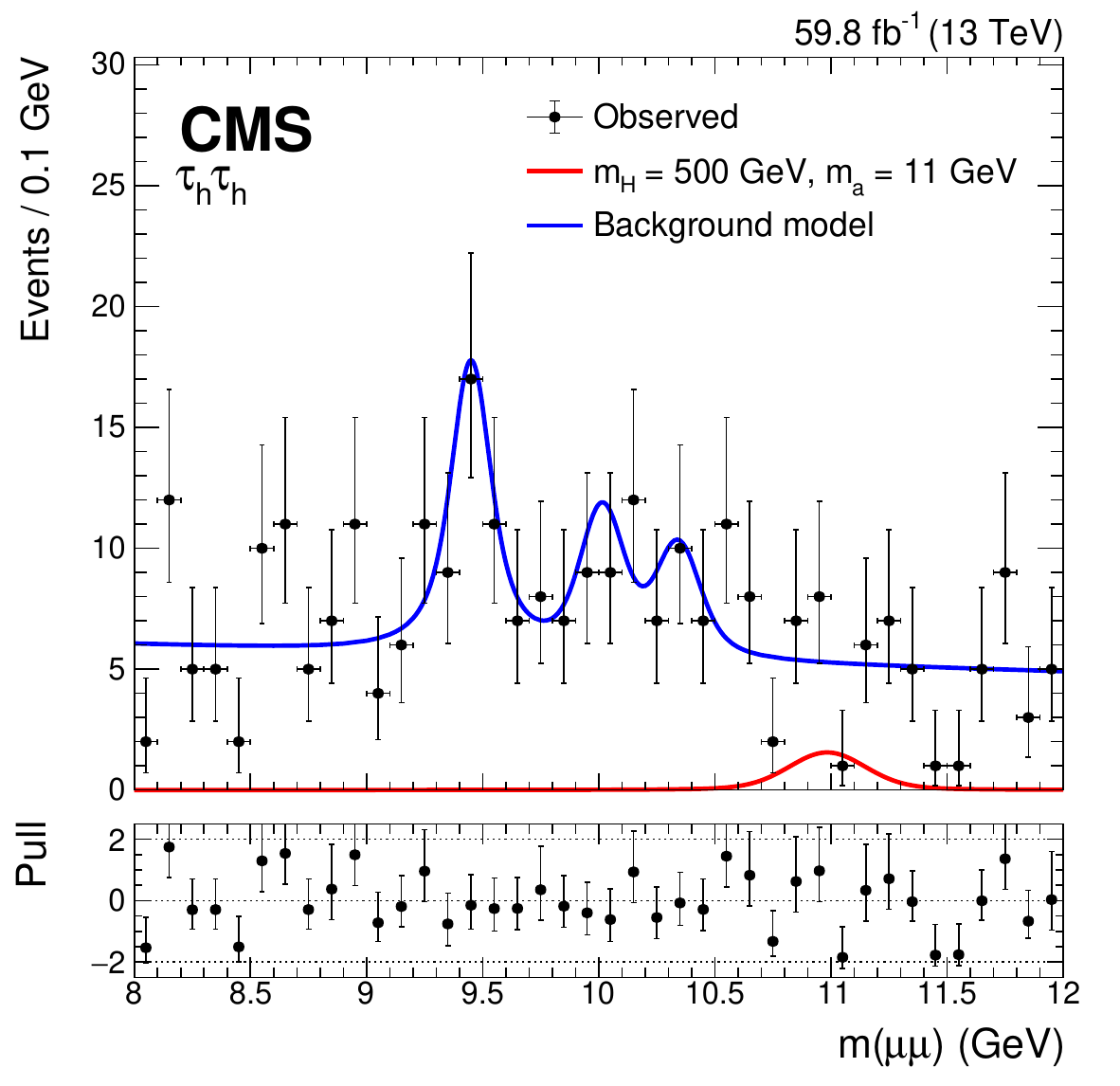} 
  \includegraphics[width=0.4\textwidth]{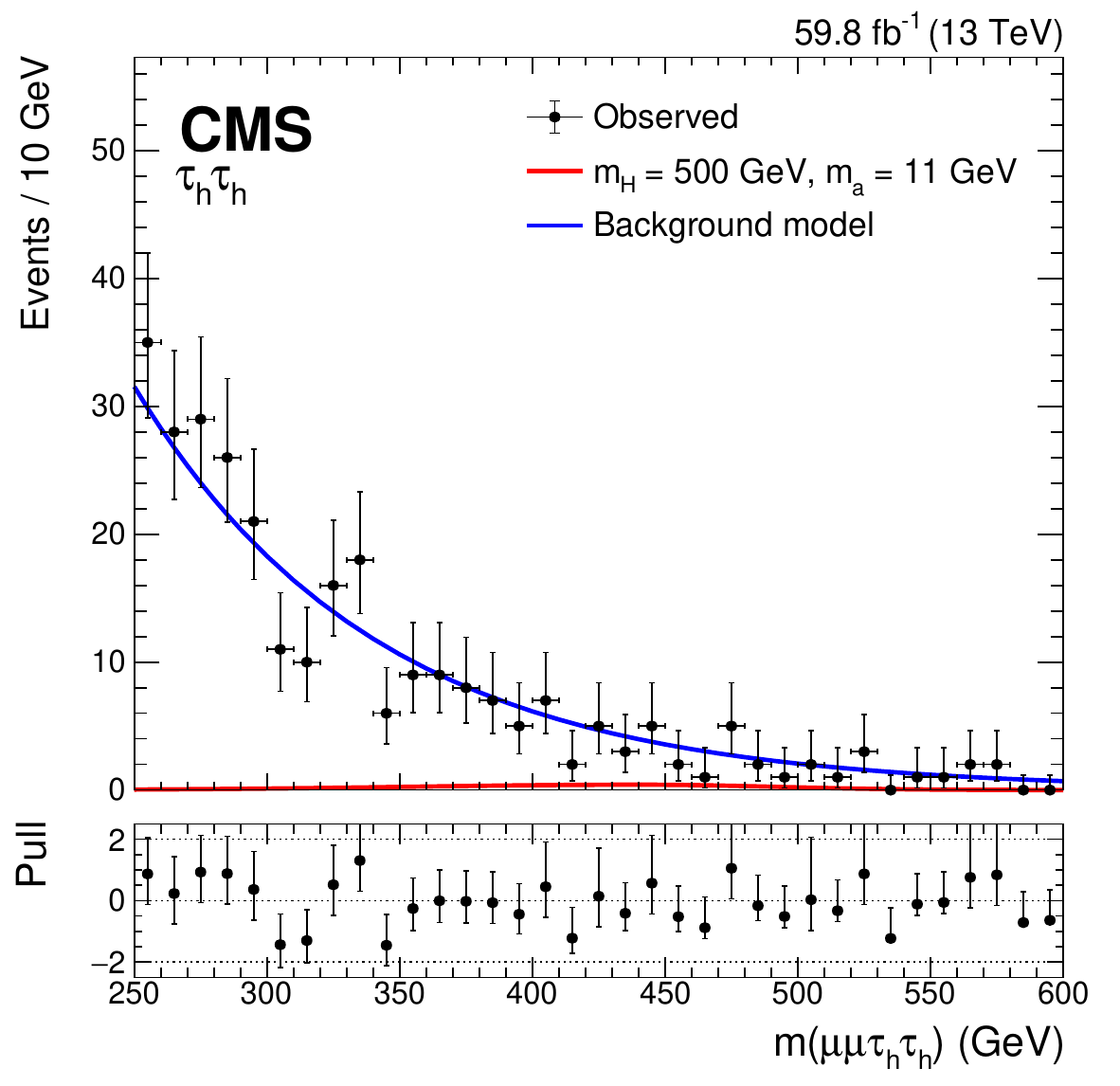} \\
  \includegraphics[width=0.4\textwidth]{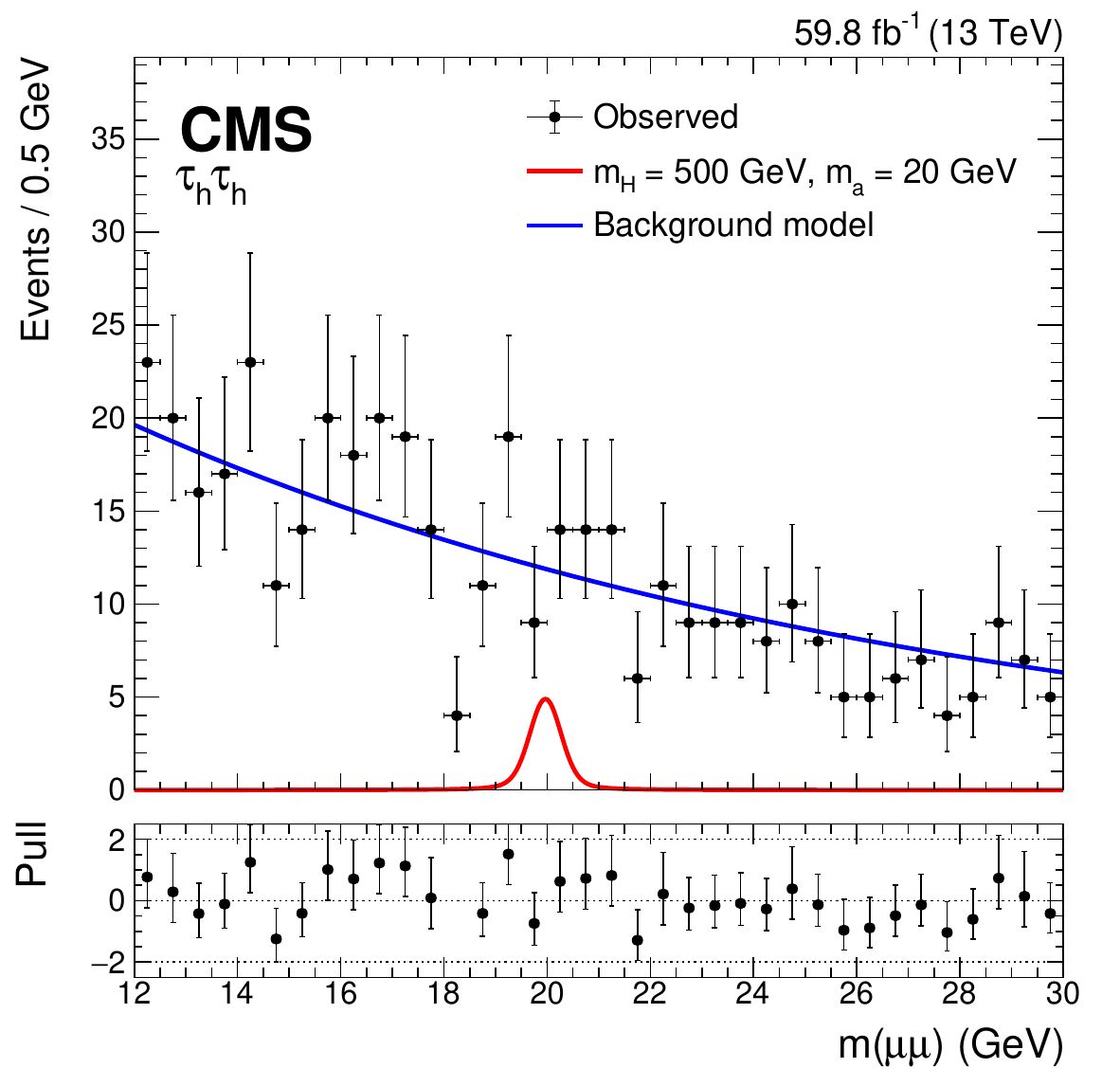} 
  \includegraphics[width=0.4\textwidth]{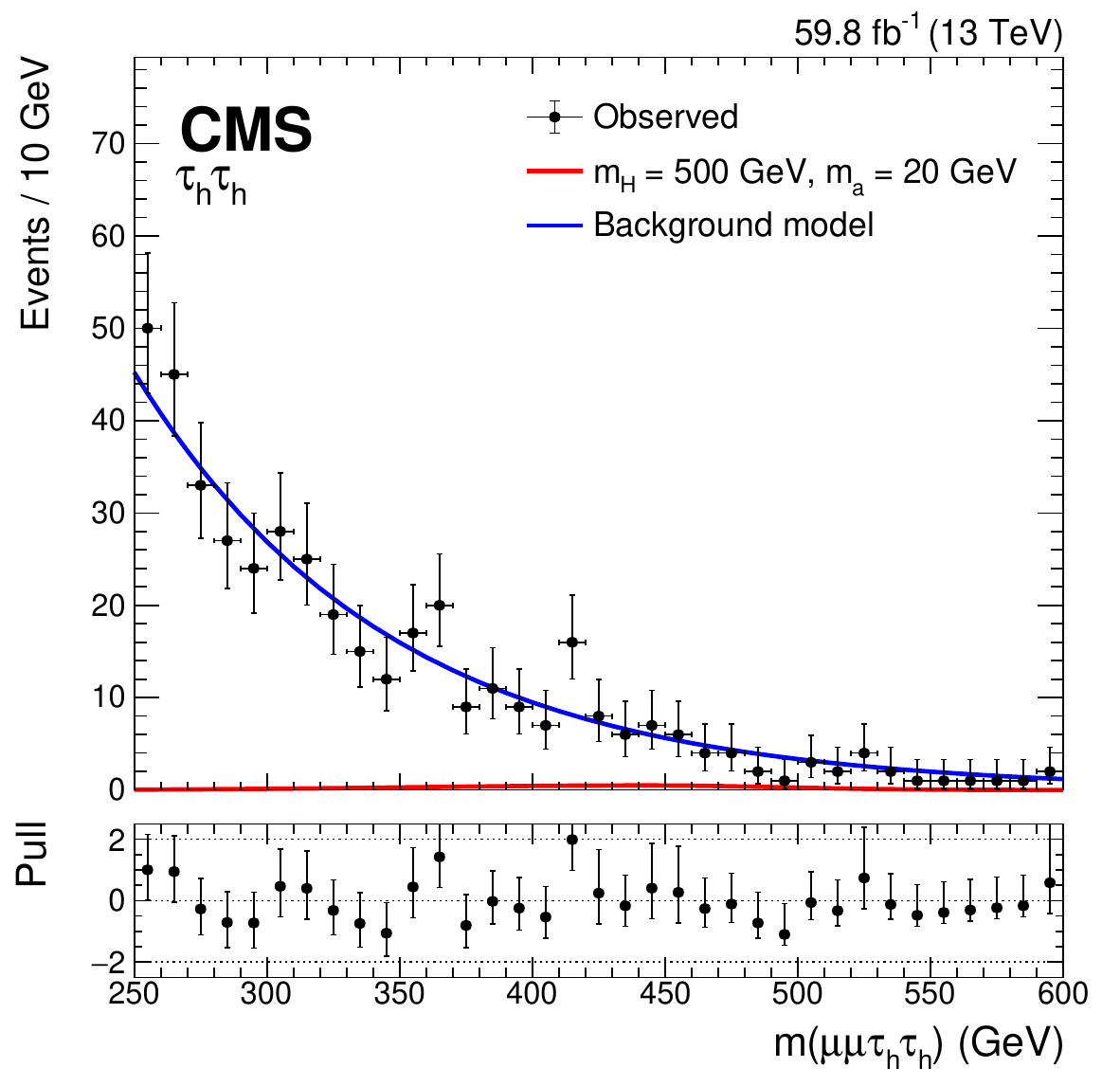} \\
  \caption{
  Projections of the post-fit two-dimensional background pdfs (blue) and the observed data onto the $m_{\PGm\PGm}$ (left) and four-body visible mass (right) axes for the $\tauh\tauh$ channel for $m_{\PH}$ = 500\GeV using the 2018 data set.
  The rows correspond to pseudoscalar masses of $m_{\pa}$ = 5, 11, and 20\GeV, from top to bottom.
  Example signal distributions (red) are overlaid assuming $\mathcal{B}(\PH\to\pa\pa\to\PGm\PGm\PGt\PGt) = 5 \times 10^{-4}$.
}
  \label{fig:postfithh_H500}
\end{figure}

\begin{figure}[htbp]
  \centering
  \includegraphics[width=0.4\textwidth]{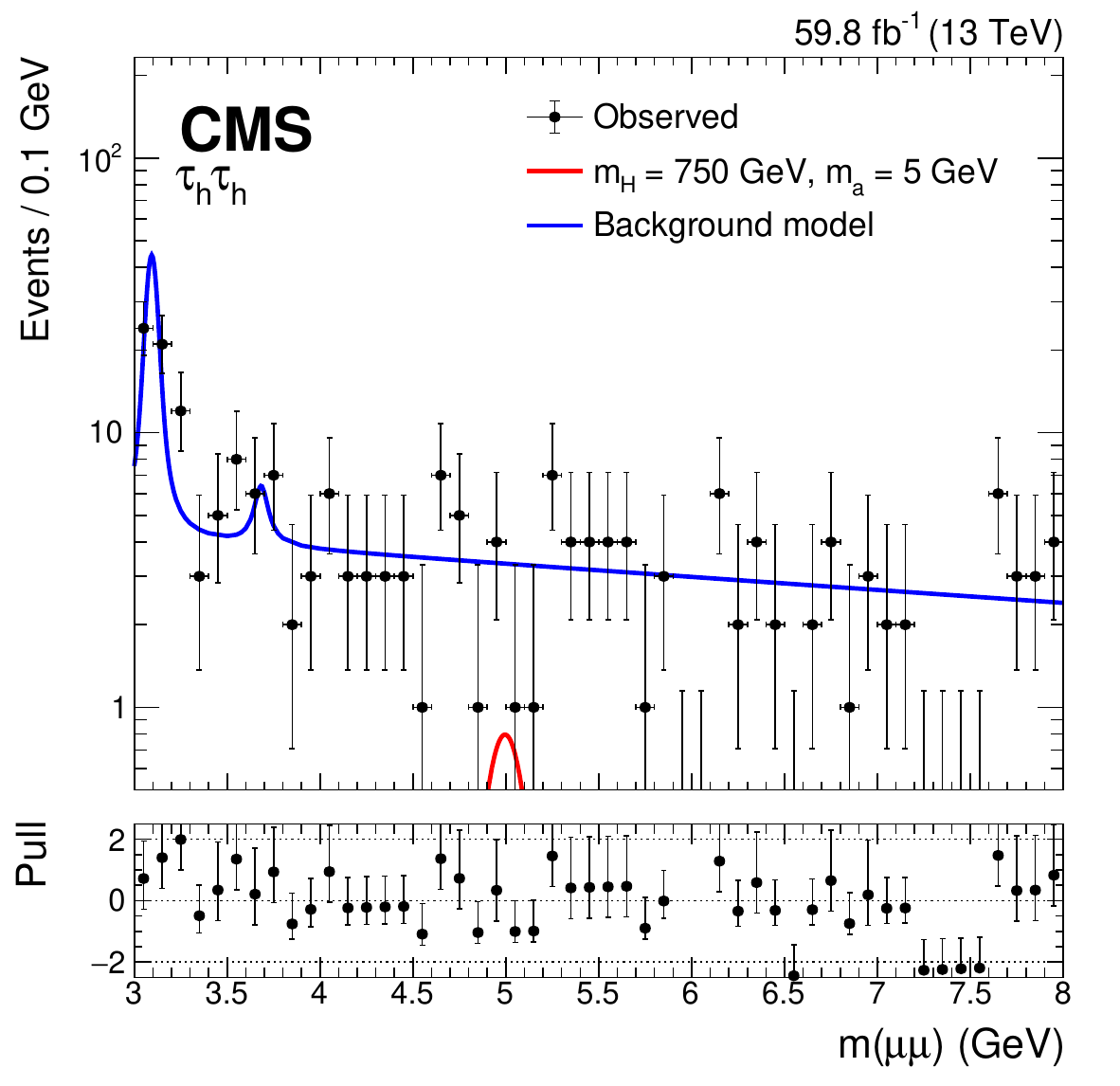}
  \includegraphics[width=0.4\textwidth]{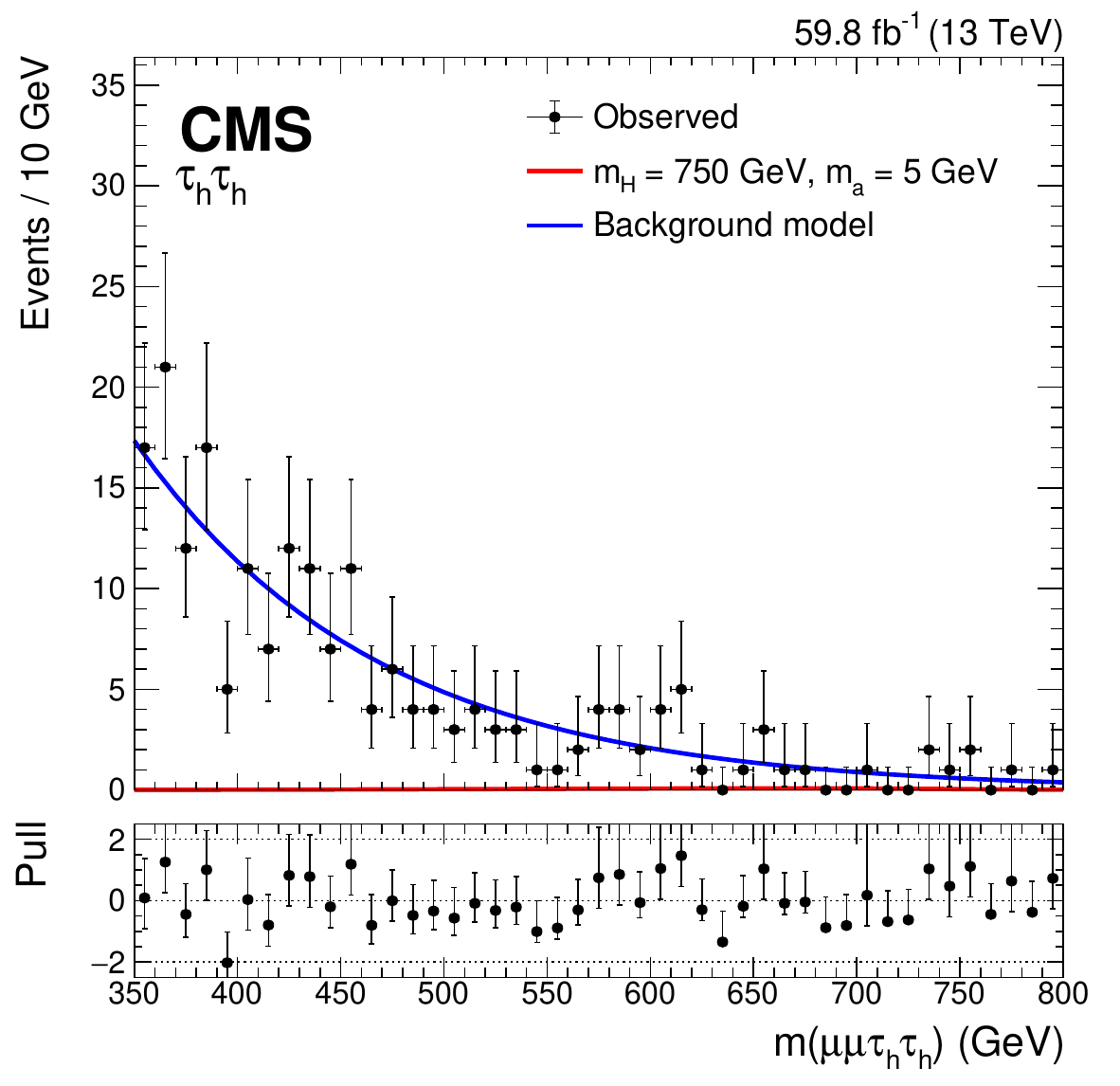} \\
  \includegraphics[width=0.4\textwidth]{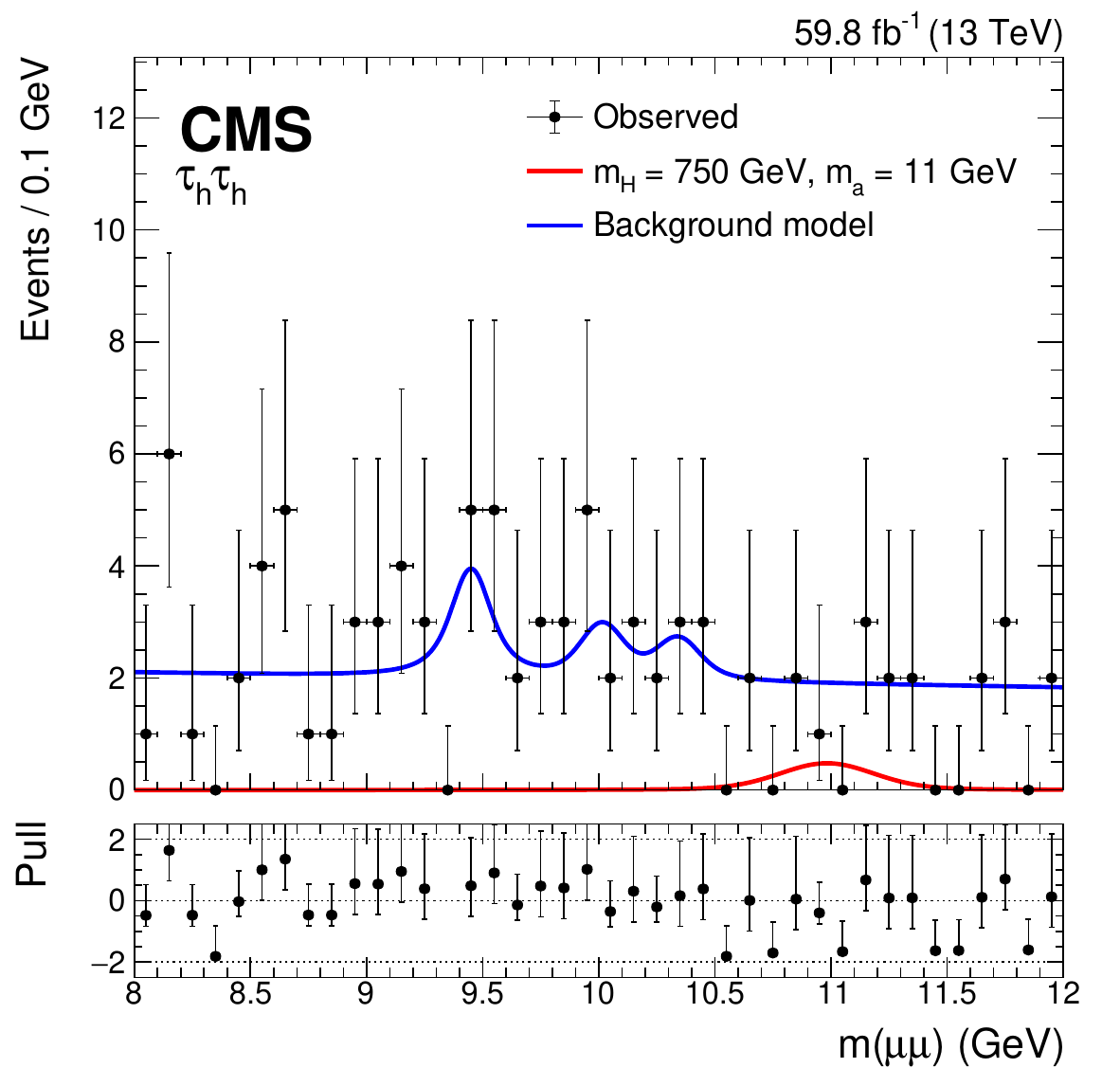} 
  \includegraphics[width=0.4\textwidth]{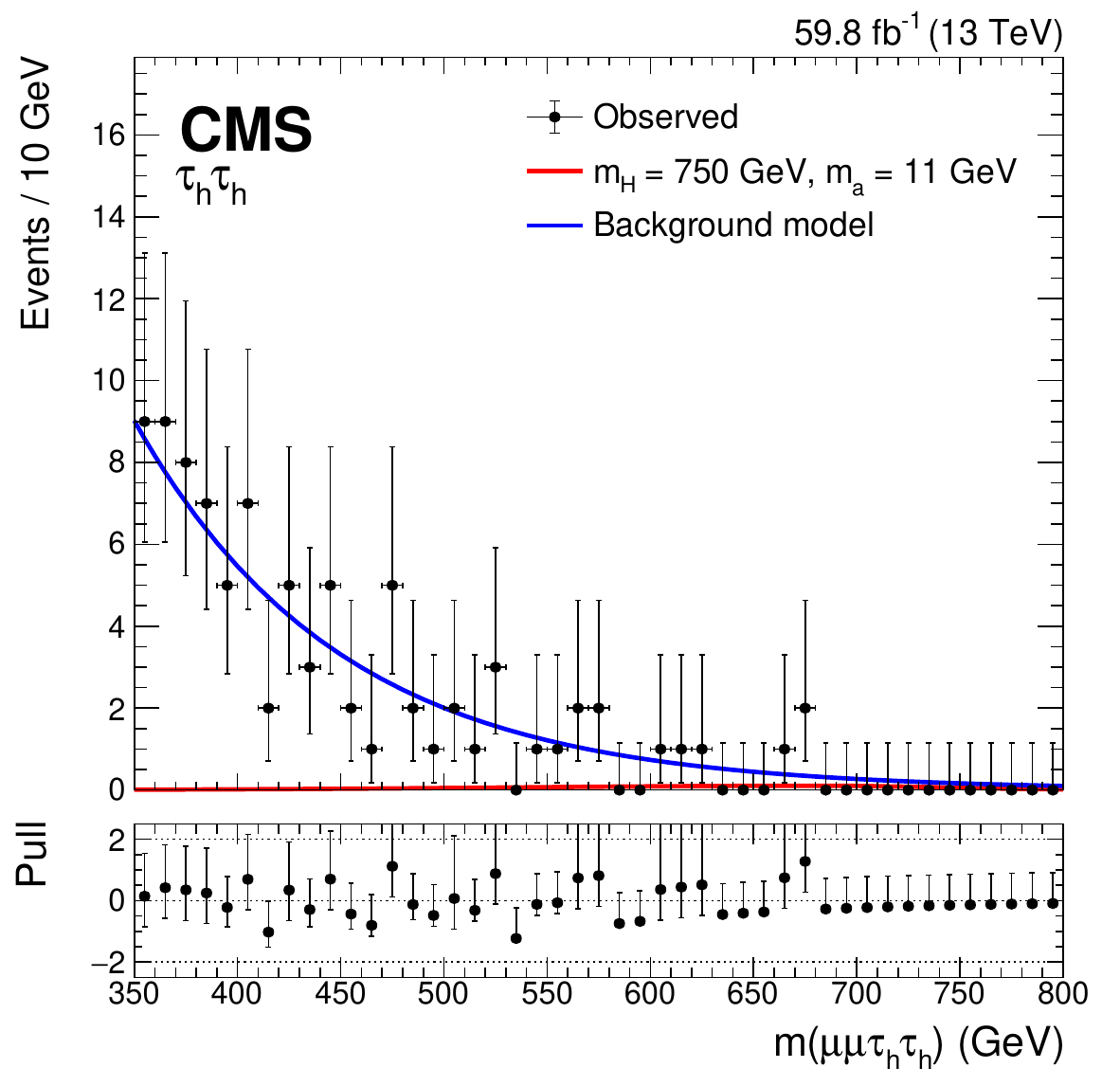} \\
  \includegraphics[width=0.4\textwidth]{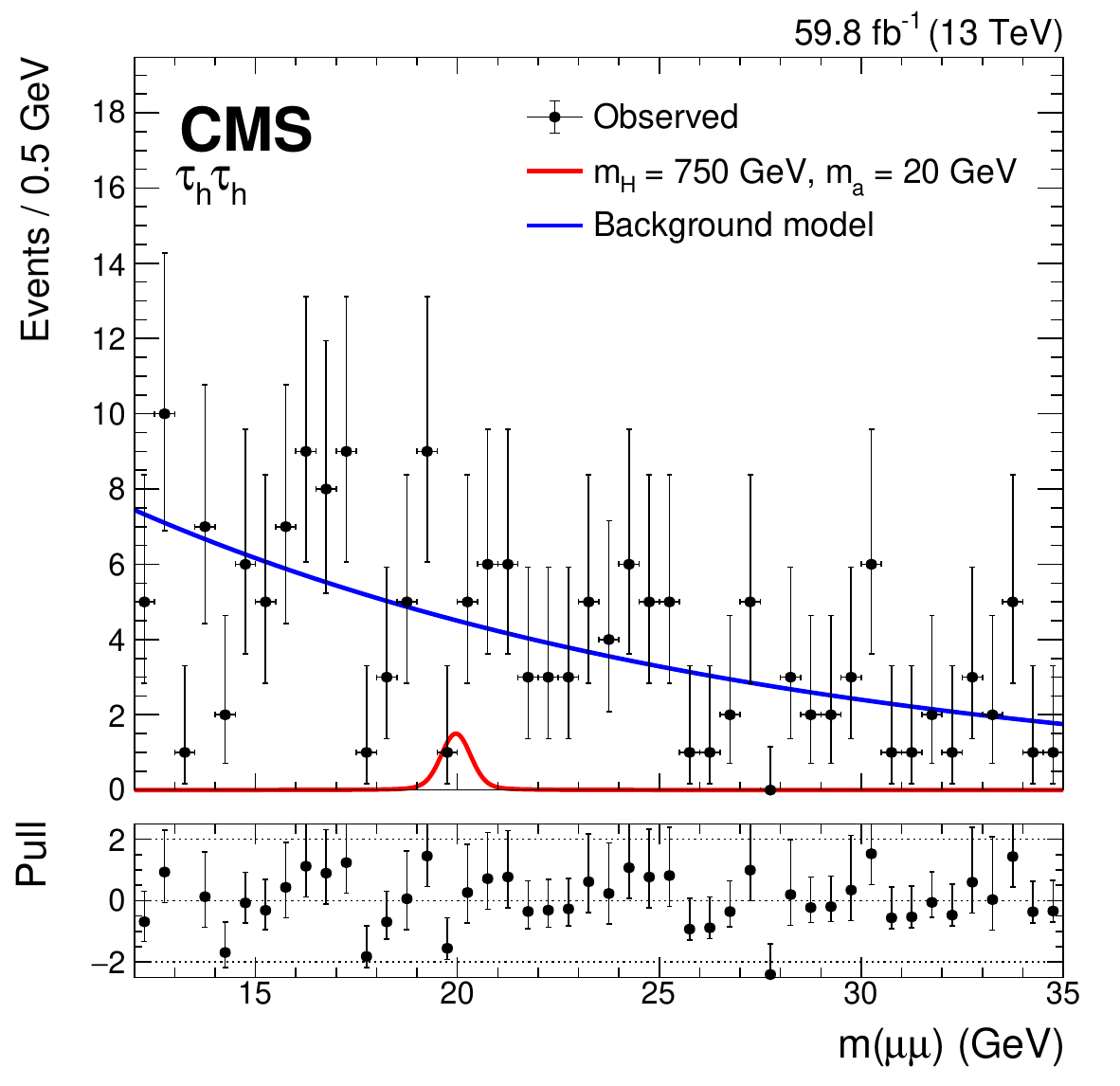} 
  \includegraphics[width=0.4\textwidth]{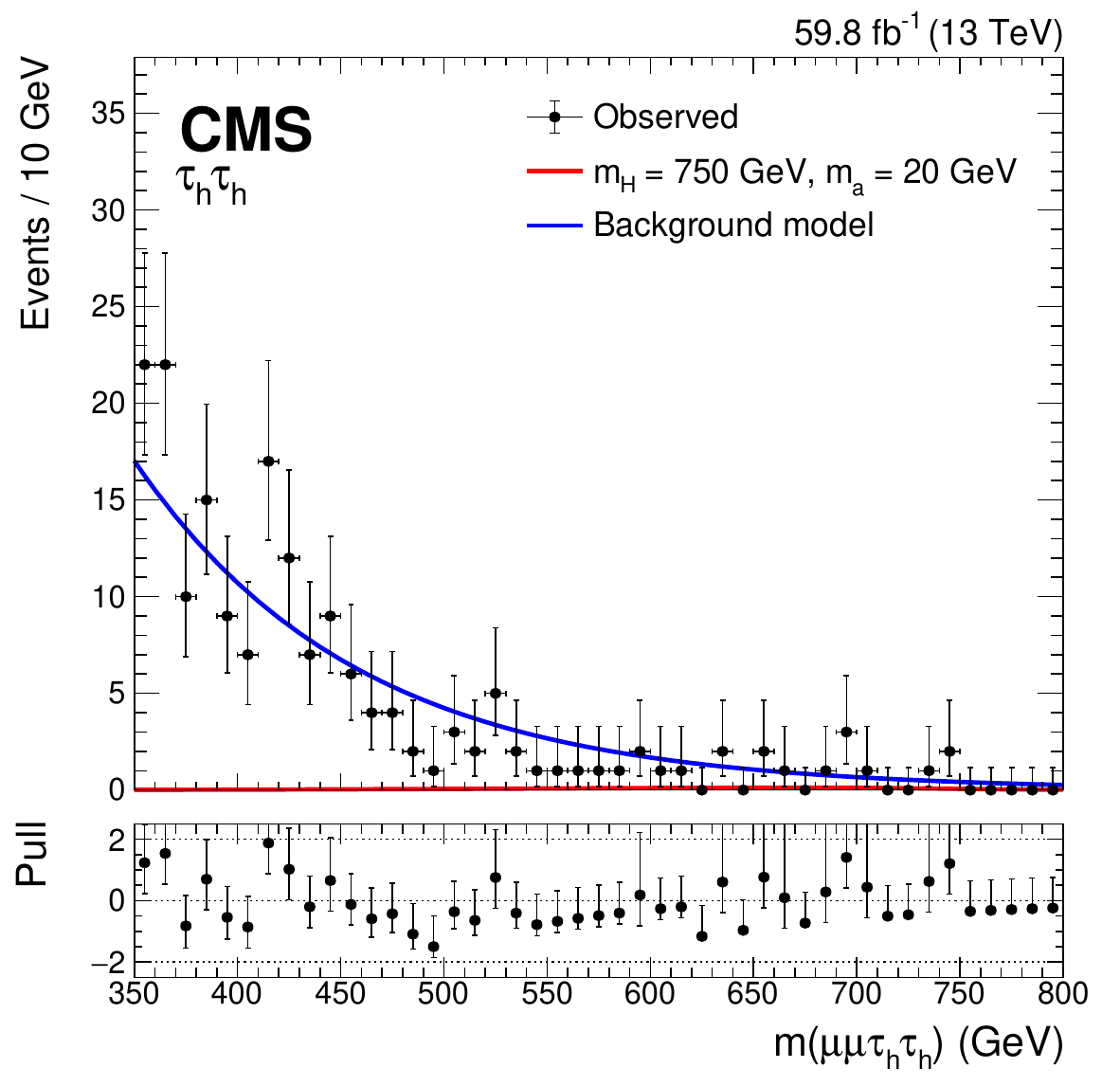} \\
  \caption{
  Projections of the post-fit two-dimensional background pdfs (blue) and the observed data onto the $m_{\PGm\PGm}$ (left) and four-body visible mass (right) axes for the $\tauh\tauh$ channel for $m_{\PH}$ = 750\GeV using the 2018 data set.
  The rows correspond to pseudoscalar masses of $m_{\pa}$ = 5, 11, and 20\GeV, from top to bottom.
  Example signal distributions (red) are overlaid assuming $\mathcal{B}(\PH\to\pa\pa\to\PGm\PGm\PGt\PGt) = 5 \times 10^{-4}$.
  }
  \label{fig:postfithh_H750}
\end{figure}

\begin{figure}[htbp]
  \centering
  \includegraphics[width=0.4\textwidth]{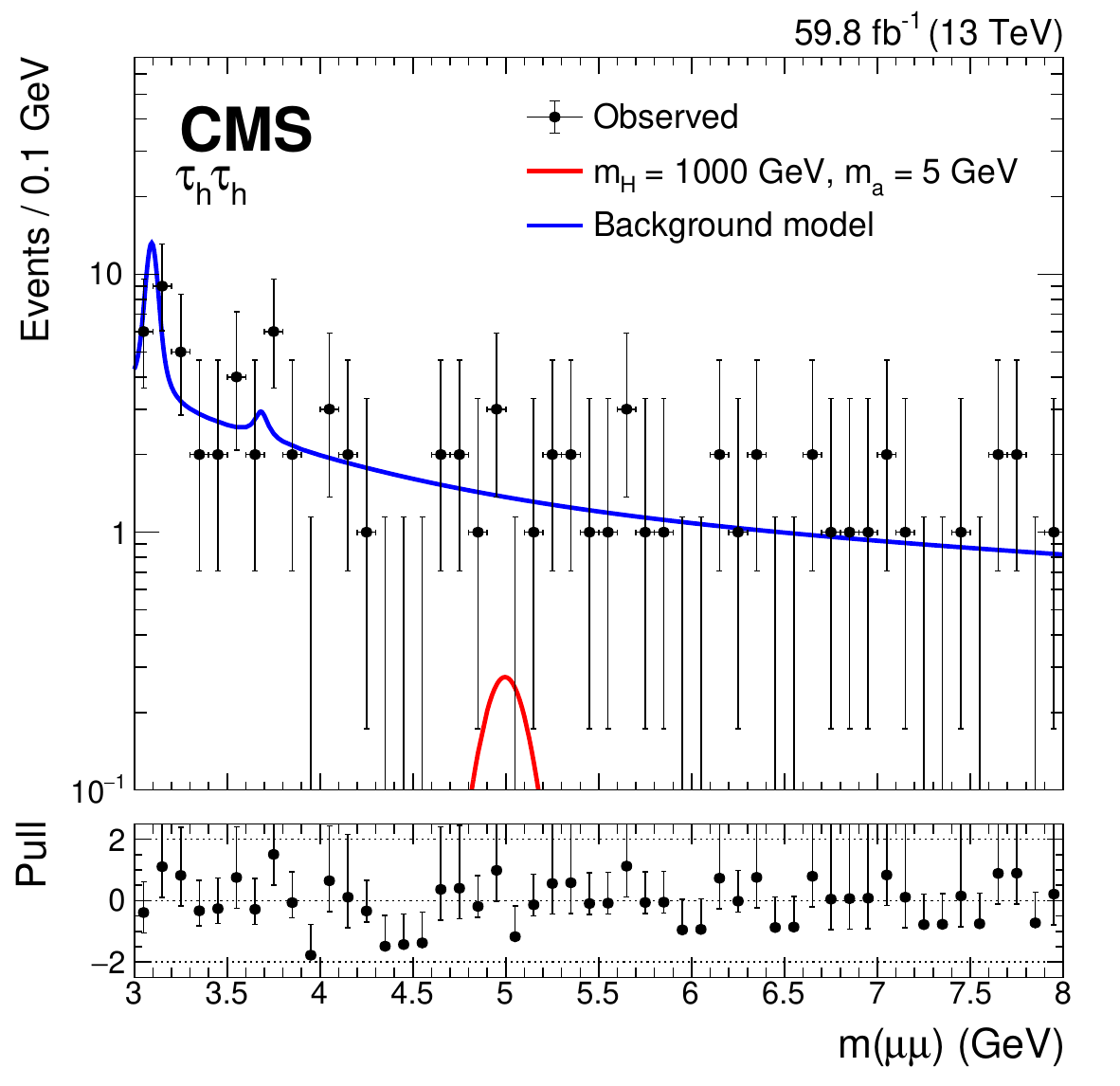} 
  \includegraphics[width=0.4\textwidth]{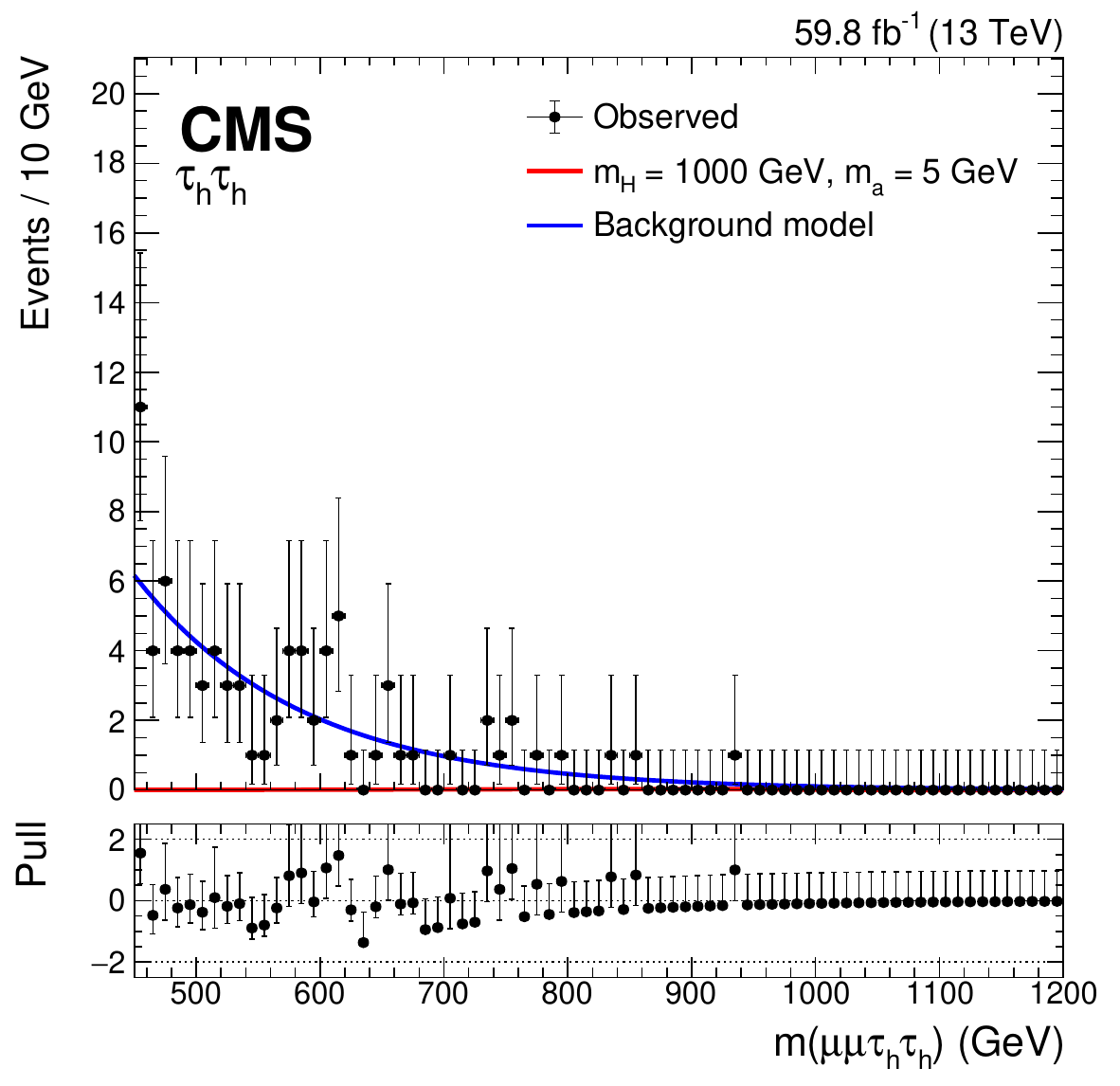} \\
  \includegraphics[width=0.4\textwidth]{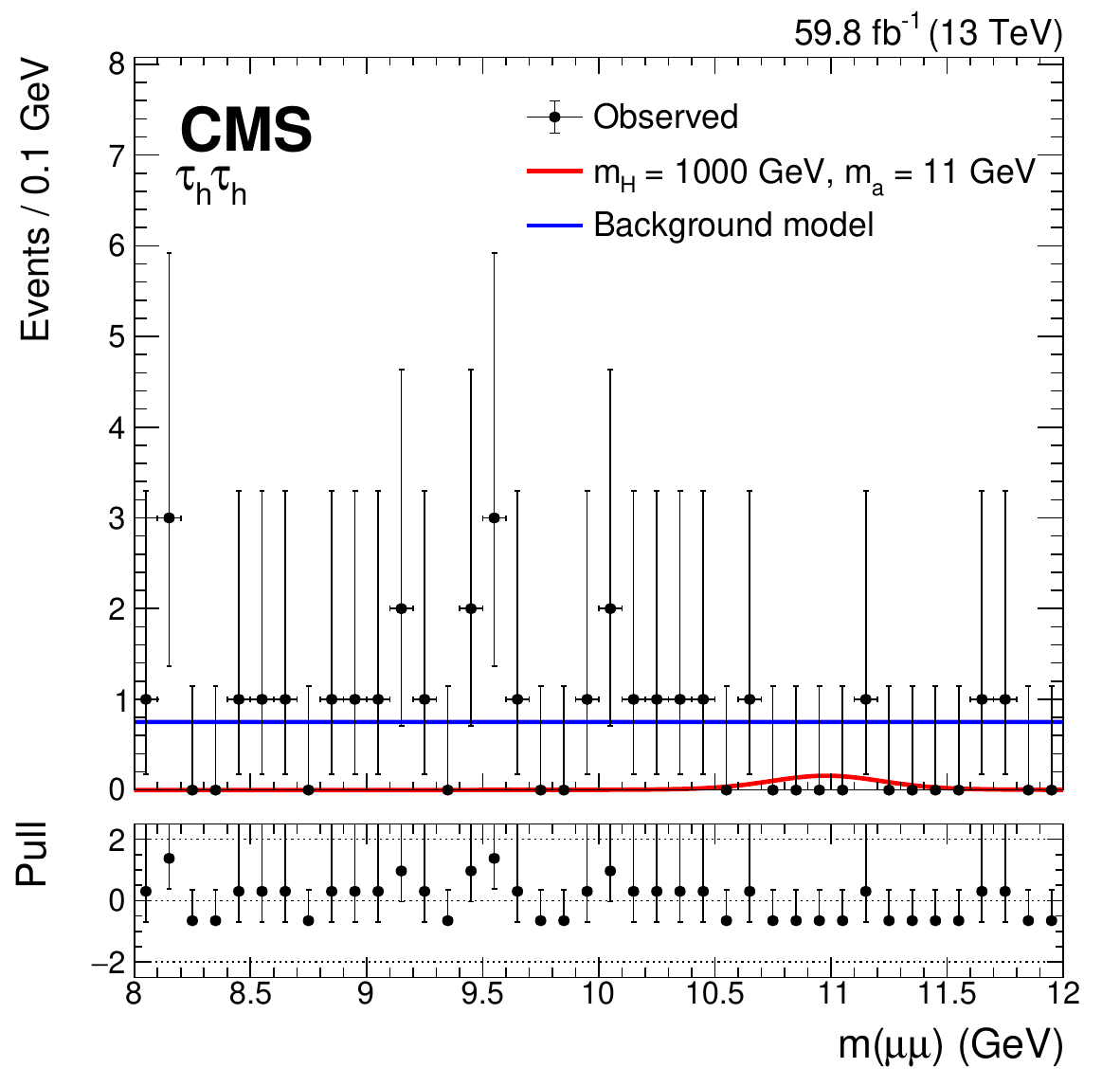} 
  \includegraphics[width=0.4\textwidth]{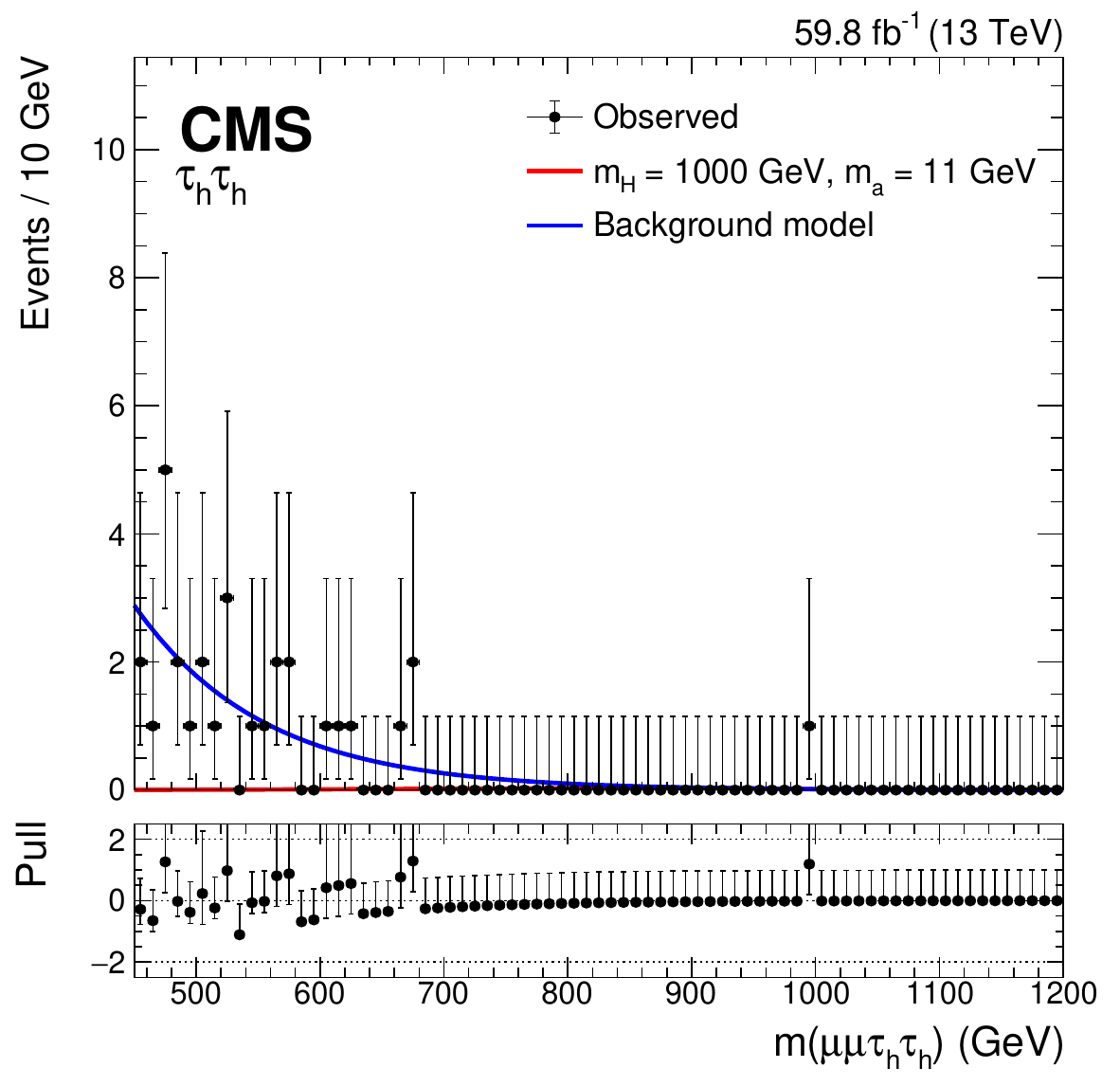} \\
  \caption{
  Projections of the post-fit two-dimensional background pdfs (blue) and the observed data onto the $m_{\PGm\PGm}$ (left) and four-body visible mass (right) axes for the two lower $m_{\PGm\PGm}$ ranges 
($ 3 < m_{\PGm\PGm} < 8 \GeV$ and $8 < m_{\PGm\PGm} < 12 \GeV$)
in the $\tauh\tauh$ channel for $m_{\PH}$ = 1000\GeV using the 2018 data set.
  The rows correspond to pseudoscalar masses of $m_{\pa}$ = 5 and 11\GeV, from top to bottom.
  Example signal distributions (red) are overlaid assuming $\mathcal{B}(\PH\to\pa\pa\to\PGm\PGm\PGt\PGt) = 5 \times 10^{-4}$.
  }
  \label{fig:postfithh_H1000}
\end{figure}

\begin{figure}[htbp]
  \centering
  \includegraphics[width=0.4\textwidth]{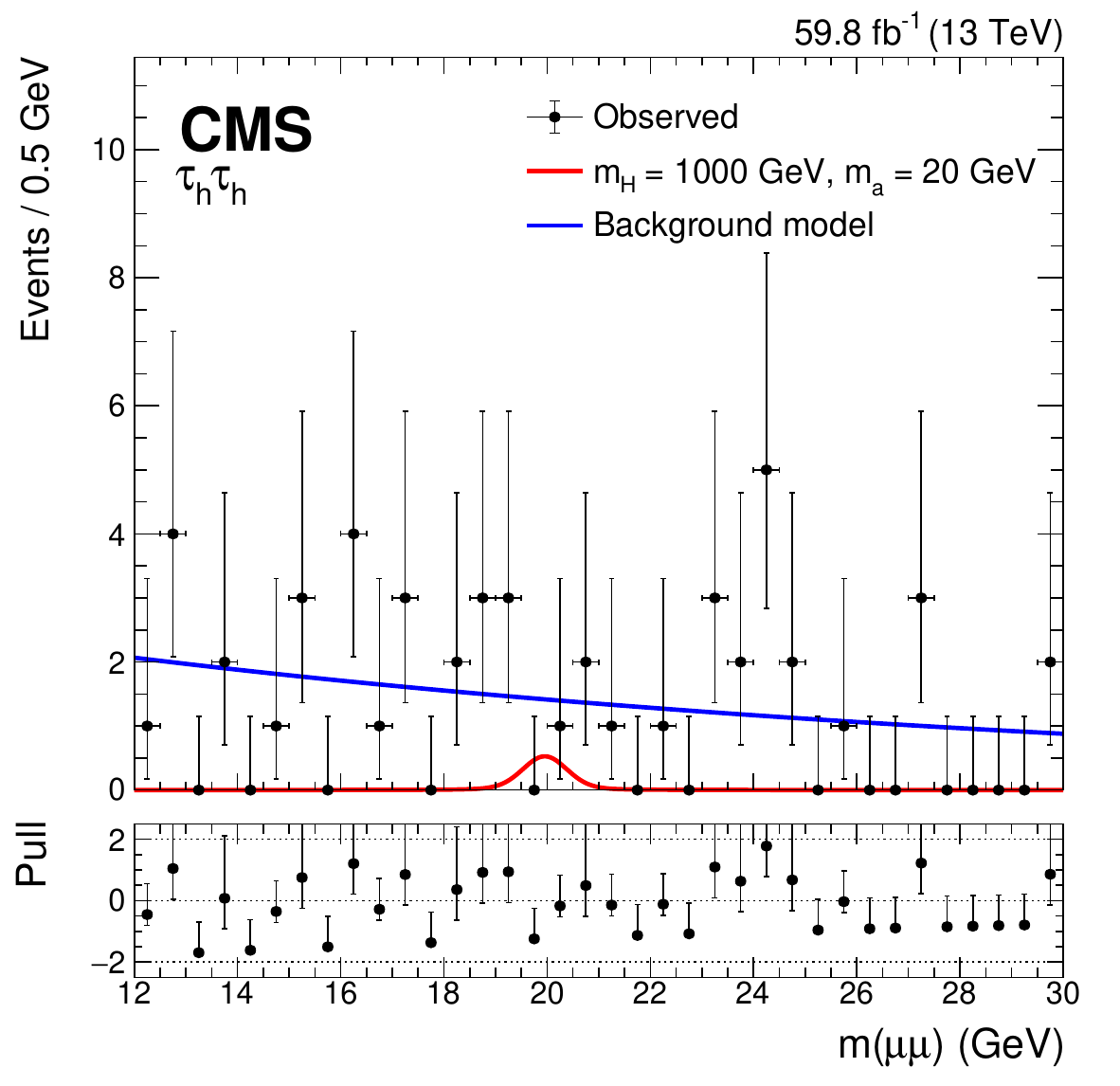} 
  \includegraphics[width=0.4\textwidth]{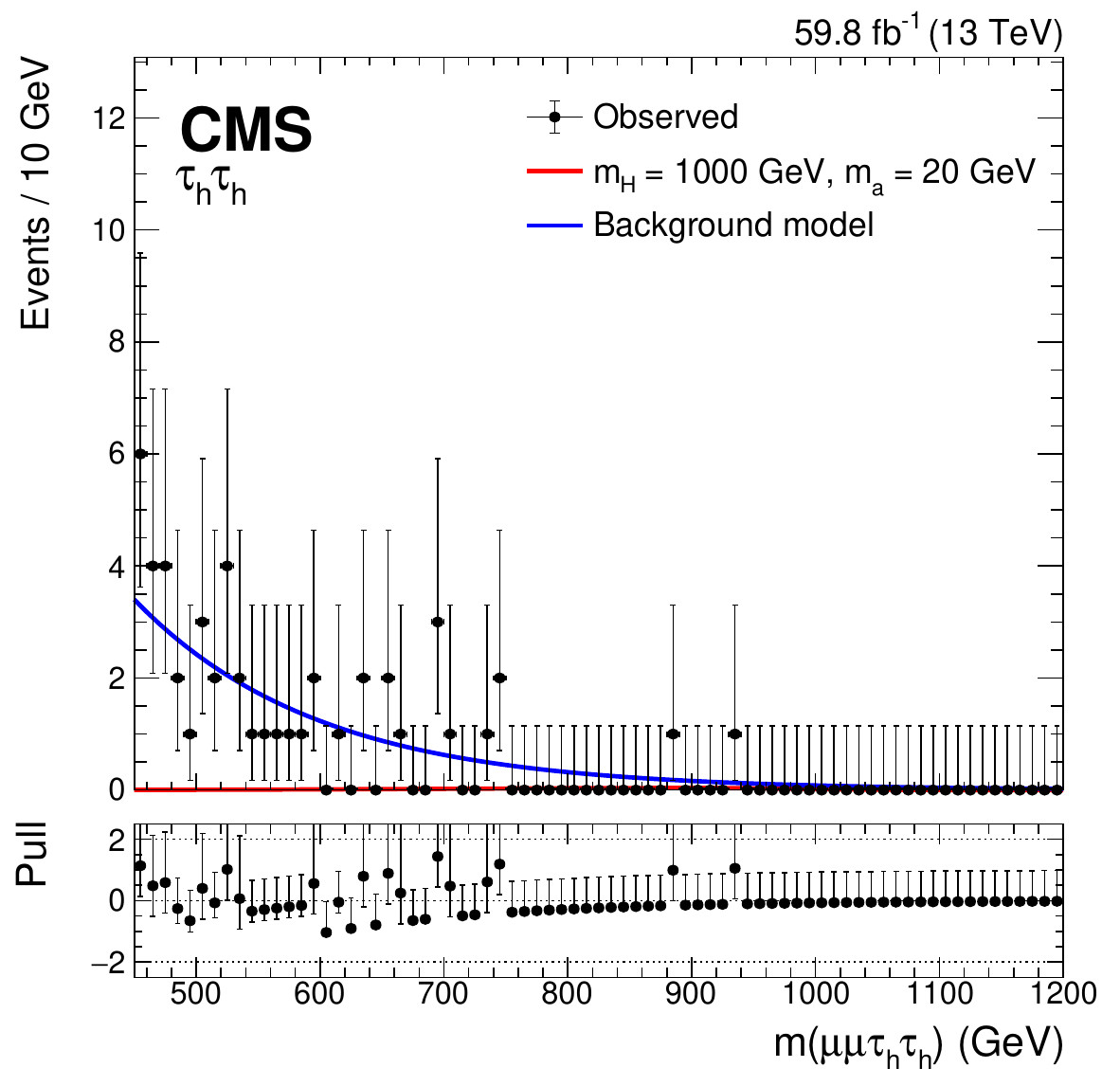} \\
  \includegraphics[width=0.4\textwidth]{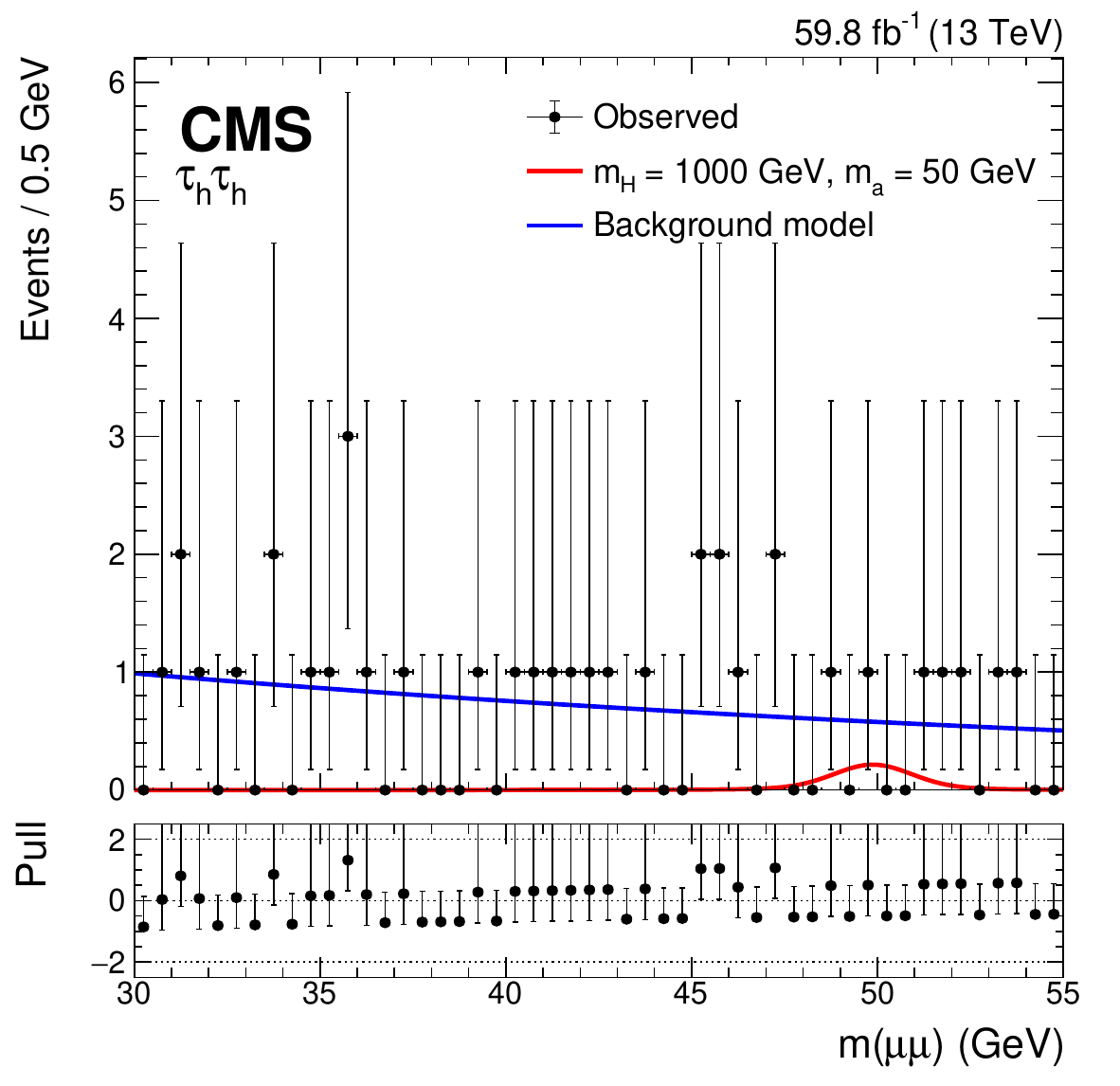} 
  \includegraphics[width=0.4\textwidth]{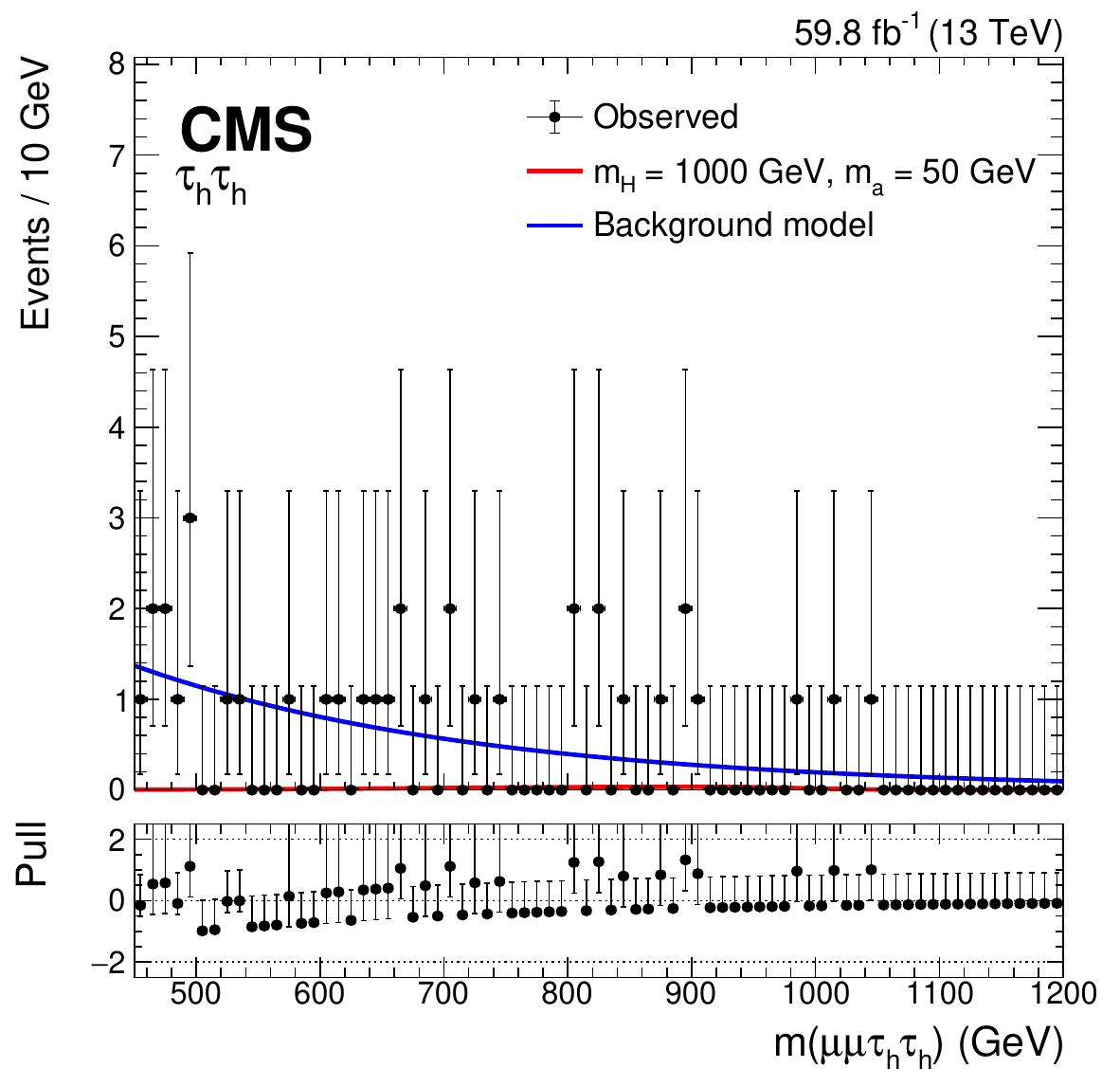} \\
  \caption{
  Projections of the post-fit two-dimensional background pdfs (blue) and the observed data onto the $m_{\PGm\PGm}$ (left) and four-body visible mass (right) axes for the two higher $m_{\PGm\PGm}$ ranges
  ($12< m_{\PGm\PGm}<30\GeV$ and $30< m_{\PGm\PGm}<55\GeV$)
   in the $\tauh\tauh$ channel for $m_{\PH}$ = 1000\GeV using the 2018 data set.
  The rows correspond to pseudoscalar masses of $m_{\pa}$ = 20 and 50\GeV, from top to bottom.
  Example signal distributions (red) are overlaid assuming $\mathcal{B}(\PH\to\pa\pa\to\PGm\PGm\PGt\PGt) = 5 \times 10^{-4}$.
  }
  \label{fig:postfithh_H1000_2}
\end{figure}

The fitting procedure is performed separately for the 2016, 2017, and 2018 data sets.
As an example, projections of the two-dimensional background pdfs onto the $m_{\PGm\PGm}$ and $m_{\PGm\PGm\PGt\PGt}$ axes, obtained from fits to the observed data in the 2018 data set, are shown in Figs.~\ref{fig:postfitmh_H125}--\ref{fig:postfithh_H1000_2}.
These figures also include example signal distributions for each fit range in the SR for the $\taum\tauh$ and $\tauh\tauh$ channels.
The signal distribution is scaled assuming the SM Higgs boson production cross section~\cite{LHCHiggsCrossSectionWorkingGroup:2016ypw} and a branching fraction of $5 \times 10^{-4}$ for $\PH\to\pa\pa\to\PGm\PGm\PGt\PGt$.

\section{Systematic uncertainties}
\label{sec:unc}

The main systematic uncertainties arise from the parametric modeling of the background and signal processes, as well as from the tight-to-loose ratio.  
To assess the systematic uncertainties from the tight-to-loose ratio, the observed data distributions in the validation region described in Section~\ref{sec:bkg} are compared with those in the validation sideband weighted by the tight-to-loose ratio.
The associated systematic uncertainty is 10\% for the $\tauh\tauh$ final state, while for the $\taum\taue$, $\taum\tauh$, and $\taue\tauh$ final states, it ranges from 10--30\% for $\pt(\tauh) <$ 40\GeV and 30--50\% for $\pt(\tauh) >$ 40\GeV depending on the decay mode (1 or 3 prongs) of the \tauh.
For each era, the statistics of the $\taum\taue$ and \taue\tauh final states, and the remaining events at $\pt(\tauh) >$ 40\GeV of the \taum\tauh final state, are rather limited in the validation region.
Systematic uncertainties are mostly treated as uncorrelated across the three years, except for the signal acceptance and theoretical uncertainties, which are treated as correlated.

The statistical uncertainties obtained from the fitting procedure described in Section~\ref{sec:bkg} are treated as systematic uncertainties in the parametric modeling.
The corresponding statistical uncertainty is included as an independent nuisance parameter in the limit-setting procedure for each shape parameter in the signal and background models.
These uncertainties are $\leq$ 20\%.

The data-to-simulation correction factors for the muon trigger, reconstruction, identification, and isolation efficiencies are estimated using a tag-and-probe method in Drell-Yan data and simulated samples.
These corrections contribute an uncertainty of 5--10\% to the total signal yield, depending on the final state.
The correction applied to account for early first-level muon trigger decisions introduces a 1\% uncertainty in the signal yield.
The uncertainty in the muon momentum scale ranges from 0.2--5.0\%.  Most muons have $\pt < 100\GeV$ and therefore an uncertainty of 0.2\%~\cite{Sirunyan:2018fpa}, while muons from heavy $m_{\PH}$ hypotheses populate the higher \pt range.
Electron reconstruction and identification contribute a 6--10\% uncertainty to the total signal acceptance.
The uncertainty in the electron momentum scale is 1\%.
For the \tauh candidate in the $\taum\tauh$ and $\taue\tauh$ final states, the uncertainty in the \tauh identification efficiency ranges from 6--25\%, depending on $\pt^{\tauh}$.
In addition, the uncertainty in the \tauh energy scale ranges from 1.2--3.0\%~\cite{Sirunyan:2018pgf}, depending on the number of charged and neutral hadrons in the \tauh decay.
The jet energy scale introduces an uncertainty of 2--5\% in the signal yield for the $\PGm\PGm\tauh\tauh$ final state.

The uncertainty in the integrated luminosity affects the normalization of simulated signal samples and ranges from 1.2--2.5\%~\cite{CMS:2021xjt,CMS-PAS-LUM-17-004,CMS-PAS-LUM-18-002,CERN-EP-2026-134} and is considered uncorrelated across the three run years.
The uncertainty from pileup effects arises from a 4.6\% uncertainty~\cite{Sirunyan:2018nqx} in the total inelastic cross section of $\Pp\Pp$ interactions.
This results in a 1--5\% uncertainty in the signal yield.

As described in Section~\ref{sec:samples}, a correction to the simulated ggF signal samples accounts for the small difference in acceptance between the ggF and VBF Higgs boson production modes, contributing a 0.5\% uncertainty in the signal yield.
Theoretical uncertainties in the Higgs boson production cross section are evaluated by varying the renormalization ($\mu_{\mathrm{R}}$) and factorization ($\mu_{\mathrm{F}}$) scales independently by a factor of two relative to their default values.
The condition $0.5 \leq \mu_{\mathrm{R}}/\mu_{\mathrm{F}} \leq 2$ is imposed.
The resulting uncertainties, combined with the PDF and $\alpha_{\mathrm{S}}$ uncertainties from Ref.~\cite{LHCHiggsCrossSectionWorkingGroup:2016ypw}, contribute less than 4--7\% to the overall signal yield uncertainty.

\section{Results}
\label{sec:results}

A modified frequentist approach based on the \CLs criterion~\cite{Junk:1999kv,Read:2002hq} is used to calculate asymptotic upper limits~\cite{Cowan:2010js} with the LHC test statistic~\cite{CMS-NOTE-2011-005}, as implemented in the CMS statistical analysis tool \textsc{Combine}~\cite{CMS:2024onh}.
Systematic uncertainties are represented by nuisance parameters.
A log-normal pdf is assumed in the likelihood fit for uncertainties in the expected yields, while a Gaussian pdf is used for uncertainties in the signal and background model parameters.

\begin{figure}[htbp]
  \centering
  \includegraphics[width=0.47\textwidth]{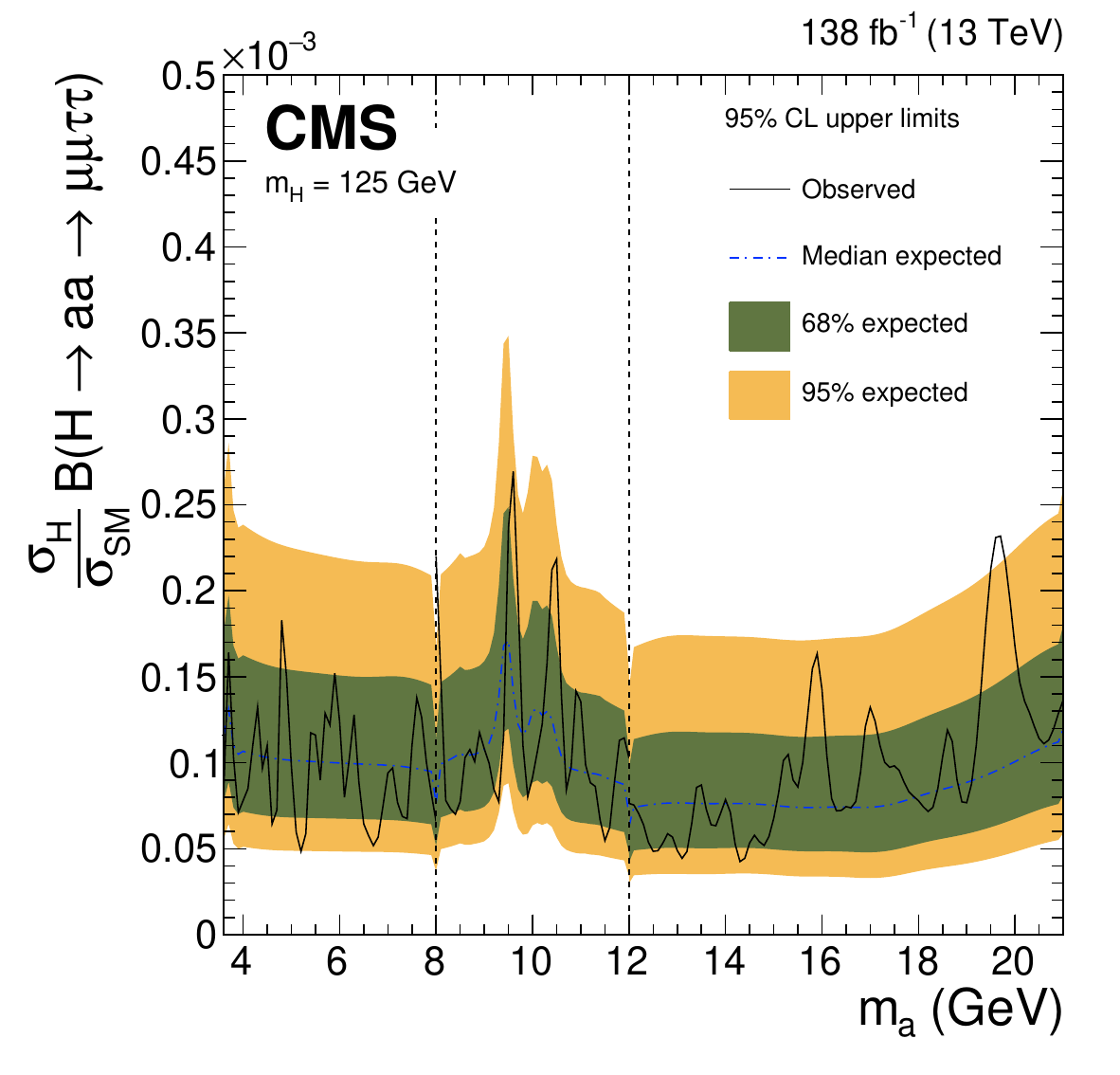}
  \includegraphics[width=0.47\textwidth]{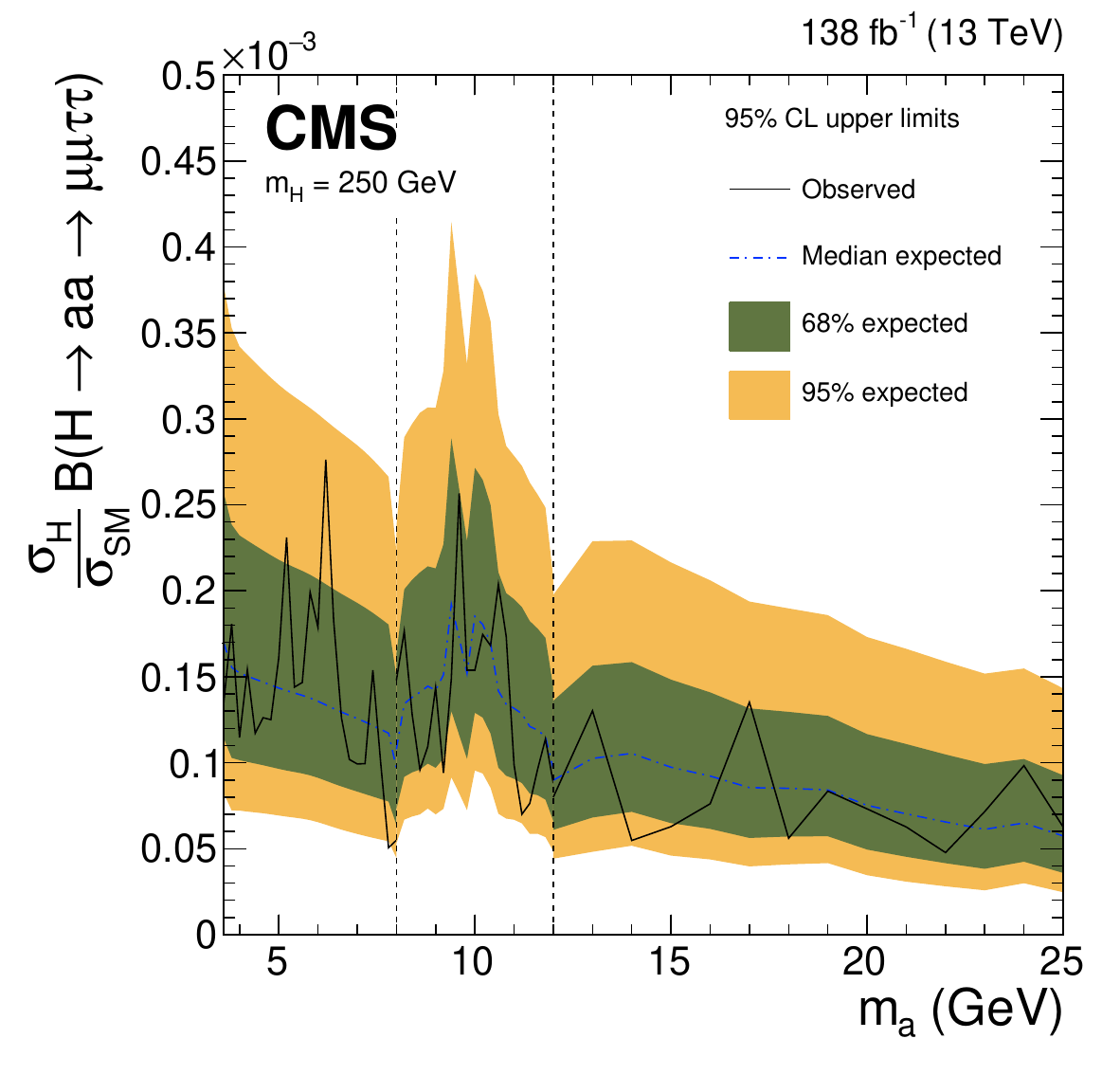} \\
  \includegraphics[width=0.47\textwidth]{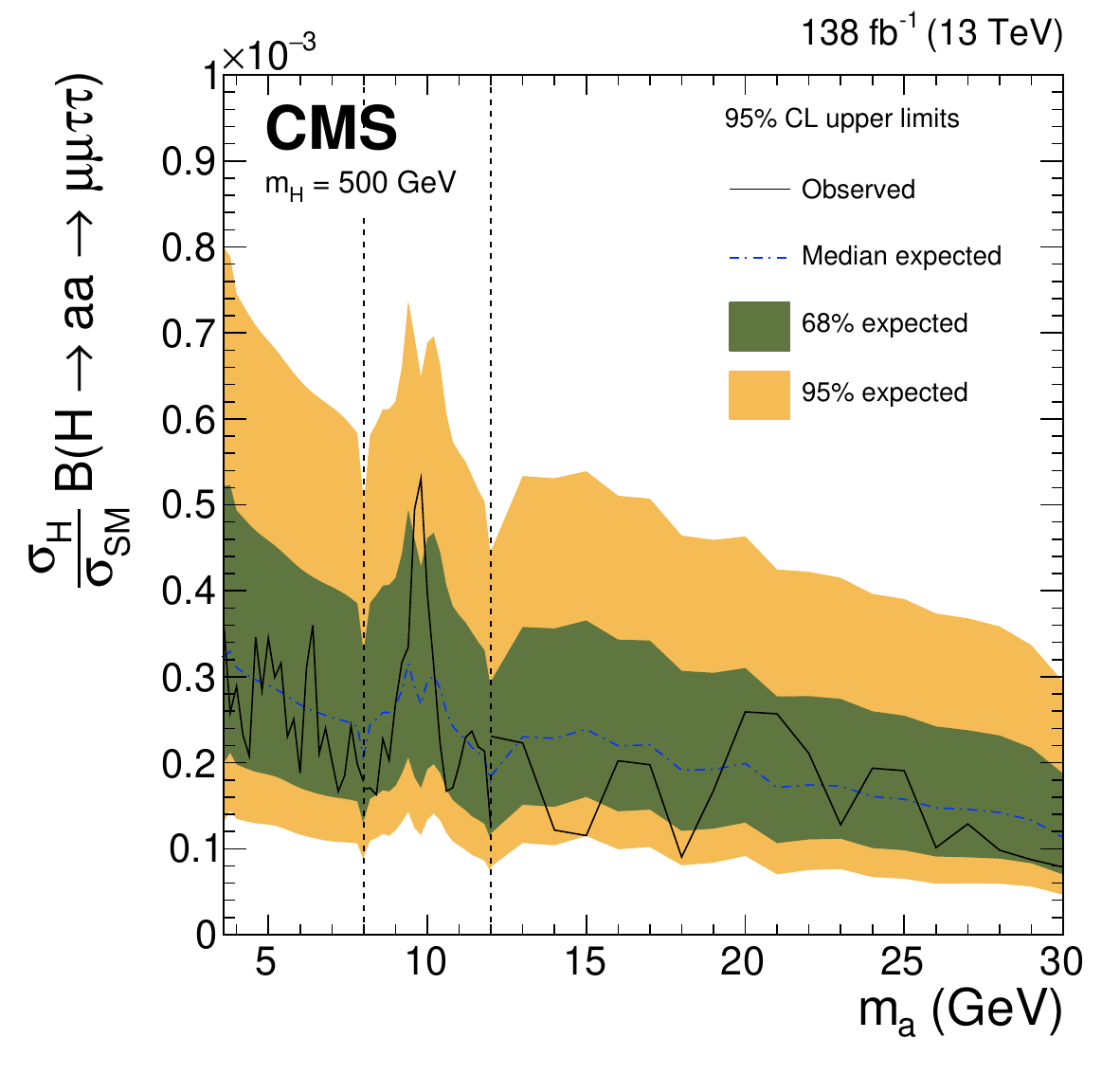}
  \includegraphics[width=0.47\textwidth]{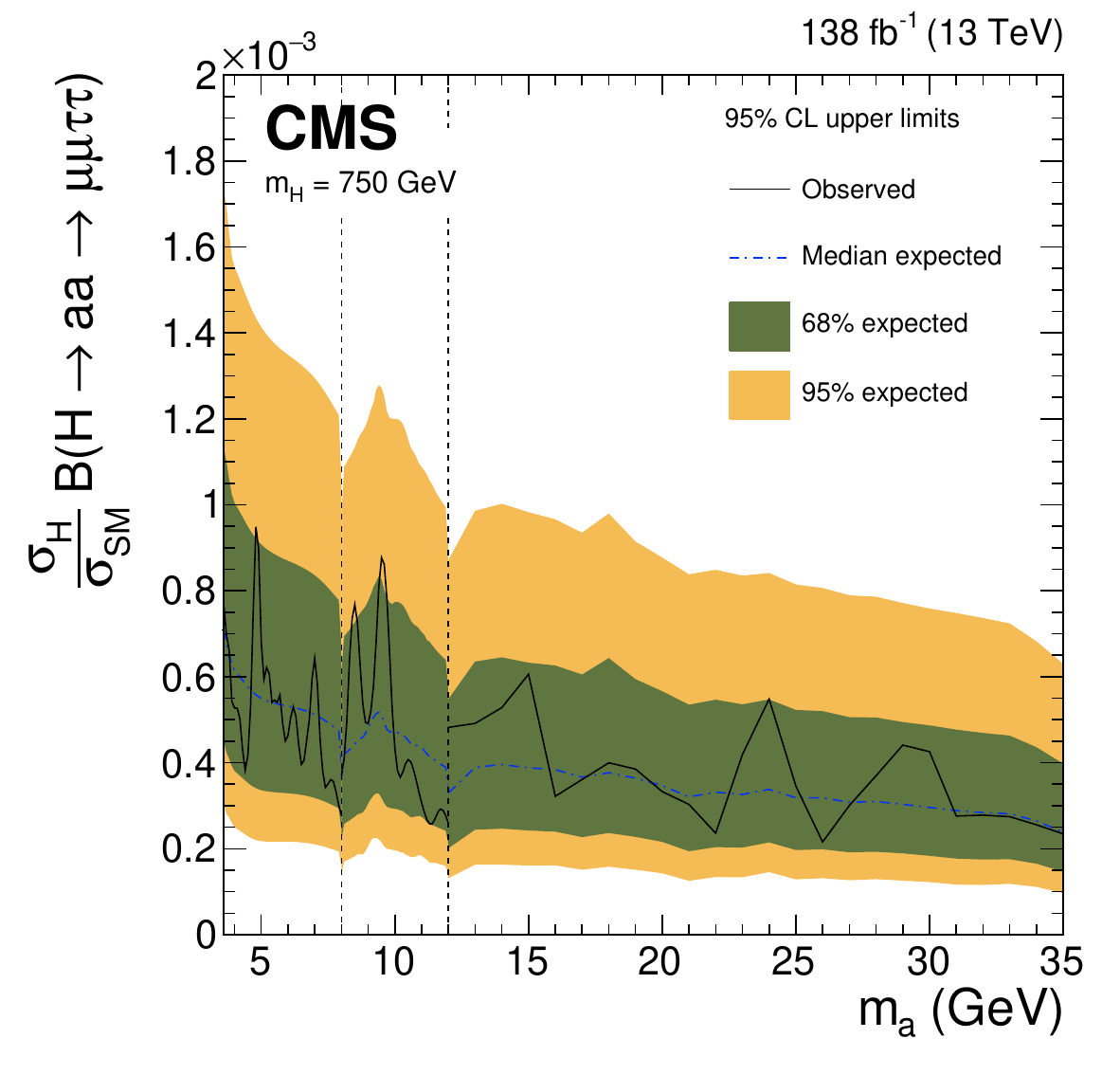} \\
  \includegraphics[width=0.47\textwidth]{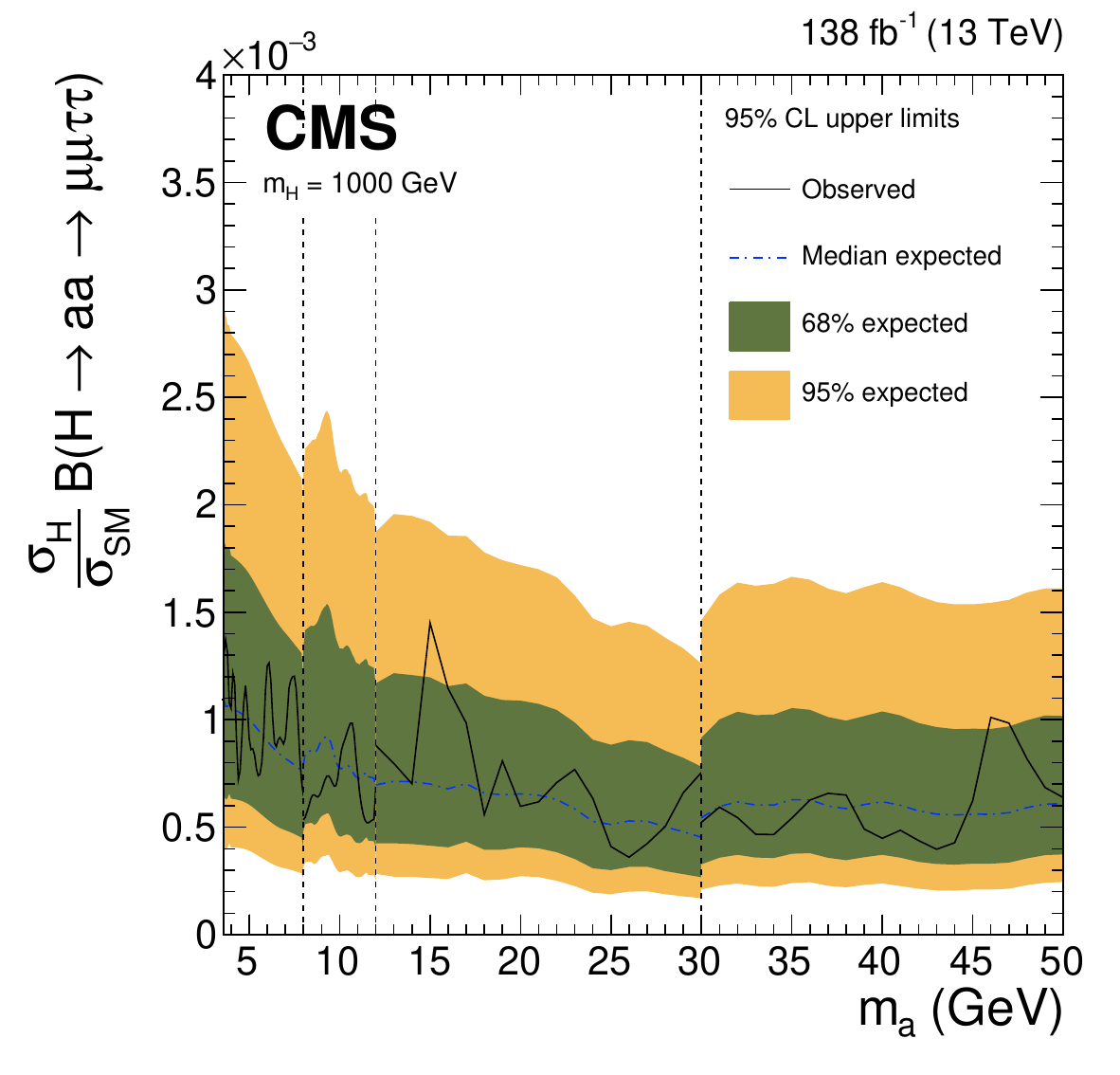}

  \caption{
  Expected and observed limits on $\sigma_{\PH} \times \mathcal{B}(\PH\to\pa\pa\to\PGm\PGm\PGt\PGt)/\sigma_{\mathrm{SM}}$, combining the $\taum\taue$, $\taum\tauh$, $\taue\tauh$, and $\tauh\tauh$ final states, for $m_{\PH} = 125$, 250, 500, 750, and 1000\GeV. The vertical dotted lines indicate the boundaries of the $m_{\PGm\PGm}$ regions used in the background fit, as described in Section~\ref{sec:bkg}.
  }
  \label{fig:limits}
\end{figure}

\begin{figure}[htbp]
  \centering
  \includegraphics[width=0.49\textwidth]{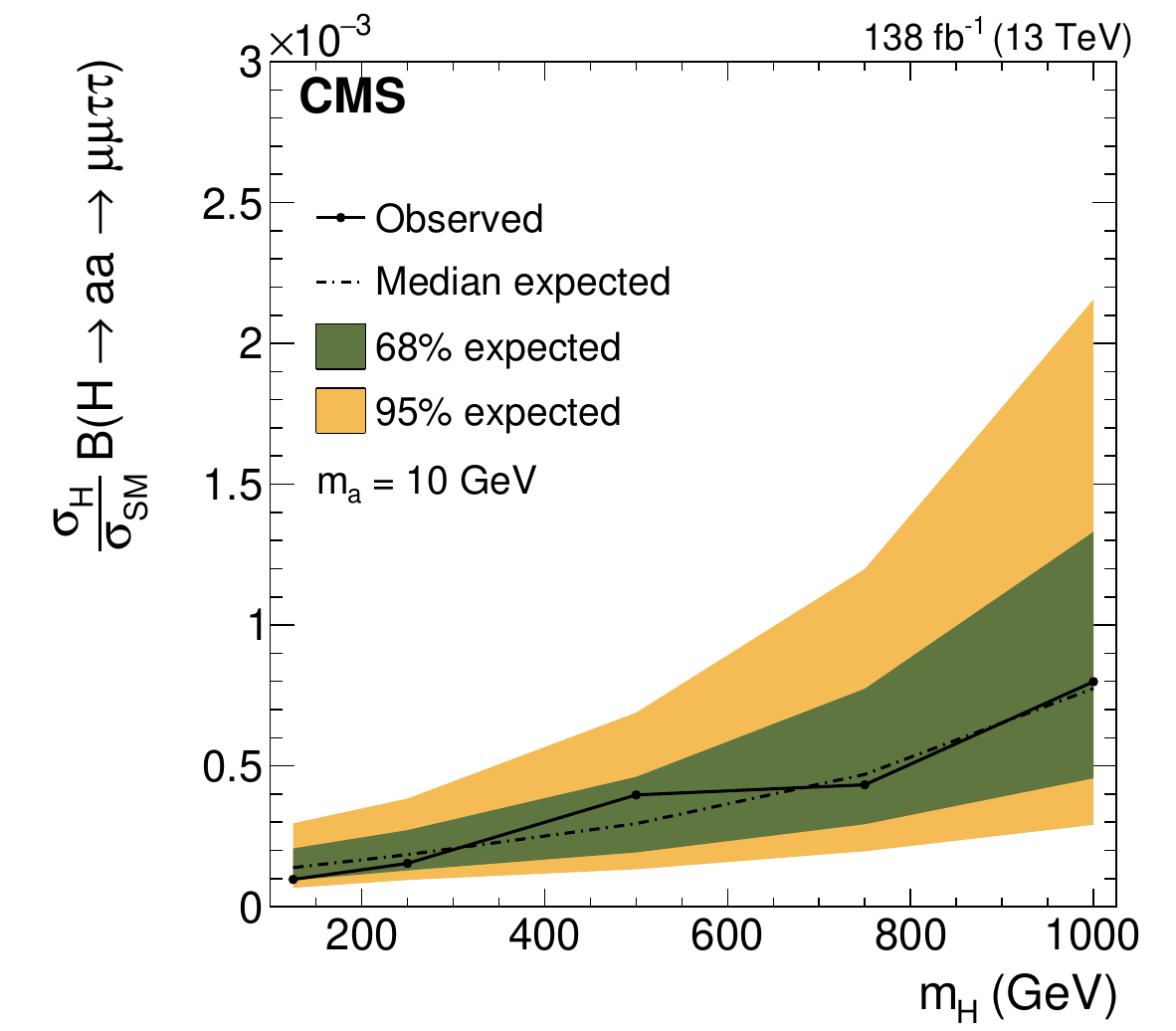} 
  \includegraphics[width=0.49\textwidth]{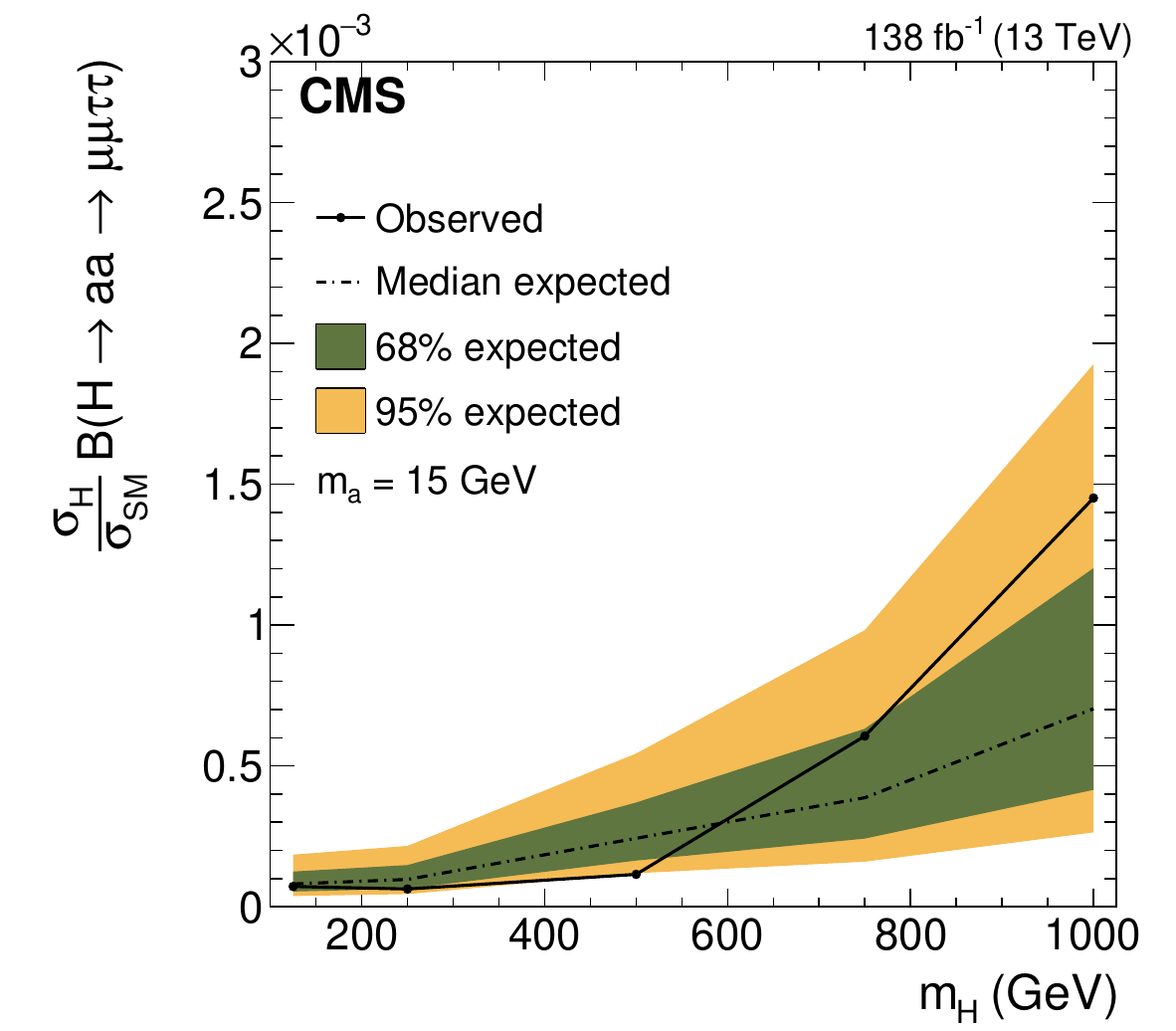}
  \caption{
  Expected and observed limits on $\sigma_{\PH} \times \mathcal{B}(\PH\to\pa\pa\to\PGm\PGm\PGt\PGt)/\sigma_{\mathrm{SM}}$, combining the final states for different Higgs boson mass hypotheses, for the pseudoscalar masses of 10\GeV (left) and 15\GeV (right).  The straight lines between points are there to guide the eye.
  }
  \label{fig:interpolation}
\end{figure}

Model-independent upper limits at the 95\% \CL are set on the ratio of the product of the production cross section and branching fraction for a Higgs boson decaying to a pair of light pseudoscalars in the $\PGm\PGm\PGt\PGt$ final state to the standard model Higgs boson production cross section, $\sigma_{\PH} \times \mathcal{B}(\PH\to\pa\pa\to\PGm\PGm\PGt\PGt)/\sigma_{\mathrm{SM}}$, as shown in Fig.~\ref{fig:limits}.
Here, $\sigma_{\mathrm{SM}}$ refers to the SM Higgs boson production cross section or, for $m_{\PH} = 250$, 500, 750, and 1000\GeV, to the SM-like Higgs boson production cross section, including the ggF and VBF production modes~\cite{LHCHiggsCrossSectionWorkingGroup:2016ypw}.
The range of tested $m_{\pa}$ values corresponds to the simulated signal samples listed in Table~\ref{tab:a_mass_h_mass}.
In general, the sensitivity of this exclusion decreases at low values of $m_{\pa}$ because of reconstruction inefficiencies that arise when the decay products of the $\PGt\PGt$ pair become fully collimated.
In addition, for $m_{\PH}$ = 125\GeV, the Lorentz boost decreases as $m_{\pa}$ increases, causing the decay products to become more widely separated and fail the requirement $\Delta R(\PGt\PGt) < 0.8$.
The peaking structures around $m_{\pa}$ = 10\GeV arise from the \PgU resonances.
A comparison with an earlier CMS result at $\sqrt{s}$ = 13\TeV~\cite{Sirunyan:2018mbx}, which targets resolved $\PGt\PGt$ decay products, is possible for SM Higgs boson decays with $15 < m_{\pa} < 20\GeV$.
In this region, the two approaches have similar sensitivity.
An example of the interpolation in $m_{\PH}$ is shown in Fig.~\ref{fig:interpolation} for $m_{\pa} = 10$ and 15\GeV.

\begin{figure}[htbp]
  \centering
  \includegraphics[width=0.49\textwidth]{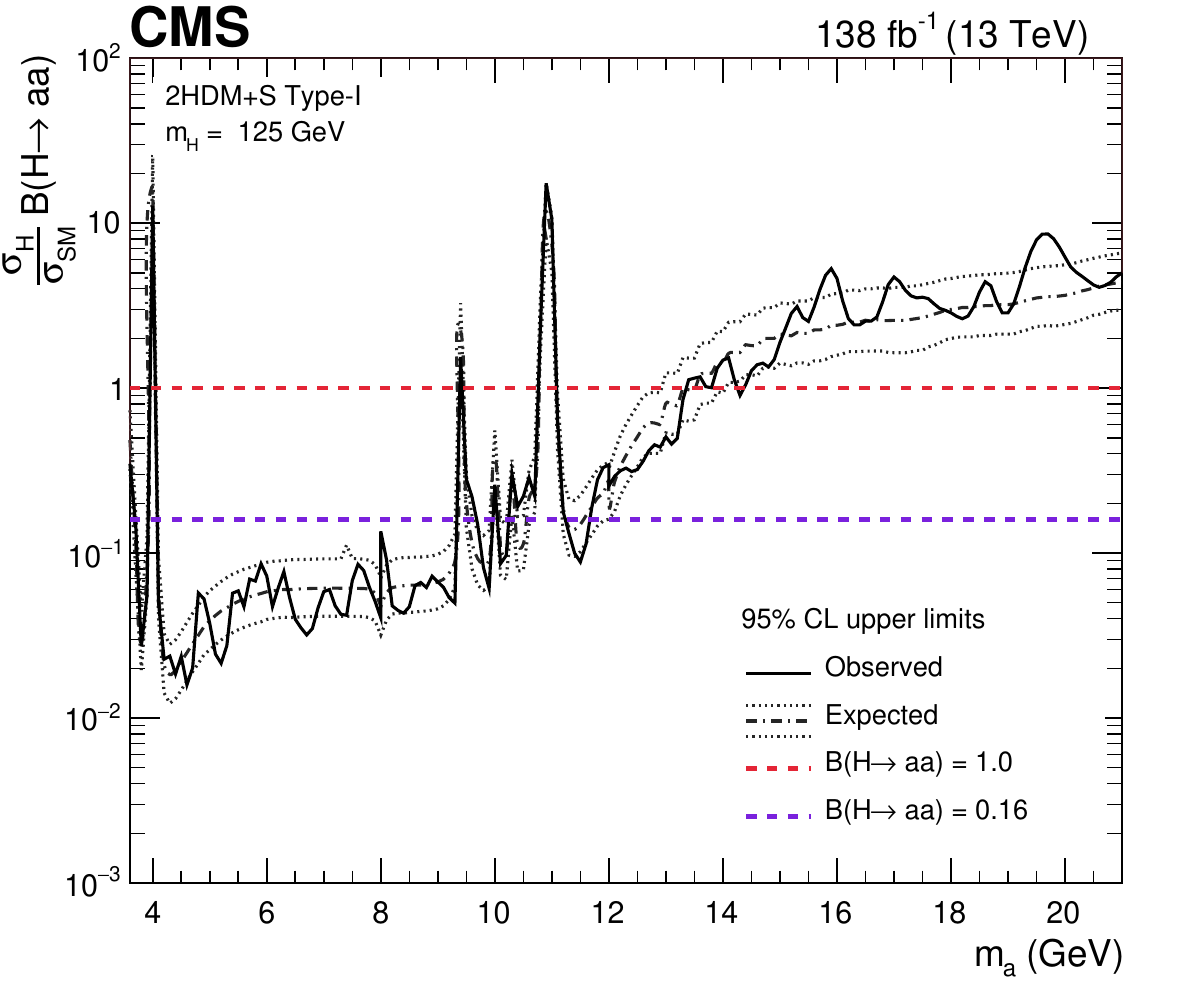} 
  \includegraphics[width=0.49\textwidth]{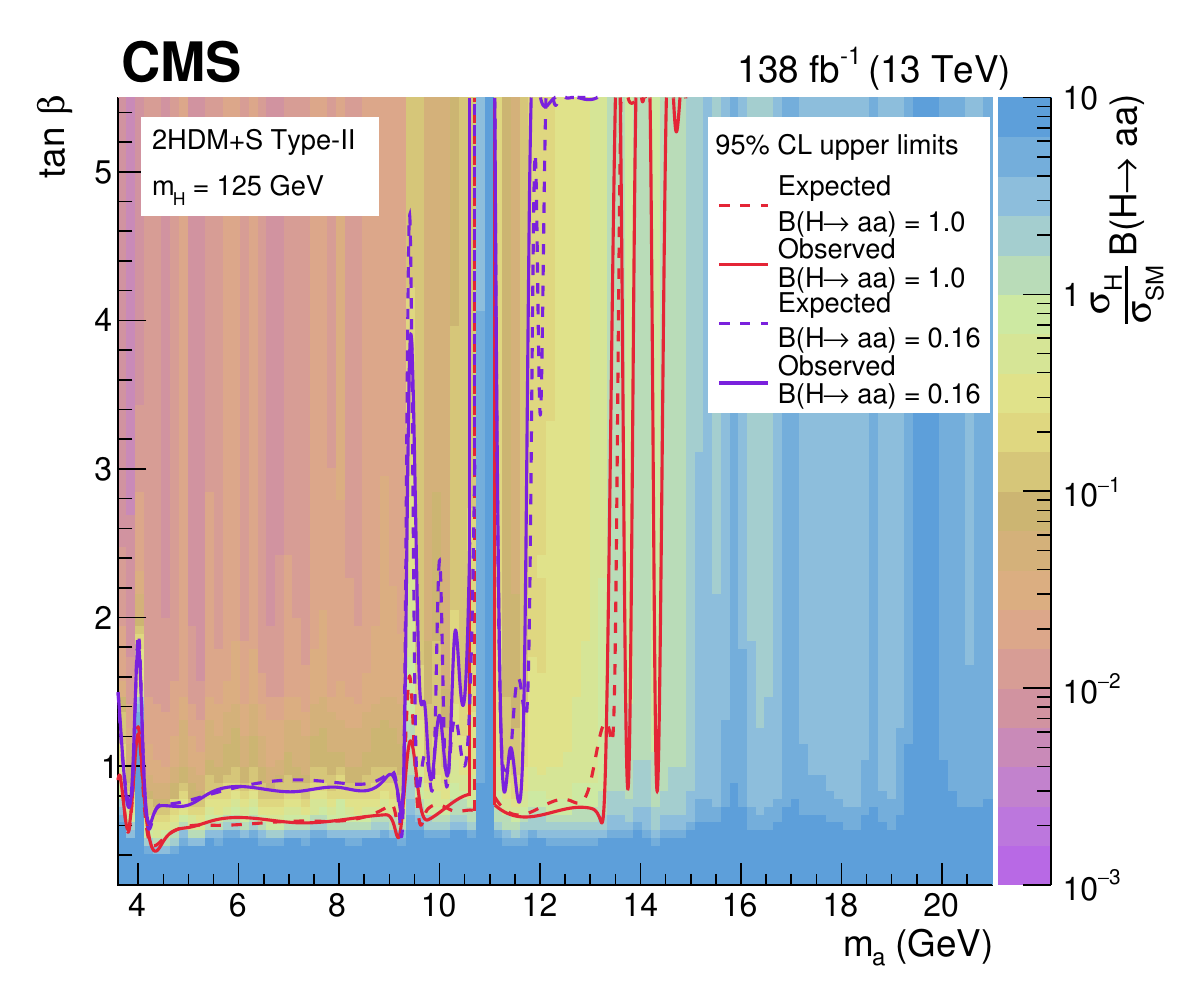} \\
  \includegraphics[width=0.49\textwidth]{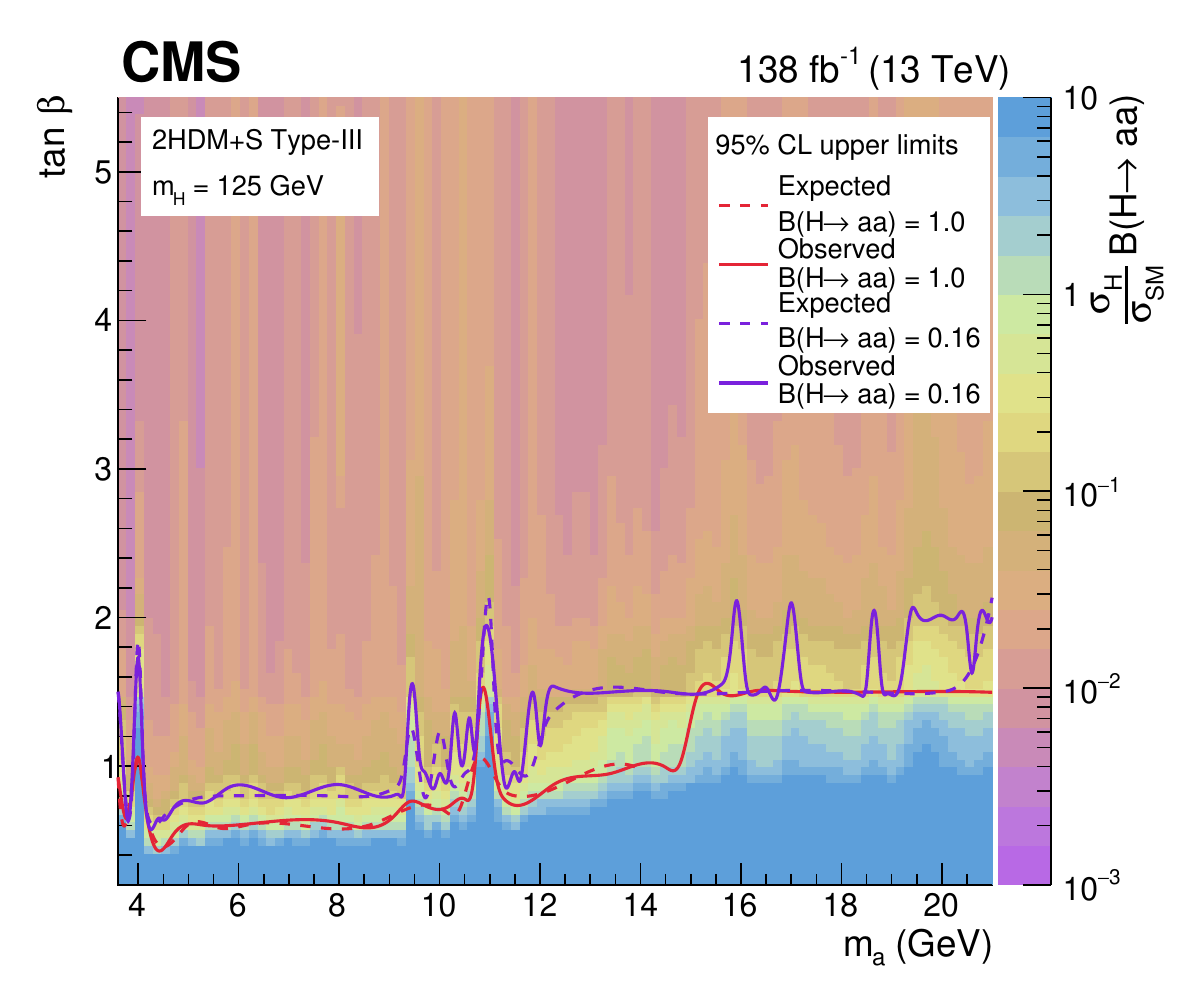}
  \includegraphics[width=0.49\textwidth]{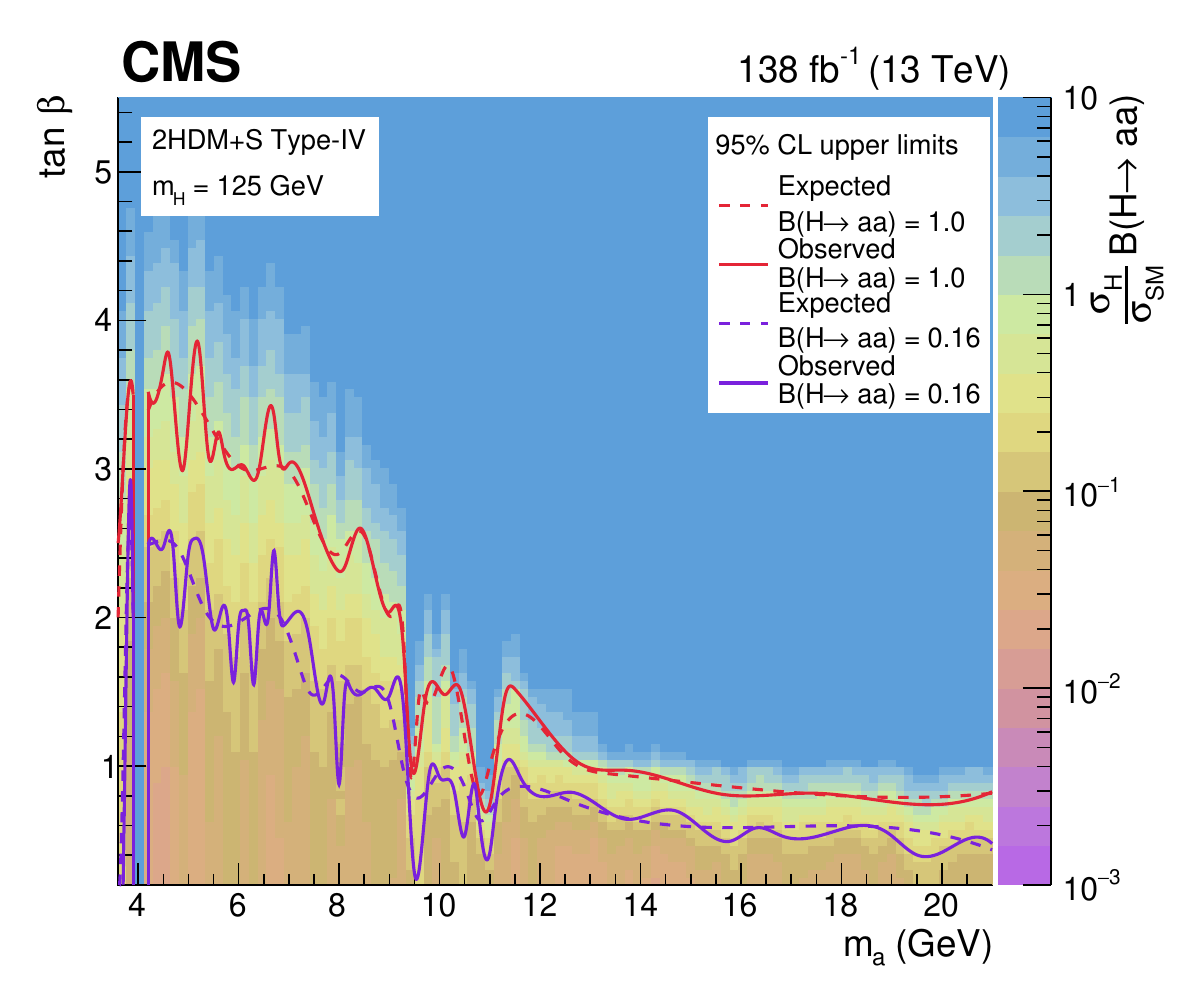}
  \caption{
  Observed and expected limits in 2HDM+S models of Types I through IV.
  The contours at a branching fraction of 0.16 are motivated by a CMS combination that sets limits on Higgs boson decays to undetected modes, including modes not covered by the standard channels, such as the boosted topologies considered in this analysis~\cite{CMS:2022dwd}.  In the Type-I model, $\mathcal{B}(\PH\rightarrow\pa\pa)$ is independent of $\tanb$.
  }
  \label{fig:limits2HDMS}
\end{figure}

Upper limits on $\sigma_{\PH}  \times  \mathcal{B}(\PH\to\pa\pa)/\sigma_{\mathrm{SM}}$ for the 2HDM+S Type-I to Type-IV models, as a function of \tanb and $m_{\pa}$, are shown in Fig.~\ref{fig:limits2HDMS}.
For each model type, the pseudoscalar branching fractions $\mathcal{B}(\pa\to\PGm\PGm)$ and $\mathcal{B}(\pa\to\PGt\PGt)$ entering this interpretation are computed following Ref.~\cite{Haisch:2018kqx}.
These limits extend previous LHC results to lower values of $m_{\pa}$ and broaden the excluded range of \tanb.
The contours indicate $\mathcal{B} = 1.0$ and 0.16.
The latter corresponds to the 95\% \CL upper limit on Higgs boson decays to undetected modes extracted from the CMS Higgs boson measurements~\cite{CMS:2022dwd}.  

\section{Summary}
\label{sec:summary}

A search for decays of the Higgs boson, \PH, to a pair of light pseudoscalar bosons, \pa, is presented for both the standard model (SM) Higgs boson at 125\GeV and an SM-like scalar with masses in the range 250--1000\GeV.
The light pseudoscalars decay to $\PGm\PGm$ and $\PGt\PGt$ with substantial overlap within each pair because of the large Lorentz boosts.
This nonstandard topology motivates the development of dedicated $\taum\tauh$, $\taue\tauh$, and $\tauh\tauh$ reconstruction and tagging methods to increase the acceptance.
A dedicated neural network enables, for the first time, the inclusion of the high-rate but high-background $\tauh\tauh$ final state.
Data collected by the CMS experiment at $\sqrt{s} = 13\TeV$, corresponding to an integrated luminosity of 138\fbinv, are examined.
No significant excess over SM processes is observed.
Model-independent upper limits at the 95\% confidence level are set on the product of the Higgs boson production cross section and branching fraction, divided by the standard model cross section, $\sigma_{\PH} \times \mathcal{B}(\PH\to\pa\pa\to\PGm\PGm\PGt\PGt)/\sigma_{\mathrm{SM}}$, with the most stringent limits reaching approximately $5 \times 10^{-5}$ for $m_{\PH} = 125\GeV$ and $3.5 \times 10^{-4}$ for $m_{\PH} = 1000\GeV$.
Model-specific upper limits on $\sigma_{\PH} \times \mathcal{B}(\PH\to\pa\pa)/\sigma_{\mathrm{SM}}$ are also extracted for $m_{\PH} = 125\GeV$ in Type-I, -II, -III, and -IV two-Higgs-doublet plus singlet models.
These results significantly extend the upper limits obtained in earlier searches by the CMS and ATLAS Collaborations and are complementary to existing searches, \eg, Ref.~\cite{Sirunyan:2018mbx}, at higher $m_{\pa}$, where the $\PGm\PGm$ and $\PGt\PGt$ final states are resolved.
Considering only the previous CMS search using boosted $\taum\tauh$ reconstruction and a partial 13\TeV data set~\cite{CMS:2020ffa}, the median expected upper limits on $\sigma_{\PH} \times \mathcal{B}(\PH\to\pa\pa\to\PGm\PGm\PGt\PGt)/\sigma_{\mathrm{SM}}$ for $m_{\PH} = 125\GeV$ are stronger than those obtained in Ref.~\cite{CMS:2020ffa} by factors of 2--5, depending on $m_{\pa}$, owing to the larger dataset and inclusion of  more tau lepton decay channels in the analysis.
In the Type-I and Type-II models, this result places the most stringent limits to date on $\mathcal{B}(\PH\to\pa\pa)$ in the range $3.6 < m_{\pa} < 4.2\GeV$ when $m_{\PH} = 125\GeV$.
In the range $4.0 < m_{\pa} < 8.0\GeV$, the upper limits in this work are stronger than those of Ref.~\cite{CMS:2025hjt} based on the same final state and dataset.
Above 4.2\GeV, they are a bit weaker than those of Ref.~\cite{PerezAdan:2025faj}, which reconstructs the predicted decay of the pseudoscalar to a pair of charm quarks as well as its dimuon decay.
In the Type-III model, in which decays to leptons are enhanced over quarks, the results of this work in the range $13 < m_{\pa} < 15\GeV$ improve upon those of Ref.~\cite{CMS:2024uru}, which reconstructs the predicted decay of the pseudoscalar to a pair of bottom quarks as well as its dimuon decay, and those of Ref.~\cite{CMS:2025hjt}.
Finally, this analysis is the first to search for the decay of a heavy neutral scalar to two promptly decaying pseudoscalars in the full Run 2 dataset of the LHC. 

\begin{acknowledgments}
We congratulate our colleagues in the CERN accelerator departments for the excellent performance of the LHC and thank the technical and administrative staffs at CERN and at other CMS institutes for their contributions to the success of the CMS effort. In addition, we gratefully acknowledge the computing centres and personnel of the Worldwide LHC Computing Grid and other centres for delivering so effectively the computing infrastructure essential to our analyses. Finally, we acknowledge the enduring support for the construction and operation of the LHC, the CMS detector, and the supporting computing infrastructure provided by the following funding agencies: SC (Armenia), BMFWF and FWF (Austria); FNRS and FWO (Belgium); CNPq, CAPES, FAPERJ, FAPERGS, and FAPESP (Brazil); MES and BNSF (Bulgaria); CERN; CAS, MoST, and NSFC (China); MINCIENCIAS (Colombia); MSES and CSF (Croatia); RIF (Cyprus); SENESCYT (Ecuador); ERC PRG and PSG, TARISTU24-TK10 and MoER TK202 (Estonia); Academy of Finland, MEC, and HIP (Finland); CEA and CNRS/IN2P3 (France); SRNSF (Georgia); BMFTR, DFG, and HGF (Germany); GSRI (Greece); MATE and NKFIH (Hungary); DAE and DST (India); IPM (Iran); SFI (Ireland); INFN (Italy); MSIT and NRF (Republic of Korea); MES (Latvia); LMTLT (Lithuania); MOE and UM (Malaysia); BUAP, CINVESTAV, CONACYT, LNS, SEP, and UASLP-FAI (Mexico); MOS (Montenegro); MBIE (New Zealand); PAEC (Pakistan); MSHE, NSC, and NAWA (Poland); FCT (Portugal); MESTD (Serbia); MICIU/AEI and PCTI (Spain); MOSTR (Sri Lanka); Swiss Funding Agencies (Switzerland); MST (Taipei); MHESI (Thailand); TUBITAK and TENMAK (T\"{u}rkiye); NASU (Ukraine); STFC (United Kingdom); DOE and NSF (USA).

\begin{sloppypar}
\setlength\emergencystretch{\hsize}
\hyphenation{Rachada-pisek} Individuals have received support from the Marie-Curie programme and the European Research Council and Horizon 2020 Grant, contract Nos.\ 675440, 724704, 752730, 758316, 765710, 824093, 101115353, 101002207, 101001205, and COST Action CA16108 (European Union); the Leventis Foundation; the Alfred P.\ Sloan Foundation; the Alexander von Humboldt Foundation; the Science Committee, project no. 22rl-037 (Armenia); the Fonds pour la Formation \`a la Recherche dans l'Industrie et dans l'Agriculture (FRIA) and Fonds voor Wetenschappelijk Onderzoek contract No. 1228724N (Belgium); the Beijing Municipal Science \& Technology Commission, No. Z191100007219010, the Fundamental Research Funds for the Central Universities, the Ministry of Science and Technology of China under Grant No. 2023YFA1605804, the Natural Science Foundation of China under Grant No. 12535004, and USTC Research Funds of the Double First-Class Initiative No.\ YD2030002017 (China); the Ministry of Education, Youth and Sports (MEYS) of the Czech Republic; the Shota Rustaveli National Science Foundation (Georgia); the Deutsche Forschungsgemeinschaft (DFG), among others, under Germany's Excellence Strategy -- EXC 2121 ``Quantum Universe" -- 390833306, and under project number 400140256 - GRK2497; the Hellenic Foundation for Research and Innovation (HFRI), Project Number 2288 (Greece); the Hungarian Academy of Sciences, the New National Excellence Program - \'UNKP, the NKFIH research grants K 131991, K 138136, K 143460, K 143477, K 147557, K 146913, K 146914, K 147048, TKP2021-NKTA-64, and 2025-1.1.5-NEMZ\_KI-2025-00004, and MATE KKP and KKPCs Research Excellence and Flagship Research Groups grants (Hungary); the Council of Science and Industrial Research, India; ICSC -- National Research Centre for High Performance Computing, Big Data and Quantum Computing, FAIR -- Future Artificial Intelligence Research, and CUP I53D23001070006 (Mission 4 Component 1), funded by the NextGenerationEU program, the Italian Ministry of University and Research (MUR) under Bando PRIN 2022 -- CUP I53C24002390006, PRIN PRIMULA 2022RBYK7T (Italy); the Latvian Council of Science; the Ministry of Science and Higher Education, project no. 2022/WK/14, and the National Science Centre, contracts Opus 2021/41/B/ST2/01369, 2021/43/B/ST2/01552, 2023/49/B/ST2/03273, and the NAWA contract BPN/PPO/2021/1/00011 (Poland); the Funda\c{c}\~ao para a Ci\^encia e a Tecnologia (Portugal); the National Priorities Research Program by Qatar National Research Fund; MICIU/AEI/10.13039/501100011033, ERDF/EU, ``European Union NextGenerationEU/PRTR", projects PID2022-142604OB-C21, PID2022-139519OB-C21, PID2023-147706NB-I00, PID2023-148896NB-I00, PID2023-146983NB-I00, PID2023-147115NB-I00, PID2023-148418NB-C41, PID2023-148418NB-C42, PID2023-148418NB-C43, PID2023-148418NB-C44, PID2024-158190NB-C22, RYC2021-033305-I, RYC2024-048719-I, CNS2023-144781, CNS2024-154769 and Plan de Ciencia, Tecnolog{\'i}a e Innovaci{\'o}n de Asturias, Spain; the Chulalongkorn Academic into Its 2nd Century Project Advancement Project, the National Science, Research and Innovation Fund program IND\_FF\_68\_369\_2300\_097, and the Program Management Unit for Human Resources \& Institutional Development, Research and Innovation, grant B39G680009 (Thailand); the Eric \& Wendy Schmidt Fund for Strategic Innovation through the CERN Next Generation Triggers project under grant agreement number SIF-2023-004; the Kavli Foundation; the Nvidia Corporation; the SuperMicro Corporation; the Welch Foundation, contract C-1845; and the Weston Havens Foundation (USA).
\end{sloppypar}
\end{acknowledgments}\section*{Data availability} Release and preservation of data used by the CMS Collaboration as the basis for publications is guided by the  \href{https://doi.org/10.7483/OPENDATA.CMS.1BNU.8V1W}{CMS data preservation, re-use and open access policy}.
\bibliography{auto_generated}

\cleardoublepage \appendix\section{The CMS Collaboration \label{app:collab}}\begin{sloppypar}\hyphenpenalty=5000\widowpenalty=500\clubpenalty=5000\cmsinstitute{Yerevan Physics Institute, Yerevan, Armenia}
{\tolerance=6000
A.~Gevorgyan\cmsorcid{0000-0003-2751-9489}, A.~Hayrapetyan, V.~Makarenko\cmsorcid{0000-0002-8406-8605}, A.~Tumasyan\cmsAuthorMark{1}\cmsorcid{0009-0000-0684-6742}
\par}
\cmsinstitute{Marietta Blau Institute for Particle Physics, Vienna, Austria}
{\tolerance=6000
W.~Adam\cmsorcid{0000-0001-9099-4341}, L.~Benato\cmsorcid{0000-0001-5135-7489}, T.~Bergauer\cmsorcid{0000-0002-5786-0293}, M.~Dragicevic\cmsorcid{0000-0003-1967-6783}, S.~Gundacker\cmsorcid{0000-0003-2087-3266}, A.K.~Guven\cmsorcid{0009-0004-5670-5138}, P.S.~Hussain\cmsorcid{0000-0002-4825-5278}, M.~Jeitler\cmsAuthorMark{2}\cmsorcid{0000-0002-5141-9560}, N.~Krammer\cmsorcid{0000-0002-0548-0985}, A.~Li\cmsorcid{0000-0002-4547-116X}, D.~Liko\cmsorcid{0000-0002-3380-473X}, M.~Matthewman, J.~Schieck\cmsAuthorMark{2}\cmsorcid{0000-0002-1058-8093}, R.~Sch\"{o}fbeck\cmsAuthorMark{2}\cmsorcid{0000-0002-2332-8784}, M.~Shooshtari\cmsorcid{0009-0004-8882-4887}, M.~Sonawane\cmsorcid{0000-0003-0510-7010}, N.~Van~Den~Bossche\cmsorcid{0000-0003-2973-4991}, W.~Waltenberger\cmsorcid{0000-0002-6215-7228}, C.E.~Wulz\cmsAuthorMark{2}\cmsorcid{0000-0001-9226-5812}
% ...
\par}
$^{1}$Also at Yerevan State University, Yerevan, Armenia\\
$^{2}$Also at Technische Universit\"{a}t Wien, Vienna, Austria\\
% ...
\end{sloppypar}
\end{document}